\documentclass[]{jfm}

\usepackage{graphicx}
\usepackage{newtxtext}
\usepackage{newtxmath}
\usepackage{romanbar}
\usepackage{natbib}
\usepackage{hyperref}
\hypersetup{
	colorlinks = true,
	urlcolor   = blue,
	citecolor  = black,
}

\newcommand{\RomanNumeralCaps}[1]

\usepackage[dvipsnames]{xcolor}
\usepackage{amsmath}
\usepackage{amsfonts}
\usepackage{mathrsfs}
\usepackage{amssymb}
\usepackage{relsize}
\usepackage{cancel}
\usepackage{siunitx}
\usepackage{float}
\usepackage{enumitem}
\usepackage{tikz}
\usepackage{pgfplots}
\pgfplotsset{compat=1.17}
\usepackage{import}
\usepackage{verbatim}
\usepackage{subcaption}
\newcommand{\B}[1]{\mathbf{#1}}
\newcommand{\DPS}{\displaystyle}

\newcommand{\Derp}   [2] { \frac { \partial #1 } { \partial #2 } }

\newcommand{\eref}[1] {Equation~(\ref{#1})}
\newcommand{\fref}[1] {Figure~\ref{#1}}
\newcommand{\secref}[1] {\S~\ref{#1}}
\newcommand{\tabref}[1] {Table~\ref{#1}}

\title{Three-dimensional global stability analysis of turbulent screeching jets}

\author{Alessandro Franchini\aff{1}
	\corresp{\email{a.franchini14@gmail.com}},
	Nicolas Alferez\aff{2} \and Jean-Christophe Robinet\aff{1}}

\affiliation{\aff{1}DynFluid Laboratory, Arts et Métiers Institute of Technology, \\
	151 Bd. de L'Hopital, 75013 Paris, France.
	\aff{2}DynFluid Laboratory, Conservatoire national des arts et métiers, \\
	292 Rue Saint-Martin, 75003, Paris, France.}

\begin{document}
	\maketitle

\begin{abstract}
A three dimensional global stability analysis is performed to investigate the problem of screeching jets under turbulent conditions. The study employs an Unsteady Reynolds-Averaged Navier--Stokes (URANS) framework, in which the compressible flow equations are discretised using the high-fidelity solver dNami, the linearised discrete system is obtained through the automatic differentiation tool Tapenade, and the global stability problem is solved in a time-stepping framework.
The fixed-point solutions of the URANS equations are first validated against experimental and numerical data, then a three dimensional global stability analysis is performed around fixed points solutions at different levels of under-expanded regimes.
The extracted modes are spatially analysed and examined in terms of acustic radiation and validated against experimental data.
Comparison with experimental POD data shows that the linear modes reproduce the main wavenumber content and spatial organisation of the screech resonance loop, even at high levels of under-expansion. The staging behaviour is also recovered from the interaction between the Kelvin--Helmholtz wave and the dominant wavenumbers of the shock-cell structure.
Finally, a Helmholtz decomposition is applied to the velocity perturbation in order to separate the vortical and irrotational parts of the modes. An energy budget of the wave components is then used to quantify the repartition of the relative feedback-loop energy perturbation. Notably, the Mach-number effects on energy partition vary depending on the type of staged mode. This insight could prove valuable for interpreting receptivity mechanisms at nozzle lips and shocks in future research.

\end{abstract}


	\section{Introduction} \label{sec:introduction}
	When jets are in an under-expansion regime, i.e. when the pressure at the nozzle exit is higher than the ambient pressure, a shock-cell structure forms outside the jet. The interaction of the shear layer with this structure can lead to various sources of noise. Within this family of noises, the screech tone, an intense monofrequency noise with pronounced upstream directivity, has been identified and extensively studied over the last fifty years.
	The first attempt at explaining the mechanism underlying the screech tone generation comes from the work of~\citet{powell1953}, where he proposes that screech arises from an acoustic feedback mechanism in which the downstream propagating hydrodynamic instability that generates in the near-nozzle shear layer, (Kelvin-Helmholtz type instability), is amplified and interacts with the quasi-periodic shock-cell structure present in the potential core of underexpanded jets. This interaction generates a free-stream acoustic wave that propagates upstream, exciting the near-lip shear layer and closing the feedback loop. If at each stage of the resonance the starting energy of the perturbation is not dissipated, the process is self-sustained, and the screech tone is observed.
	Powell then formalised this mechanism by introducing two criteria for the feedback loop to be self-sustained:
	A phase criterion and a gain criterion. Following the notation of \citet{EdgingtonMitchell2019}, the phase criteria can be defined mathematically as:
	\begin{equation}\label{eq:phase_criteria}
		\frac{N}{f} = \frac{h}{U_1} + \frac{h}{U_2} + \psi
	\end{equation}
	where $U_1$ is the average speed of the downstream propagating perturbation, $U_2$ is the average speed of the upstream propagating perturbation, $h$ is the estimated distance in which the interaction between the downstream and upstream process happens, $N$ is the total number of upstream and downstream propagating disturbances at any moment, and $\psi$ accounts for any delay that might occur during the transfer mechanism. This criterion formalises the fact that for a feedback loop to be self-sustained, the phases of the downstream and upstream travelling waves must be synchronised so that a precise acoustic frequency $f$ arises from the interaction.
	The gain criteria reads:
	\begin{equation}\label{eq:Powell_gain}
		q_d \eta_g \eta_u \eta_r \ge 1 ,
	\end{equation}
	where $q_d$ is the spatial gain of the downstream process, $\eta_g$ represents the efficiency of the energy transmission from the downstream disturbance to the upstream one, $\eta_u$ is the efficiency coefficient that models the dissipation of energy during the upstream propagation, and $\eta_r$ is the efficiency of the receptivity mechanism.
	The gain criteria impose that the dissipation between wave interactions is small enough so that the resonance has sufficient energy to be self-sustained.
	Adding the observation of maximum upstream directivity, Powell arrives at the following formula to predict screech frequency:
	\begin{equation}\label{eq:powell}
		f_s = \frac{U_c}{s(1+M_c)},
	\end{equation}
	where $U_c$ is the phase velocity of the downstream disturbance, $s$ is the shock spacing, and $M_c$ is the convective Mach number of the downstream disturbance.
	While Powell’s formulation provides a robust phenomenological framework and a useful prediction of the screech frequency, it does not explicitly describe the interaction between instability waves and the shock-cell system. This limitation motivated alternative interpretations based on broadband shock-associated noise.

	In the work of \cite{Tam1986}, the authors proposed the weakest link feedback model as an alternative explanation of the mechanism underlying screech tone. The idea is based on the intuition that screech appears as a special case of broadband shock-associated noise (BBSAN). BBSAN belongs to a family of broadband noise present in supersonic jets \citep{tam1995}  that arises from the interaction of downstream propagating perturbations and the shock-cell structure of the jet.  One characteristic of BBSAN is that, within the broadband frequency spectrum, it presents a peak that shifts to lower values when the observer angle $\theta$ (with respect to the jet centerline) moves upstream the jet, reaching maximum amplitude at $\theta = 180^{\circ}$ (upstream).
	The sound generation mechanism of BBSAN has been described via a complete three-dimensional analysis in \cite{tam1987stochastic}. For simplicity, here is reported a one-dimensional model of the BBSAN frequency peak based on the Fourier decomposition components of the shock-cell structure $k_{s_n}$, following the notations of \citet{EdgingtonMitchell2021} and \citet{tam1995}.

	Based on the vortex-sheet approach of \citet{prandtl1904} and \cite{Pack1950}, the spatial distribution of the pressure oscillations of the shock-cell structure induces a modulation of the velocity $u_s$ that can be expressed as:
	\begin{equation}\label{eq:wave_fp}
		u_s=\sum_{n=1}^{\infty} A_n \left[(\exp(i k_{s_{n}} x) + \exp(-i k_{s_{n}}x)\right],
	\end{equation}
	where $A_n$ is the amplitude of the shock-cell modes, and $k_{s_n}$ are the associated wave numbers.
	The details of the calculation are omitted here and can be found in the original work by Tam \citep{tam1987stochastic}, but the authors arrived at the following formula for the peak frequency of BBSAN:
	\begin{equation}\label{eq:fbbsan}
		f_{BBSAN} = \frac{U_c k_{s_n}}{2 \pi(1 - M_c \cos(\theta))}, \qquad n = 1,2,3,\dots
	\end{equation}
	This formula describes the frequency peak of BBSAN with respect to the observer angle $\theta$ measured from the downstream direction. If the first shock-cell Fourier wave number $k_{s_1}$ is considered ($k_{s_1}$ is the wave number linked to the shock-cell spacing $s$) and when the observer direction tends to the upstream direction ($\theta \rightarrow \pi$), equation \eqref{eq:fbbsan} tends to \eqref{eq:powell}, showing that screech can be seen as a particular case of BBSAN.
	
	If the large turbulence structures that interact with the shocks are modeled as a downstream-travelling Kelvin-Helmholtz (KH) wave-packet, they can be expressed as:
	\begin{equation}
		u_{kh}=\psi(x,r) \exp(i k^+_{kh}x - \omega t)
	\end{equation}
	where $\psi(x,r)$ represents the wave amplitude, $k^+_{kh}$ the KH wave number and $\omega$ the angular frequency. As in \citet{Tam_tanna1982}, the interactions between these wave structures are represented by the product of the two wave expressions. This concept have been clearly presented in \cite{EdgingtonMitchell2021}, where the interaction of the KH wave-packet with the first shock-cell mode of wave number $k_{s_1}$ is expressed as:
	\begin{equation}\label{eq:wave_sum}
		u_{kh}u_s \varpropto \exp\left(i (k^+_{kh} +  k_{s_1})x - i \omega t\right) + \exp\left(i (k^+_{kh} - k_{s_1})x - i \omega t\right)
	\end{equation}
	From this expression is clear that two wave-like disturbances of wave number $k^+_{kh}+k_{s_1}$ and $k^+_{kh}-k_{s_1}$ appears. If the KH wave number is smaller than $k_{s_1}$, the resulting disturbance has negative phase velocity. Following the notation from \citet{EdgingtonMitchell2018,towne2017acoustic,schmidt2017wavepackets,EdgingtonMitchell2021}, the positive phase velocity wave is referred to $k_t^+ = k^+_{kh}+k_{s_1}$, while the negative phase velocity wavepacket is $k^-=k^+_{kh}-k_{s_1}$.

	As mentioned earlier, by considering the screech tone as a particular case of BBSAN, the prediction formula is the same as Powell's, but the explanation mechanism for frequency selection differs and involves a weak interaction between KH wave packets and the first wave number of the shock-cell structure.
	This explanation is consistent for some jet configurations, but it is unable to correctly predict the mode-staging phenomenon that has been observed in circular jets in many experimental studies~\citep{powell1953,sherman1976jet,davies1962tones,davies1962tones2,merle1956}.

	Mode staging is a phenomenon in which a sudden frequency shift appears with small variation of the jet's nozzle pressure ratio. This frequency shift can also be linked to a change in the azimuthal wave number of the acoustic wave.
	For a low level of under-expansion, the screech phenomenon experiences two toroidal modes (A1 and A2) that are axisymmetric. In contrast, for a higher level of under-expansion, screech experiences a flapping/helical mode denoted by B and C modes \citep{Norum_83,powell1992observations,edgington2015staging}.
	A first explanation of the staging behaviour was proposed in \citet{ShenTam2002}, where the authors suggested that the staging behaviour could be linked to different ways of closing the acoustic feedback loop: one with a free stream acoustic wave and the other via an upstream travelling neutral acoustic mode that was already described in previous works from Tam~\citep{Tam_tanna1982,Tam1986,Tam_Hu_1989}.
	It is generally agreed that the neutral acoustic mode, also known as guided-jet mode (G-JM) or $k^-$ is responsible for closing the screech feedback loop \citep{EdgingtonMitchell2018,nogueira2024guided}. Recently, analysing the wave structures of experimental data, \cite{nogueira2022closure} found that the guided jet mode responsible for the closure of the A1 mode appears from the interaction between the KH wave $k^+_{kh}$ and the first wave number $k_{s_1}$ of the shock-cell structure, while the G-JM that closes the A2 modes comes from the interaction with the second wave number $k_{s_2}$. This results have been generalised for the B and C modes \citep{edgington2022unifying} so that two family of modes can be defined: The modes A1 and C are generated from the interaction $k^+_{kh} - k_{s_1}$, while the modes A2 and B by the interaction $k^+_{kh} - k_{s_2}$. These results suggest that mode staging in screeching jets is governed by a change in the dominant shock-cell harmonic participating in the interaction with the Kelvin--Helmholtz wave packet, leading to distinct families of resonant modes characterised by different azimuthal symmetries.

	Historically, the upstream propagating wave responsible for the closure mechanism was believed to be a free-stream acoustic wave. Eventually, recent works proved that the G-JM was the main responsible for the upstream closure mechanism, and  proved its importance in many resonant mechanisms like subsonic impinging jets~\citep{tam1990theoretical}, supersonic impinging jets~\citep{bogey2017feedback,jaunet2019dynamics}, installed jets \citep{jordan2018jet} and also screeching jets \citep{mancinelli2019screech,EdgingtonMitchell2018,mancinelli2021complex,gojon2018oscillation}.
	The guided jet mode was first described through local stability analysis in \citep{Tam_Hu_1989}, and had already been visualised before in \citep{oertel1980mach}. This wave belongs to the subsonic family of waves described in \citet{Tam_Hu_1989} that appears over an extensive range of Mach number configurations. The range of frequencies in which these waves propagate upstream is narrow, corresponding to the frequencies of many of the resonant mechanisms in which they participate. The importance of the G-JM was also highlighted in the studies of \citep{bogey2021acoustic} within a high-fidelity simulation framework. In a recent work \citep{nogueira2024guided}, the nature and characteristics of guided-jet waves have been extensively studied and analysed, elucidating many of the mechanisms that regard this family of waves, an important one being their upstream propagation behaviour.

	\subsection{Stability analysis of turbulent flows and the screech phenomenon}
	Many of the insights on screech come from extensive experimental studies conducted over the last century, encompassing a vast range of jet configurations, with the most important early work being that of Powell \citep{powell1953,Powell_1953,APowell_1953}. A series of detailed study on the screeching modes, the dependence on nozzle geometry and on the noise amplitude have been conducted by the group at NASA Langley Research Center, in which a large amount of experimental data have been collected in a series of studies (a list of the most important experiments can be found in the review from \citet{raman1998}).
	In the meantime, significant advancements in the analytical modelling of screech have been pursued by Tam and coauthors in a series of papers regarding wave propagation in supersonic jets \citep{Tam_tanna1982,tam1985,Tam1986}. Within the context of numerical simulation, a series of Unsteady RANS simulations has been conducted, aiming at exploring the behaviour of the axisymmetric modes of screeching jets \cite{shen1998numerical,ShenTam2002}. In another work \citep{shen2000}, the authors studied the effect of jet temperature on screech frequency. In the work of \cite{berland2007}, the authors explored the sound generation mechanism of screech, known as shock leakage, using a high-fidelity LES approach on rectangular jets. This mechanism has been studied extensively from experimental observations in \citet{EdgingtonMitchell2021_shockleakage}.

	Stability analysis has provided valuable insights into the comprehension of screech, particularly in describing the amplification behaviour of the downstream propagating Kelvin-Helmholtz perturbation (see the review in \cite{EdgingtonMitchell2019}). In \citet{beneddine2015}, global stability analysis has been applied to a two-dimensional laminar jet, recovering global modes that presented some of the characteristics of screech.
	Recently, in the work of \citet{EdgingtonMitchell2021}, the authors showed that global stability analysis performed over the mean flow could correctly predict the wave number properties of the $k^-, k_t^+$ and $k^+_{kh}$.

	Stability analysis can be an efficient tool for reconstructing coherent fluctuations, and it is recently being used more and more within turbulent flows. Since a proper fixed-point solution of the turbulent Navier-Stokes equations cannot be defined due to their chaotic behaviour, most recent studies are based on the mean flow as a base flow over which the Navier-Stokes equations are linearised. Different approaches can be chosen depending on how the Reynolds-stress tensor is treated after the average operation. If no eddy-viscosity is employed, the Reynolds stresses are neglected, and stability analysis is performed using the laminar linear Navier-Stokes equations. These equations are obtained by linearising over a mean field that can be extrapolated from high-fidelity data simulation or experimental data \cite{schmidt2017wavepackets,EdgingtonMitchell2021,hildebrand2015simulation}. If the non-linear terms are not neglected, under some conditions of validity \citep{beneddine2016},  they can be modelled as an external forcing to perform classical resolvent analysis over the mean flow \citep{mckeon2010critical}.
	This approach is widely used due to its simplicity and can provide good results in practical cases \citep{EdgingtonMitchell2021}. However, it lacks a methodological formal background, as the mean flow is not a fixed point solution of the Navier-Stokes equations.
	This inconsistency can influence the stability analysis result, specifically the linear growth rate of the stability mode \citep{barkley2006}. Moreover, in the work of \citet{karban2020ambiguity}, the authors found that resolvent analysis over the mean field of a turbulent flow can yield different results depending on whether the equations are written in conservative or primitive form.
	In addition, from a physical standpoint, it does not discriminate between the effects of coherent and non-coherent fluctuations as in the triple decomposition framework \citep{reynolds1972mechanics}.
	To overcome these limitations, an eddy-viscosity-based approach can be adopted, in which the turbulent closure is explicitly included in the linearised equations. If the eddy-viscosity is not perturbed, the method is called \textit{frozen-eddy-viscosity} \citep{cossu2009optimal,del2006linear,penet:thesis}, that is, the turbulent eddy-viscosity is employed in the linearised equations, but it is considered stationary and not perturbed. This approach yields superior results compared to the no-eddy-viscosity approach, with minimal effort required for modelling the eddy-viscosity \citep{alizard2015optimal,alizard2019restricted}.

	The approach followed in this work is an extension of the former. The averaged Navier-Stokes equations are closed via an eddy-viscosity model, forming the URANS system of equations. Then, the URANS equations are linearised, together with the turbulence closure model, and the eddy-viscosity is perturbed in the linear analysis. The mean flow calculated within this framework is then a fixed point of the URANS system, allowing us to study the linear behaviour of the large-scale turbulent structures and their interactions with the turbulent base flow.
	Firstly employed by \citet{crouch2007,crouch2009origin} to predict the onset of buffet in transonic flow, it is also applied on a larger scale in \cite{sansica2023} and in an over-expanded nozzle in \cite{morisco2020nonlinear}.
	Recently, \cite{Sarras2024} employed a data-driven approach is employed to enhance the eddy-viscosity modelling. In this work the goal is to combine the use of automatic differentiation over a turbulence model to perform a fully matrix free three dimensional global stability analysis on compressible turbulent flows with strong shocks.

	In \secref{sec:methodology}, the mathematical and computational framework for global stability analysis over URANS modelling is detailed, focusing on the construction of the Jacobian in a general setting. This description outlines the mathematical basis of the analysis, along with the chosen computational strategy. The physical and numerical configurations are presented in \secref{sec:configuration}, along with an overview of the different pressure ratios studied. \secref{sec:fixedpoint} focuses on the URANS fixed point solutions, calculated via Newton-Krylov methods, over which global stability analysis is performed. In \secref{sec:stability}, results from global stability analysis over a three-dimensional turbulent fixed point of the URANS model are presented, focusing on the spatial distribution and azimuthal symmetries of the found modes. In \secref{sec:waves}, the waves that co mpose the screech modes are analysed by looking at the spectral signature of the complex mode, revealing the downstream and upstream propagating waves, their link to the spectral structure of the fixed point shocks and an analysis of the staging behaviour. In \secref{sec:helmholtz_waves}, an Helmholtz decomposition is performed and the relative energy of the feedback loop wave component is analysed.

	\section{Methodology} \label{sec:methodology}
	\subsection{Global stability analysis of URANS equations}
	In this section, the stability of a turbulent flow is described within the framework of the Unsteady RANS (URANS) equations.
	The Navier-Stokes equations are formally written under the conservative form:
	\begin{equation}\label{eq.NS}
		\frac{\partial \B{q}}{\partial t} = \mathcal{N}(\B{q})
	\end{equation}
	where $\mathcal{N}$ is the compressible Navier-Stokes operator and $ \B{q} = \left[ \rho, \rho \B{u}, \rho E\right]^T $  is the state vector of composed of the fluid density $\rho$, the vector of velocity components in the three Cartesian directions $\mathbf{u}=\left[u,v,w\right]^T$ and the total energy of the fluid $E$. 
	The set of physical quantities is decomposed as the sum of a Reynolds-Favre average $\B{\widetilde{q}}(\B{x},t) $ and a finite amplitude fluctuation $ \B{q}^*(\B{x},t)$:
	\begin{equation}\label{eq.DRF}
		\B{q}(\B{x},t) = \B{\widetilde{q}}(\B{x},t) + \B{q}^*(\B{x},t).
	\end{equation}
	This decomposition verifies the usual Reynolds-Favre rules \citep{wilcox1993}.
	Here, it should be noted that the mean flow is allowed to be time-dependent, modeling slow-varying properties of the fluid that are not turbulent in nature. The conditions for the URANS equations to be well posed require that the characteristic time scales of the large-scale structures be much longer than those of the unresolved small-scale structures, whose effects are modelled by the turbulent eddy-viscosity. More precisely, let $T_U$ denote the time scale of the unsteady mean-flow oscillations and $T_t$ the time scale of the turbulent fluctuations. If $T$ is the integration time scale of the averaging procedure that defines an unsteady RANS solution, one must have $T_t \ll T \ll T_U$ for the URANS framework to be well posed~\citep{wilcox1993}.
	In doing so, it is implicitly assumed that the time scales of the turbulent fluctuations $T_t$ are much smaller than those of the mean flow, $T_U$.
	In this sense, the average $\widetilde{\bullet}$ is not a long-time average but a form of temporal low-pass filtering allowing the separation of turbulence scales from other scales, generally low frequencies, not directly linked to turbulence.  Whithin these assumptions, \eref{eq.DRF} is injected into \eref{eq.NS}, and the resulting equations are averaged with the usual Reynolds-Favre average, obtaining the URANS equations:
	\begin{equation}
		\Derp{\widetilde{\B{q}}}{t}=\mathcal{N}(\widetilde{\B{q}}) + \widetilde{\mathcal{R}(\B{q}^*,\B{q}^*)}.
	\end{equation}
	The various Reynolds-Favre tensors that are represented by the spatio-temporal varying operator $\widetilde{\mathcal{R}(\B{q}^*,\B{q}^*)}$ are assumed to be proportional to the mean-field strain tensor, according
	to the Boussinesq relation. The proportionality coefficient is a scalar with the dimensions of a viscosity,
	called turbulent viscosity $\mu_t$:
	\begin{equation}\label{eq:boussinesq}
		\widetilde{\mathcal{R}(\B{q}^*,\B{q}^*)} = -\mu_t \mathcal{H}(\widetilde{\B{q}},\boldsymbol{\phi}),
	\end{equation}
	where $\mathcal{H}$ represent the operator that links turbulence fluctuation to mean flow gradients and $\boldsymbol{\phi}$ represents the vector of turbulence variables. The URANS equations are then written:
	\begin{equation}\label{eq.URANS}
		\Derp{\widetilde{\B{q}}}{t}= \mathcal{N}(\widetilde{\B{q}}) -\mu_t \mathcal{H}(\widetilde{\B{q}},\boldsymbol{\phi})=\mathcal{F}(\widetilde{\B{q}},\boldsymbol{\phi}) .
	\end{equation}
	These equations are closed by one or more transport equations of the turbulent vector field $\boldsymbol{\phi}$, (e.g. in the case of a $k-\omega$ turbulence model $\boldsymbol{\phi} = \left(\rho k, \rho \omega\right)^T$ ). These closure equations are written as:
	\begin{equation}\label{eq.TE}
		\frac{\partial \boldsymbol{\phi}}{\partial t} = \mathcal{G}(\widetilde{\B{q}},\boldsymbol{\phi}),
	\end{equation}
	where $\mathcal{G}$ is the operator that represents the turbulence model, and that usually contains the turbulent production, dissipation and diffusion terms for the quantity $\boldsymbol{\phi}$. The turbulence variable vector $\boldsymbol{\phi}$ is then linked to the turbulent eddy-viscosity $\mu_t$ via an algebraic expression that can be written as:
	\begin{equation}\label{eq.Mmut}
		\mu_t = \mathcal{M}(\widetilde{\B{q}},\boldsymbol{\phi}).
	\end{equation}
	So now the URANS system with turbulence model reads:
	\begin{equation}\label{eq:URANS}
		\begin{aligned}
			\begin{cases}
				\DPS \frac{\partial \widetilde{\B{q}}}{\partial t} = \mathcal{F}(\widetilde{\B{q}},\boldsymbol{\phi}) \\[10pt]
				\DPS \frac{\partial \boldsymbol{\phi}}{\partial t} =\mathcal{G}(\widetilde{\B{q}}, \boldsymbol{\phi}).
			\end{cases}
		\end{aligned}
	\end{equation}

		Within this framework, the large-scale structures are resolved, the smaller-scale structures are modelled via a turbulent viscosity, and the dynamics of the large-scale structures can be analysed linearly using global stability analysis.

	It is from the URANS system (\ref{eq:URANS}) that the stability problem for a fully turbulent flow is formulated.
	The state vector is written using the following decomposition:
	\begin{equation}
		[\widetilde{\B{q}},\boldsymbol{\phi}]^T=  [\B{q}_0,\boldsymbol{\phi}_0]^T+ [\B{q}',\boldsymbol{\phi}']^T
	\end{equation}
	where $[\B{q}',\boldsymbol{\phi}']^T$ is an infinitesimal perturbation and $[\B{q}_0,\boldsymbol{\phi}_0]^T$ is a stationary fixed point solution of \eref{eq:URANS} which corresponds to a statistically steady turbulent mean field.
	In this context, the perturbation vector $[\B{q}',\boldsymbol{\phi}']^T$ does not model the turbulent fluctuations but the linear growth and oscillation frequency of some temporally self-sustained instabilities \citep{reynolds1972mechanics}.
	This perturbation vector $[\B{q}',\boldsymbol{\phi}']^T$ is a solution of the linearised URANS system, which is obtained by developing the first-order Taylor expansion of \eref{eq:URANS} and neglecting the higher-order terms.
	Formally, the linearised URANS system reads:
	\begin{equation}\label{eq1:linearurans}
		\begin{aligned}
			\begin{cases}
				\DPS \frac{\partial \mathbf{q}'}{\partial t} = \frac{\partial \mathcal{F}}{\partial \mathbf{q}}(\mathbf{q}_0,\boldsymbol{\phi}_0) \mathbf{q}' + \frac{\partial \mathcal{F}}{\partial \boldsymbol{\phi}}(\mathbf{q}_0,\boldsymbol{\phi}_0) \boldsymbol{\phi}' \\[10pt]
				\DPS \frac{\partial \boldsymbol{\phi}'}{\partial t} =\frac{\partial \mathcal{G}}{\partial \mathbf{q}}(\mathbf{q}_0,\boldsymbol{\phi}_0) \mathbf{q}' + \frac{\partial \mathcal{G}}{\partial \boldsymbol{\phi}}(\mathbf{q}_0,\boldsymbol{\phi}_0) \boldsymbol{\phi}'
			\end{cases}
		\end{aligned}
	\end{equation}
	where the Jacobian operator $\mathcal{J}$ of system (\ref{eq1:linearurans}) reads:
	\begin{equation}
		\mathcal{J} =
		\begin{pmatrix}
			\displaystyle &\frac{\partial \mathcal{F}}{\partial \mathbf{q}}(\mathbf{q}_0,\boldsymbol{\phi}_0) \ &\frac{\partial \mathcal{F}}{\partial \boldsymbol{\phi}}(\mathbf{q}_0,\boldsymbol{\phi}_0) \\[10pt]\displaystyle &\frac{\partial \mathcal{G}}{\partial \mathbf{q}}(\mathbf{q}_0,\boldsymbol{\phi}_0) \ &\frac{\partial \mathcal{G}}{\partial \boldsymbol{\phi}}(\mathbf{q}_0,\boldsymbol{\phi}_0)
		\end{pmatrix}
	\end{equation}
	Since the problem is described by a system of ordinary partial differential equations, starting from an
	initial condition $\left[ \B{q}'(\B{x},0), \boldsymbol{\phi}'(\B{x},0) \right]^T $,
	the perturbation at time $t$ is given by:
	\begin{equation} \label{linearsys}
		\begin{pmatrix}
			\DPS  \B{q}'(\B{x},t) \\[10pt]
			\DPS  \boldsymbol{\phi}'(\B{x},t)
		\end{pmatrix}
		= e^{\mathcal{J}t}
		\begin{pmatrix}
			\DPS  \B{q}'(\B{x},0) \\[10pt]
			\DPS  \boldsymbol{\phi}'(\B{x},0)
		\end{pmatrix}
	\end{equation}
	By looking at the spectral information (eigenvalues and eigenvectors) of the Jacobian matrix, one can deduce the asymptotic stability property of the fixed point solution $[\B{q}_0,\boldsymbol{\phi}_0]^T$, as discussed in the previous section.
	
	Using this methodology, the present work aims to study the screech phenomenon through global stability analysis, to describe the spatial distribution of acoustic modes, to recover the staging effect observed in experiments, and to examine the waves that compose the feedback loop.

	In practice, especially when employing turbulence modelling for eddy-viscosity $\mu_t$, turbulent kinetic energy $k$ or dissipation ratio $\omega$, the terms in the operators $\mathcal{F}$ and  $\mathcal{G}$ are highly non-linear and, depending on the turbulence model, could even be non-differentiable. It means that the actual derivation of the Jacobian operator $\mathcal{J}$ could be tedious and must take into account non-differentiability.
	For these reasons, the approach of first discretising \eref{eq.URANS} and then performing algorithmic differentiation have been chosen, in order to have the discretised linear solver without having to derive analytically the linearised equations.

	\subsection{Computational Framework for Stability Analysis} \label{sec:computational}
	To perform three-dimensional global stability analysis of turbulent screeching jets under URANS modelling, a matrix-free framework for eigenvalue problems has been developed.
	For the computation of fixed point solutions and three-dimensional global stability analysis, a multi-language automatic differentiation framework has been used. The core of the computational framework consists of a nonlinear generative solver \texttt{dNami} (see \cite{dNami,winn2023two} for detailed information about the code), an automatic differentiation tool \texttt{Tapenade}, \citep{TapenadeRef13} and a linear stability analysis module \texttt{liNami}.
	\texttt{dNami} is a domain-specific language framework that produces solvers for  arbitrary spatio-temporal partial differential equation systems. C and FORTRAN source code are generated from symbolic descriptions of the physical problem. The framework employs a lightweight, ad hoc syntax that utilizes character strings within a Python environment. Gradients can be automatically discretized using arbitrary-order finite differences. Boundary conditions are provided with the same flexibility as for the main equations. General multiblock structured grid meshes are supported owing to \texttt{dNami}’s integration within the open-source Pre-Post processing tools \texttt{CASSIOPEE}~\citep{benoit2015cassiopee} developed at ONERA. This method is particularly useful for solving problems that require a complex, structured multiblock grid to model the physical problem correctly.
	Once the nonlinear discretised code is generated with \texttt{dNami}, the differentiated source code is built using the source transformation automatic differentiation tool \texttt{Tapenade}. This procedure allows access to the source code that solves the linearised problem defined within the \texttt{dNami} framework.
	The module \texttt{liNami} then translates the linear code from \texttt{Tapenade} and creates a module that iteratively solves the linear discretised equations in time within the optimised HPC framework of \texttt{dNami}. The module also provides a set of functions for solving fixed-point and eigenvalue problems.

\begin{figure}
		\centering
		\includegraphics[width=0.7\linewidth,trim=0 50 0 0,clip]{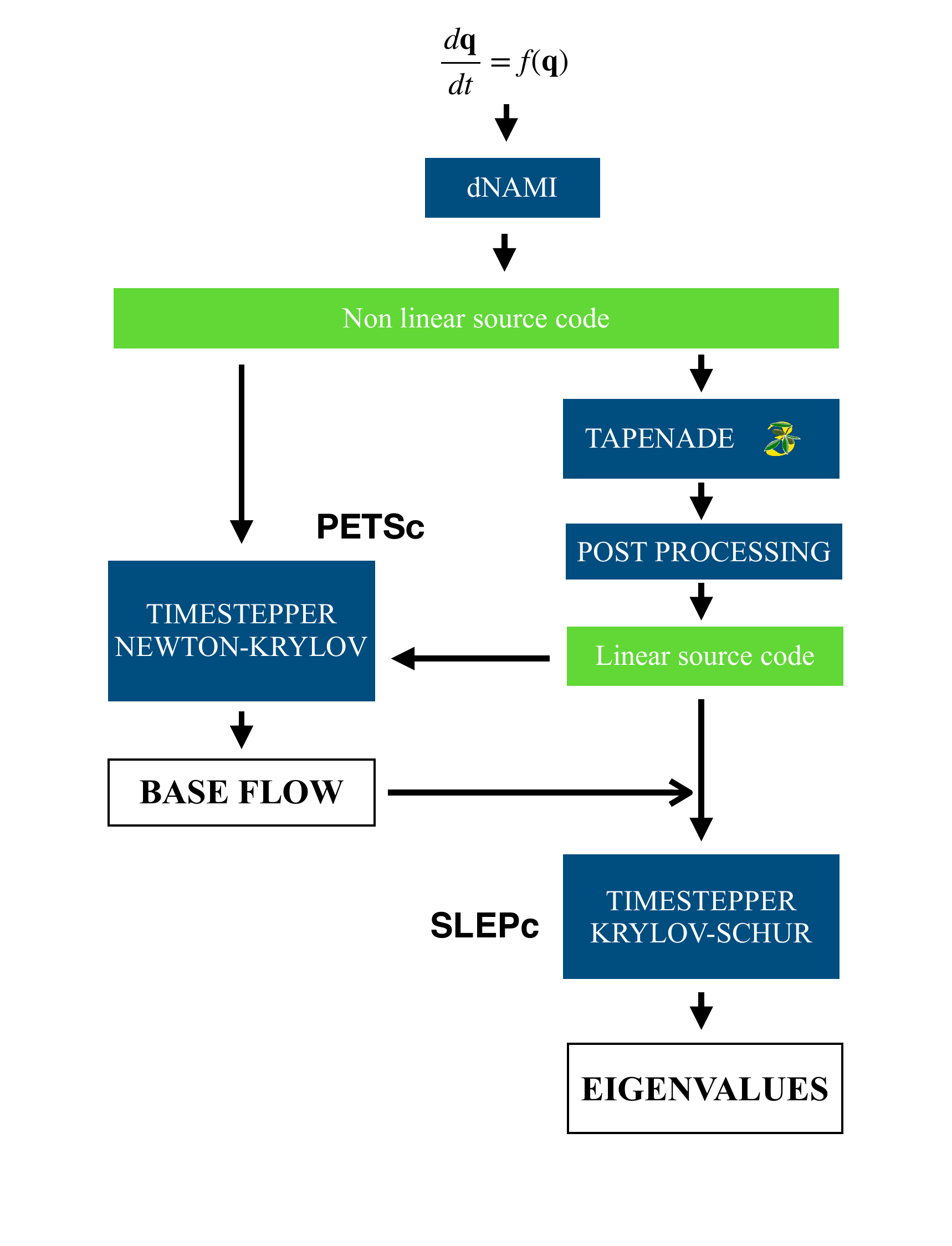}
		\caption{Scheme of the Automatic Differentiation Framework for Stability Analysis. The scheme must be read from top to bottom. A general dynamical system is written in symbolic Python within the solver \texttt{dNami}; the nonlinear source code is generated and used within \texttt{Tapenade} to generate the linear source code. Then, combining the linear and nonlinear source code, the base flow is calculated via a timestepper Newton-Krylov method. Injecting the base flow into the linear source code and using a timestepper Krylov-Schur algorithm, the eigenvalue problem is solved.
		}\label{fig:scheme_autodiff}
	\end{figure}

	The linear solver is then used within a time-stepper Newton-Krylov \citep{edwards1994,bagheri2009matrix,Frantz2023} method. Analogously, a time-stepper Krylov-Schur method \citep{loiseau2019time} has been used to calculate the eigenvalues and eigenfunctions related to the screech problem.
	\cite{poulain2022} also use \texttt{Tapenade} to perform automatic differentiation of the compressible Navier-Stokes equations for 2D base-flows by assembling the Jacobian matrix of the problem via a coloring method. On the contrary, in this work, a completely matrix-free framework has been chosen to avoid assembling the full Jacobian matrix, which in a three-dimensional stability analysis problem can reach prohibitive dimensions in terms of computer memory.
	The linear algebra problems were solved using the Python version (\texttt{petsc4py} \cite{petsc4py2011} and \texttt{slepc4py} ) of the solver \texttt{PETSc} \citep{petsc-web-page} and of the eigenvalue problem solver \texttt{SLEPc}  \citep{slepc}, respectively.
	A schematic overview of the whole computational process can be seen in \fref{fig:scheme_autodiff}

	\section{Physical and Numerical configuration} \label{sec:configuration}
	\subsection{Physical Configuration}
	\begin{figure}
		\begin{subfigure}[c]{0.49\linewidth}
			\caption{}\label{jets_config}
			\includegraphics[page=1,width=\linewidth,trim=0 0 0 0,clip]{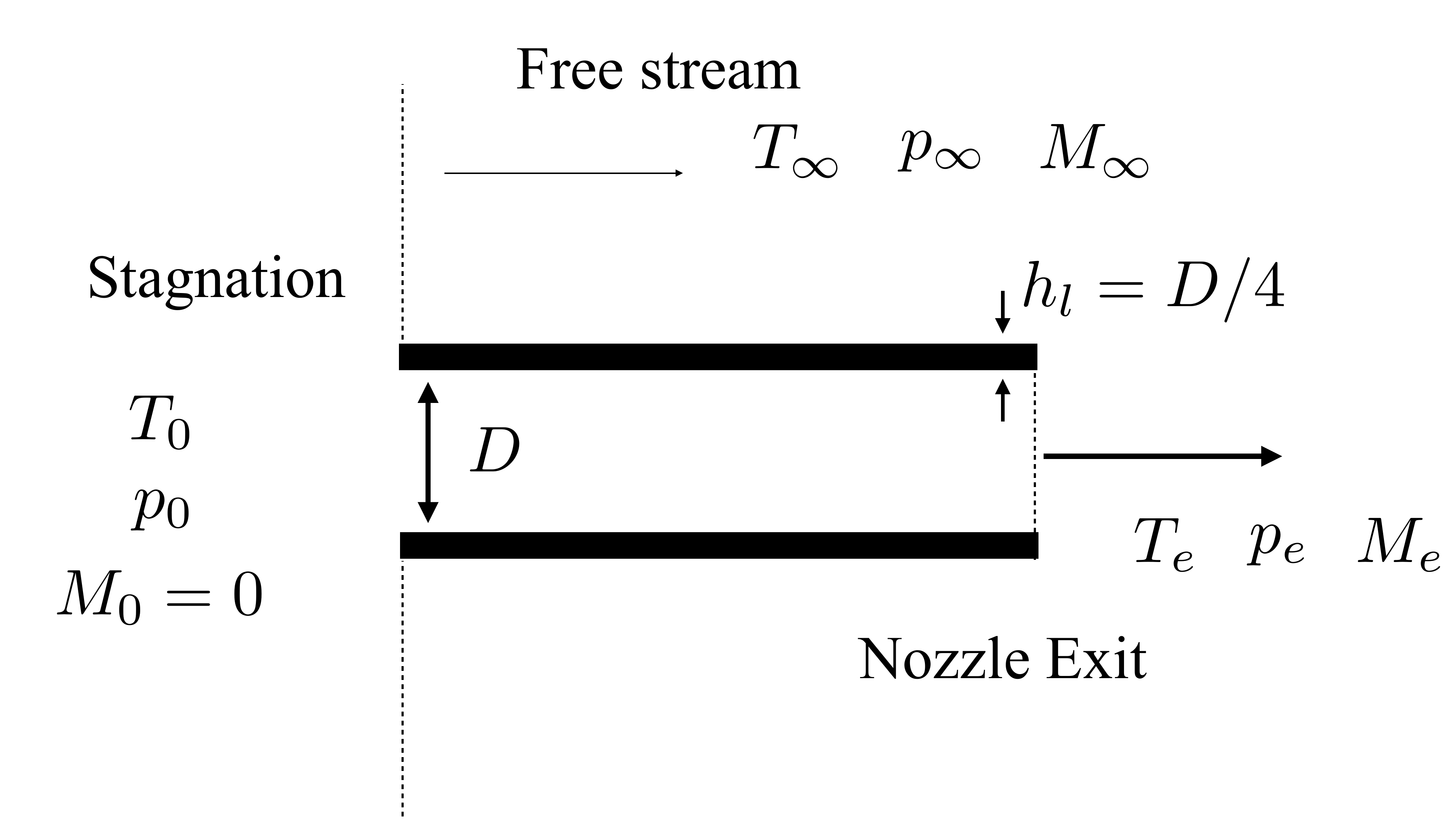}
		\end{subfigure}
		\begin{subfigure}[c]{0.49\linewidth}
			\caption{}\label{sketch_dimension}
			\includegraphics[page=2,width=\linewidth,trim=0 0 0 0,clip]{IMG/convergent_nozzle.pdf}
		\end{subfigure}
		\caption{(a) Important parameters for the studied configurations. Nozzle exit quantities are identified with the subscript $(e)$, ambient quantities with the subscript $(\infty)$ and stagnation quantities with the subscript $(0)$. (b) Sketch of the computational domain size with applied boundary conditions. The blue dashed line represent the physical domain, in the outer parts sponge layers are active.
		}\label{configurations}
	\end{figure}
	A jet coming from a convergent nozzle immersed in a coflow of static temperature $T_{\infty}$, static pressure $p_{\infty}$, and free-stream Mach number $M_{\infty}$ is considered.
	In the reservoir, a resting fluid at stagnation pressure $p_0$ and stagnation temperature $T_0$ is accelerated isentropically through a convergent nozzle, so that the nozzle exit Mach number $M_e$ is equal to one. The jet pressure ratio ($\text{JPR}=p_e/p_{\infty}$) is defined as the ratio between the static pressure at the nozzle exit $p_e$ and the static ambient pressure $p_{\infty}$.
	\fref{jets_config} shows a graphical sketch of the jet configuration considered with important parameters. In supersonic jets, another commonly used quantity is the nozzle pressure ratio ($\text{NPR}$), defined by the ratio between the stagnation pressure $p_0$ and the ambient pressure $p_{\infty}$.
	Assuming the flow in the reservoir is accelerated isentropically, the stagnation pressure $p_0$ and
	the ambient pressure $p_{\infty}$ are related by the following:
	\begin{equation}\label{eq:iso_Mj}
		\text{NPR} = \frac{p_0}{p_{\infty}}  =  \left(1 + \dfrac{\gamma -1}{2}M_j^2 \right)^{\frac{\gamma}{\gamma -1}},
	\end{equation}
	where $M_j$ is the fully expanded Mach number. For a give ambient pressure $p_{\infty}$, $M_j$ represents the
	Mach number at the nozzle exit that would result in a perfectly expanded jet.
	Each of the quantities $M_j$, JPR, and NPR is related through isentropic relations. Thus, to define the expansion conditions for a supersonic jet, it is sufficient to select a value for one of these three quantities.
	The temperature condition of the jet is defined by the nozzle temperature ratio (NTR), namely the ratio between the reservoir stagnation temperature $T_0$ and the ambient static temperature $T_{\infty}$. All cases considered here are cold jets, where a unitary nozzle temperature ratio is imposed $(\text{NTR}=T_0/T_{\infty}=1)$.

	The jet nozzle has a lip thickness of $h_l= D/4$, where $D$ is the nozzle diameter.
	Each quantity has been non-dimensionalised by its reference one measured at the nozzle exit: $D$, $U_e$, $\rho_e$, $T_e$, $\mu_e$. Imposing one value between $M_j$, JPR or NPR, the condition on pressure is imposed. The Reynolds number $\text{Re}=\rho_e U_e D/\mu$ fixes the geometrical dimensions and scales of the nozzle. Fully turbulent cases are considered, imposing a Reynolds number of $\text{Re}=5 \cdot 10^6$ at the nozzle exit for all configurations studied.

	\begin{table}
		\caption{Parameter of the different configuration studied, the NPR are calculated up to the first decimal.
		}
		\begin{center}
			\begin{tabular}{c|c|c}
				$M_j$ & NPR & JPR \\
				\hline
				1.1   & 2.1 & 1.13 \\
				1.2   & 2.4 & 1.28 \\
				1.25  & 2.6 & 1.37 \\
				1.45  & 3.4 & 1.8  \\
				\hline
			\end{tabular}\label{tab_3D_param}
		\end{center}
	\end{table}

	\subsection{Numerical Configuration}
	\begin{figure}
		\centering
		\includegraphics[width=\linewidth,trim=0 0 0 0,clip]{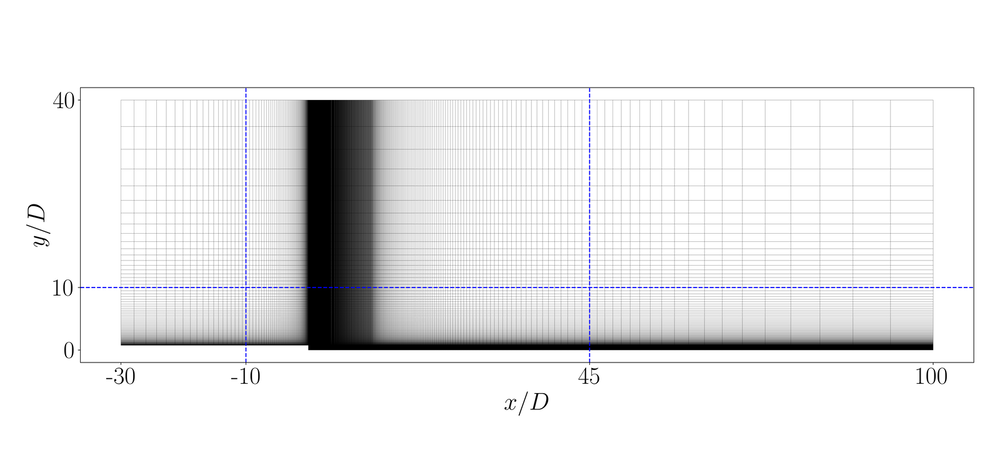}
		\caption{View of the grid before rotation. The blue dashed line represents the physical domain. The outer part of the domain is filled with sponge layers.
		}\label{grid_cut_xy}
	\end{figure}

	Each jet configuration is simulated over a three-dimensional multi-block curvilinear structured grid. The mesh is generated starting from a two-dimensional grid as shown in \fref{grid_cut_xy}.
	The mesh cells are denser near the nozzle, concentrated in the shock-cell structure part of the flow. In this zone, the points are distributed almost uniformly, with a light stretch
	of approximately $1\%$ ending at $x/D = 10$.
	\begin{figure}
		\centering
		\begin{subfigure}[c]{0.49\linewidth}
			\caption{}\label{grid_3D}
			\centering
			\includegraphics[width=\linewidth,trim=0 0 0 0,clip]{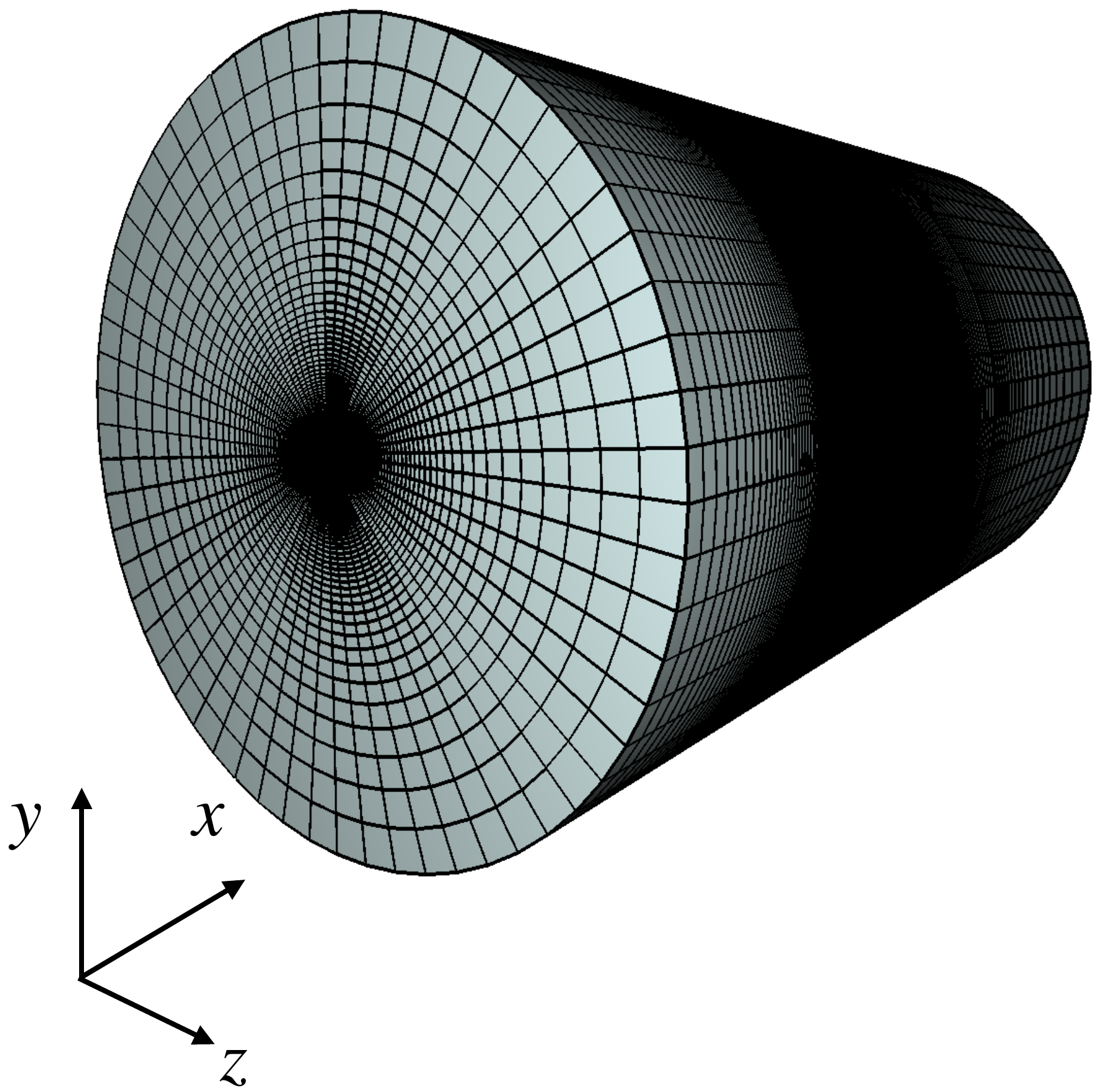}
		\end{subfigure}
		\begin{subfigure}[c]{0.49\linewidth}
			\caption{}\label{grid_zy_papillon}
			\centering
			\includegraphics[width=\linewidth,trim=0 0 0 0,clip]{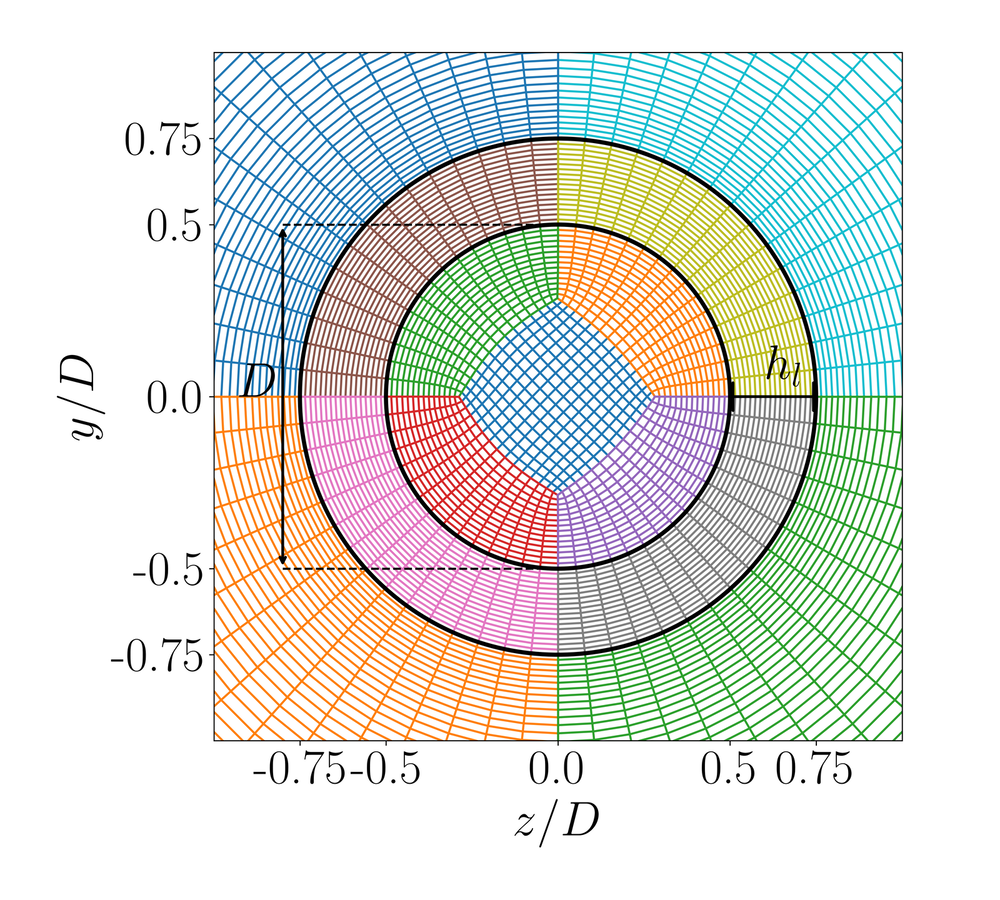}
		\end{subfigure}
		\caption{(a) General view of the three-dimensional grid used. (b) Focus on the O-H topology used to remove the singularity in the nozzle center line. Different colours of mesh represent different blocks.
		}\label{grid_3D_screech}
	\end{figure}
	 To avoid a singularity at the center line of rotation, an O-H or ``butterfly'' topology structured grid is constructed (see \fref{grid_zy_papillon}). A three-dimensional view of the resulting grid is shown in \fref{grid_3D}.
	The grid has $N_x=740$ points in the longitudinal direction, $N_y=150$ points in the
	radial direction and $N_{\theta}=60$ plane in the azimuthal direction, resulting in a grid composed of
	approximately 5.9 million points. The interior of the nozzle is not discretised, following~\cite{ShenTam2002},
	supposing that the screech phenomenon involves only large-scale turbulence structures
	appearing in the shear layer, neglecting the influence of the boundary layer developing in the nozzle.
	Sponge zones are imposed to reduce spurious reflections coming from the boundaries.
	In cylindrical coordinates, the sponge layer spans in the jet axial direction from $x/D=-30$ to $x/D=-10$
	upstream and from $x/D=45$ to $x/D=100$ downstream. In the radial direction, the sponge layer spans from $y/D=10$ to $y/D=40$ and from $\theta = 0 $ to $\theta = 2\pi$ in the azimuthal direction.
	This defines a cylindrical physical domain encapsulated inside the sponge layers that spans from $x/D=-10$ to $x/D=45$, from $y/D=0$ to $y/D=10$ and from $\theta = 0 $ to $\theta = 2\pi$ in the longitudinal, radial and azimuthal directions, respectively.
	The dashed blue lines in \fref{grid_cut_xy} define the boundaries of the physical domain.

	The three-dimensional curvilinear URANS equations are solved with a $k-\log(\omega)$ SST turbulence model \citep{menter1994, menter2003}
	on the vector of conservative state variables $\B{q} = \left[\rho, \rho u, \rho v, \rho w, \rho E,
	\rho k, \rho \omega\right]^T$ where $k$ represent the specific turbulent kinetic energy and $\omega$ the turbulent
	specific dissipation rate. The equations are discretised in space using a sixth-order finite difference centered scheme with a fifth-order Jameson scheme \citep{jameson1981,sciacovelli2021}. The time discretisation scheme used is a third-order Runge-Kutta \citep{shu1989}. Subsonic non-reflective boundary conditions \citep{POINSOT1992104} impose the coflow velocity, and supersonic inflow boundary conditions are imposed at the nozzle exit. No-slip adiabatic boundary conditions are imposed on the exterior nozzle walls and on the nozzle lips. Non-reflective subsonic boundary conditions are imposed in the outer boundary of the domain by imposing a pressure level equal to the ambient pressure $P_{\infty}$ \citep{POINSOT1992104,jiang1999}.
	For turbulent quantities, the inflow values are imposed ten times larger than the ambient ones, and the boundary conditions are chosen as presented in \secref{sec:komega}.

	\section{Fixed point solution of the URANS model} \label{sec:fixedpoint}

	\begin{figure}
		\centering
		\begin{subfigure}[c]{\linewidth}
			\centering
			\caption{}\label{fp_3D_rho_1.1}
			\includegraphics[width=\linewidth,trim=0 0 0 0,clip]{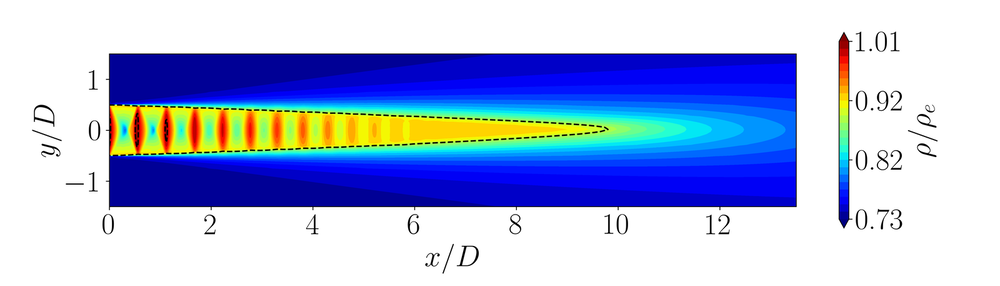}
		\end{subfigure}
		\begin{subfigure}[c]{\linewidth}
			\centering
			\caption{}\label{fp_3D_rho_1.2}
			\includegraphics[width=\linewidth,trim=0 0 0 0,clip]{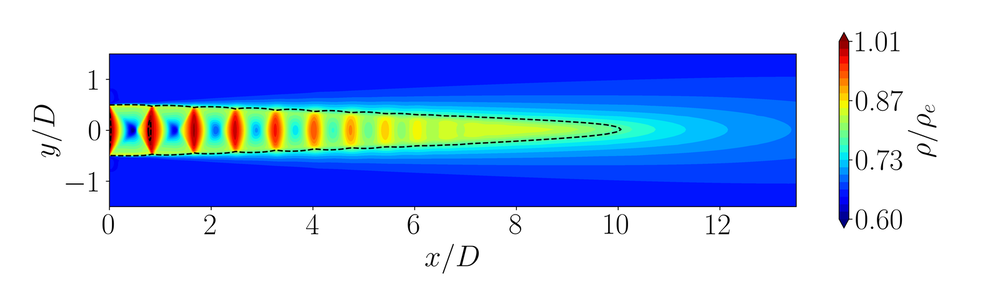}
		\end{subfigure}
		\begin{subfigure}[c]{\linewidth}
			\centering
			\caption{}\label{fp_3D_rho_1.45}
			\includegraphics[width=\linewidth,trim=0 0 0 0,clip]{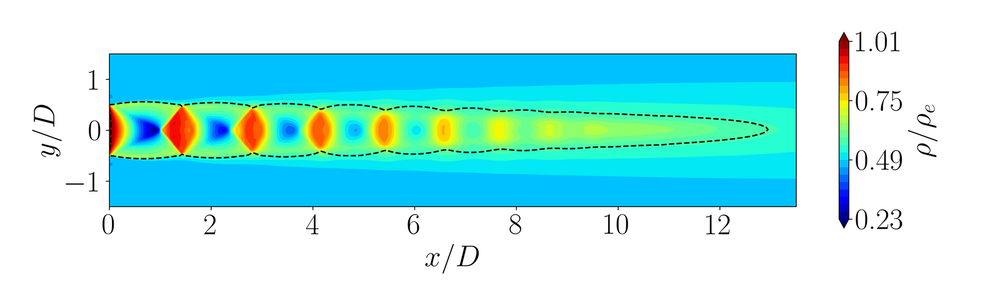}
		\end{subfigure}
		\caption{Density field of the fixed point solutions in the $xy$ plane ($z=0$). The black dashed line represents the sonic line. (a) $M_j=1.1$, (b) $M_j=1.2$, (c) $M_j=1.45$.
		}\label{fp_3D_rho}
	\end{figure}
	URANS simulations are performed at various $M_j$'s and initialised with ambient quantities everywhere. After the transients are evacuated, the simulation for $M_j=1.1$ and $M_j=1.25$ locks into a non-stationary solution, while for $M_j=1.2$ and $M_j=1.45$, the solution reaches a steady state.
	For the stationary cases, the fixed point solution is the steady RANS solutions, while for the case at $M_j = 1.1$ and $M_j = 1.25$, a Newton-Krylov method is used to converge to a fixed point solution.

	Initialising the Newton-Krylov algorithm \citep{Frantz2023} with a non-linear snapshot, the fixed point converges with a residual error of $\| r\|_2=\| \Phi(t+\tau)- \Phi(t)\|_2 < 10^{-6}$, where $\Phi(t)$ is the non-linear discretised URANS operator. The sampling time $\tau$ is chosen so that approximately ten snapshots are collected for each period. The GMRES iterations within the Newton method are solved using the dynamical residual technique as in \cite{Frantz2023}.

\begin{figure}
		\centering
		\includegraphics[width=\linewidth,trim=0 0 0 0,clip]{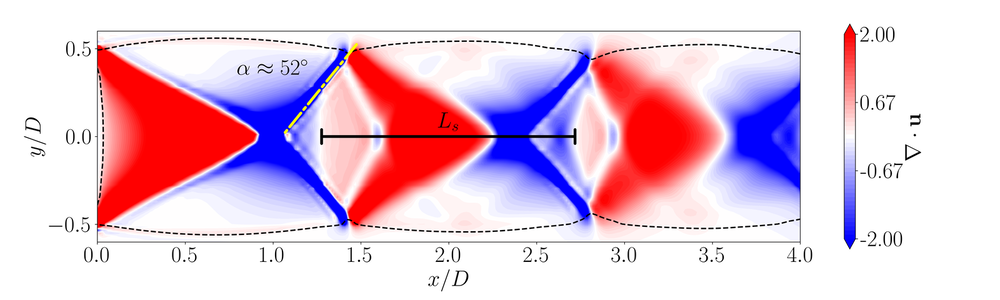}
		\caption{Close up on the velocity divergence component $\nabla \cdot \B{u}$ for $M_j=1.45$. The oblique shocks are visible with an angle $\alpha$ of approximately $\alpha \approx 52^{\circ}$ with respect to the incident flow (in agreement with the local Rankine-Hugoniot conditions).
}\label{fp_3D_div}
	\end{figure}

	The calculated density component of the fixed point solutions is shown in \fref{fp_3D_rho}. Three different levels of under-expansion are shown, $M_j=1.1$, $M_j=1.2$ and $M_j=1.45$. The typical shock-cell structure of under-expanded jets is visible in the near nozzle area, extending throughout the jet core in all jets shown. At higher levels of under-expansion (higher $M_j$'s), the shock strength increases, visible by higher gradients in the density contours. Another important parameter that depends on the $M_j$ is the shock-cell length, here named $L_s$. It is measured as the distance between two pressure peaks along the jets' central line, and it is crucial in the selection of the screech frequency (see \fref{fp_3D_div} for a visualisation of the shock-cell length $L_s$). As can be seen in \fref{fp_3D_rho}, the shock-cell length increases at higher $M_j$'s, since the expansion fans that are generated at the nozzle lips have a wider angle with respect to the jet boundary and they rebound to form compression fans at a larger distance from the nozzle. The dashed lines in both \fref{fp_3D_rho} and \ref{fp_3D_div} represent the sonic line ($M=1$), which divides the supersonic from the subsonic part of the flow.

	\fref{fp_3D_div} shows a close-up of the velocity divergence for the case $M_j=1.45$. A series of oblique shocks is present where the compression fans coalesce. The expansion fans generated in the nozzle are visible as positive values of the velocity divergence. Fans rebound on the jet sonic line and convert into compression fans, which coalesce, forming an oblique shock, clearly visible by the points of minimum divergence of velocity.

	\begin{figure}
		\centering
		\begin{subfigure}[c]{0.495\linewidth}
			\centering
			\caption{}\label{fp_3D_u_1.1}
			\includegraphics[width=\linewidth,trim=0 0 0 0,clip]{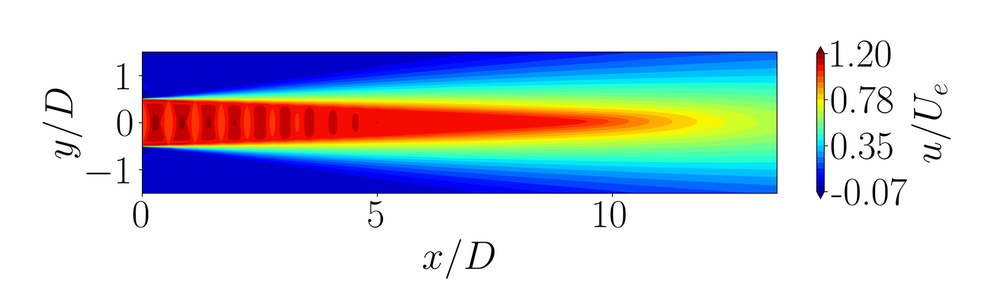}
		\end{subfigure}
		\begin{subfigure}[c]{0.495\linewidth}
			\centering
			\caption{}\label{fp_3D_v_1.1}
			\includegraphics[width=\linewidth,trim=0 0 0 0,clip]{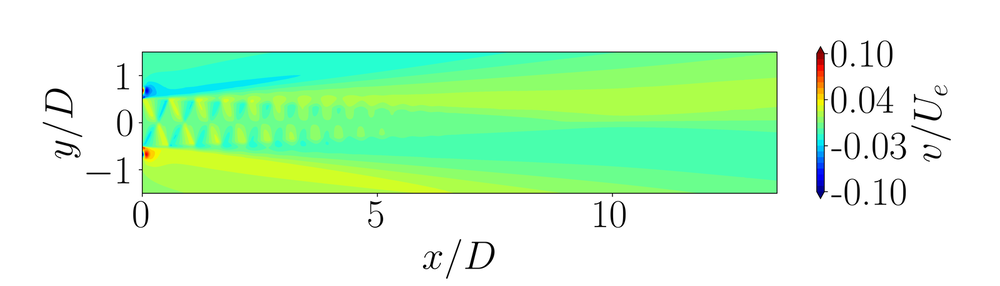}
		\end{subfigure}\\
		\begin{subfigure}[c]{0.495\linewidth}
			\centering
			\caption{}\label{fp_3D_u_1.2}
			\includegraphics[width=\linewidth,trim=0 0 0 0,clip]{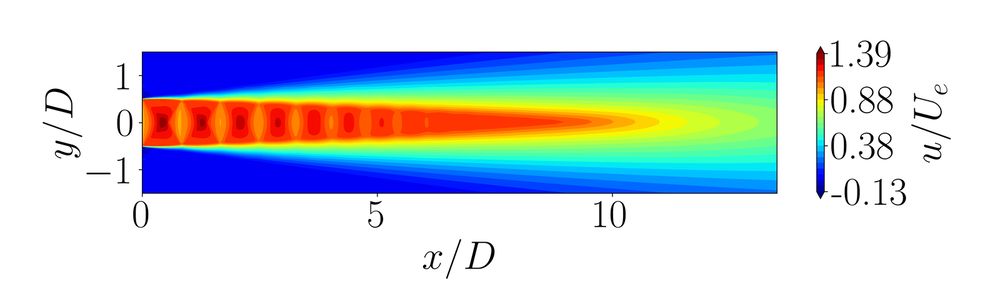}
		\end{subfigure}
		\begin{subfigure}[c]{0.495\linewidth}
			\centering
			\caption{}\label{fp_3D_v_1.2}
			\includegraphics[width=\linewidth,trim=0 0 0 0,clip]{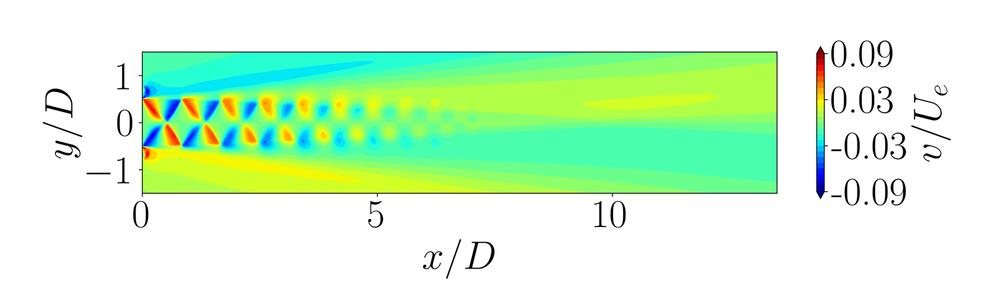}
		\end{subfigure}\\
		\begin{subfigure}[c]{0.495\linewidth}
			\centering
			\caption{}\label{fp_3D_u_1.45}
			\includegraphics[width=\linewidth,trim=0 0 0 0,clip]{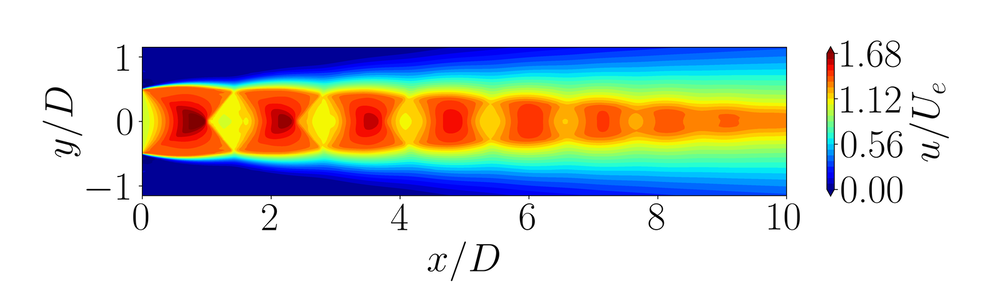}
		\end{subfigure}
		\begin{subfigure}[c]{0.495\linewidth}
			\centering
			\caption{}\label{fp_3D_v_1.45}
			\includegraphics[width=\linewidth,trim=0 0 0 0,clip]{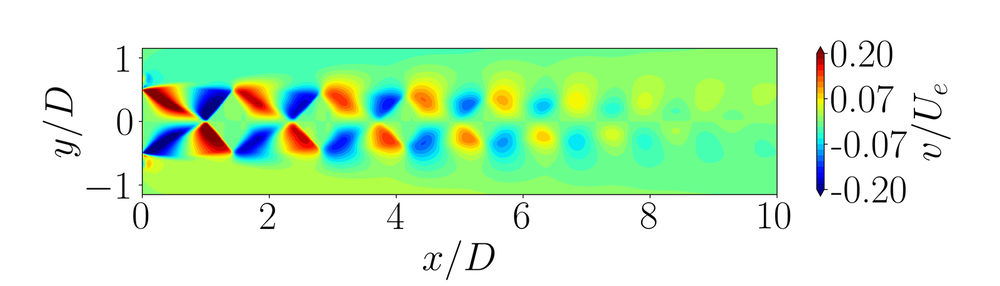}
		\end{subfigure}\\
		\caption{Velocity field of the fixed point solutions in the $xy$ plane ($z=0$). (a,b) $M_j=1.1$, (c,d) $M_j=1.2$, (e,f) $M_j=1.45$. (a,c,e) Streamwise velocity component $u$, (b,d,f) normal velocity component $v$.	}\label{fp_3D_uv}
	\end{figure}

	In \fref{fp_3D_uv}, the streamwise and radial velocity components are shown. As with the density field, the shock-cell structure is clearly visible in the streamwise velocity component. The streamwise component $u$ exhibits axial symmetry, while the radial component $v$ is anti-symmetric about the jet axis. What is relevant here from a computational point of view is that the accurate reproduction of these symmetries in the base flow is needed for obtaining clean stability modes. Bigger numerical imperfections would introduce spurious symmetry-breaking in the global spectrum. The fact that the computed fields faithfully recover the axial symmetry serves as a verification of the numerical precision of the fixed-point solution. The fields are also in qualitative agreement with the flow features reported in experimental data \citep{EdgingtonMitchell2018,EdgingtonMitchell2021,nogueira2022closure,mancinelli2021complex} and numerical simulations \citep{shen1998numerical,shen2000}.
	
	\subsection{Validation of the fixed point solutions}

	The spacing between shocks in the fixed-point solution is an important factor affecting screech frequency, which is noticeable in the initial estimate made by Powell (see \eref{eq:powell}).
	In addition to shock-cell spacing, the quasi-periodic structure of the fixed-point solution is also important for numerically predicting screech frequency. Indeed, both the first two dominant wavenumber in \eref{eq:wave_fp} play a significant role in the staging mechanism and on the selection of screech frequency for specific configurations \citep{nogueira2022closure, edgington2022unifying}.

	\begin{figure}
		\begin{subfigure}[c]{0.495\linewidth}
		\centering
		\caption{}\label{pressure_line_Mj_1.1}
		\includegraphics[width=\linewidth,trim=0 0 0 0,clip]{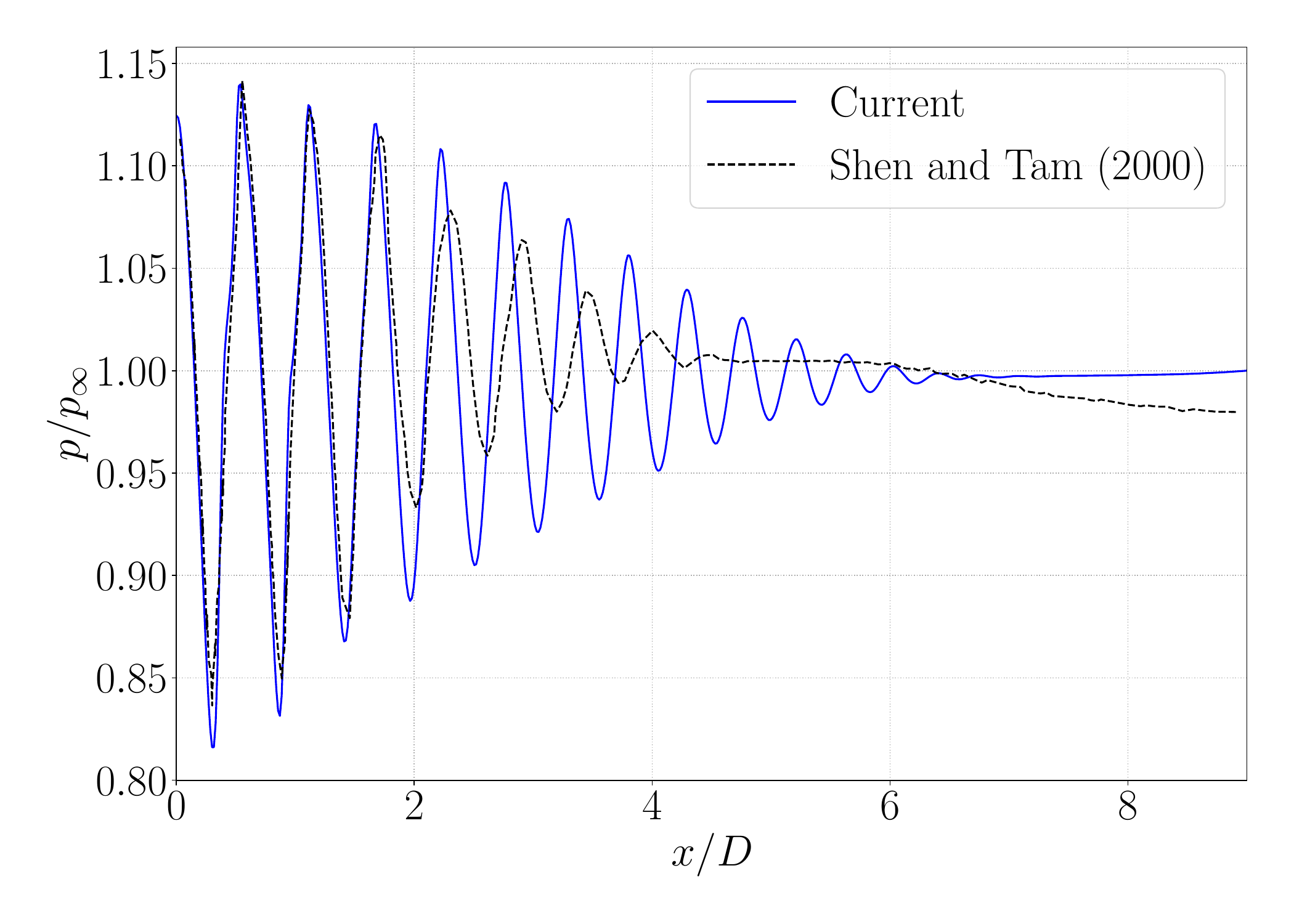}
	\end{subfigure}
	\begin{subfigure}[c]{0.495\linewidth}
	\caption{}\label{pressure_line_Mj_1.2}
	\includegraphics[width=\linewidth,trim=0 0 0 0,clip]{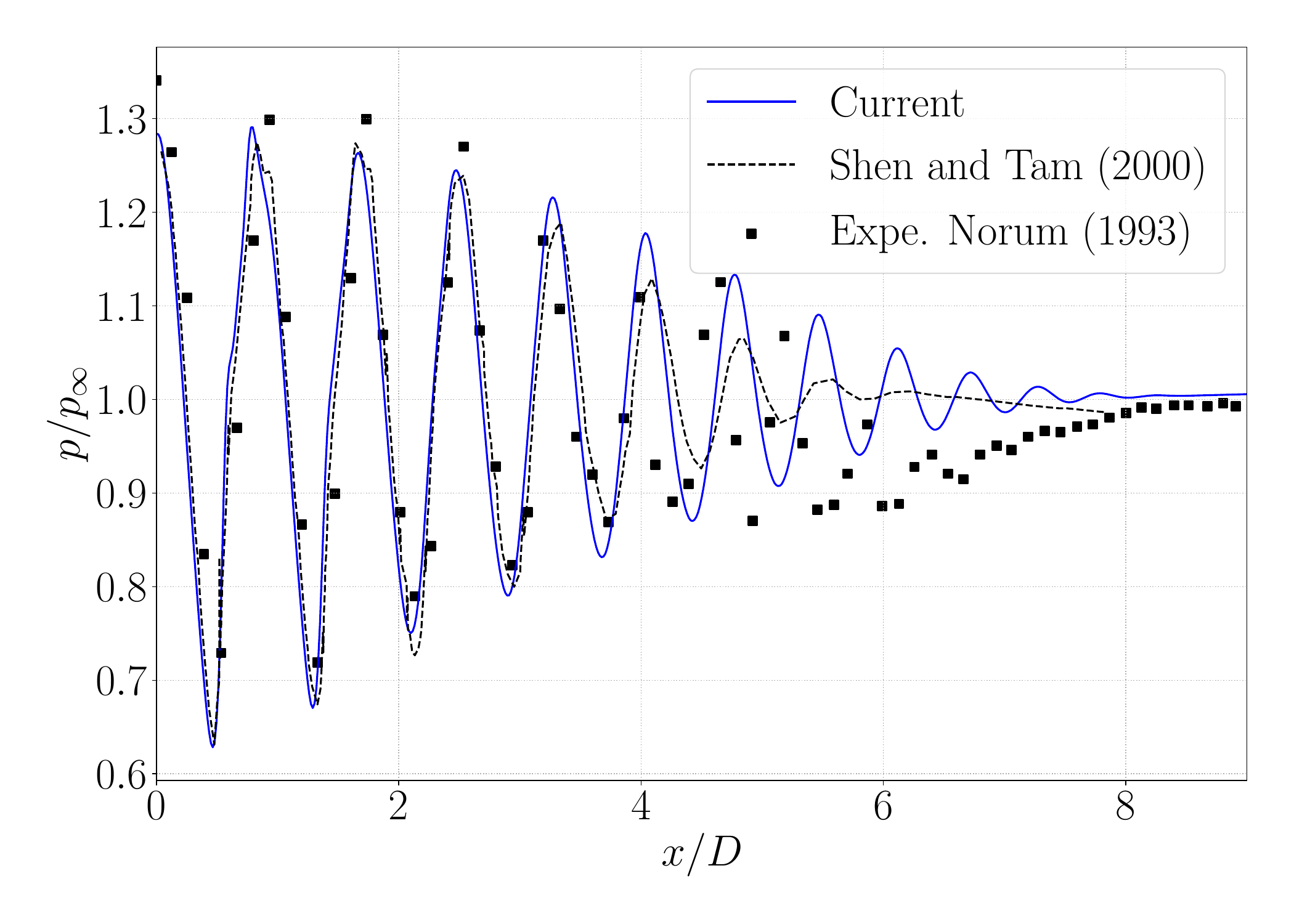}
	\end{subfigure}
	\caption{Pressure measured in the axial line passing through the jet core. (a) $M_j=1.1$, (b) $M_j=1.2$.  The values here are normalised by ambient pressure. RANS data from~\cite{shen2000}, experimental data from~\cite{norum1993}.
	}\label{pressure_line_3D}
\end{figure}

	For these reasons, before performing stability analysis, the fixed-point data obtained from the URANS equations are validated by comparing them with experimental data \citep{li2021screech,norum1993} and previously done URANS calculations \citep{shen2000}. Througout this work, a series of numerical and experimental data are analysed and compared with our calculation, so that an overview of the configurations that have been used for comparison are listed.
	The dataset from \citep{norum1993} comes from experimental data taken at the
	\textit{NASA Langley Acoustics Research Laboratory QuietFlow Facility}, with a thin lipped nozzle with 0.75 inch exit diameter.
	Dataset from \citep{li2021screech} comes from experimental data recently performed of a cold jet of $5$ mm lip thickness and $10$ mm jet diameter ($D/2$).
	URANS data comes from a previous calculation from \cite{shen2000} with a nozzle lip thickness of $D/4$ at $Re = 10^6$, where $D$ is the diameter of the nozzle. Powell experiments \citep{powell1992observations} have been performed with a circular convergent nozzle of diameter $10$ mm, with a lip thickness of $8$ mm ($4/5D$). The experimental data in \cite{edgington2022unifying}, referred in their work as Monash data, comes from a purely convergent nozzle of $10$ mm of diameter, with a lip thickness of only $0.15 D$.

	The metrics used to evaluate the quality of the URANS fixed-point solutions are the shock strength and the shock-cell length. To assess the shock strength, the normalised pressure values along the centreline are compared with experimental data. For the shock-cell length, the distance between successive pressure peaks along the centreline is examined, together with the power spectral density computed in the streamwise direction of the pressure signal.

	In \fref{pressure_line_3D}, pressure data extracted along the centreline are compared with experimental measurements from \citet{norum1993} and with URANS calculations reported by \citet{shen2000} for the cases $M_j = 1.1$ and $M_j=1.2$. The fixed-point solution correctly captures the pressure distribution observed in the experiments for the first sequence of shocks, both in terms of strength and spacing. For the oscillations associated with the downstream shocks, the present solution appears slightly cleaner and more regular, whereas the experimental data exhibit greater irregularity. When compared with the URANS simulation of \citet{shen2000}, the present calculation appears less dissipative, particularly in the downstream pressure oscillations near the end of the jet for both cases at $M_j=1.1$ and $M_j=1.2$. This behaviour can be attributed to the slightly finer mesh employed in the present study.

	\begin{figure}
		\centering
		\begin{subfigure}[c]{0.49\linewidth}
			\caption{}\label{FFT_line_fp_Mj_1.25}
			\includegraphics[width=\linewidth,trim=0 0 0 0,clip]{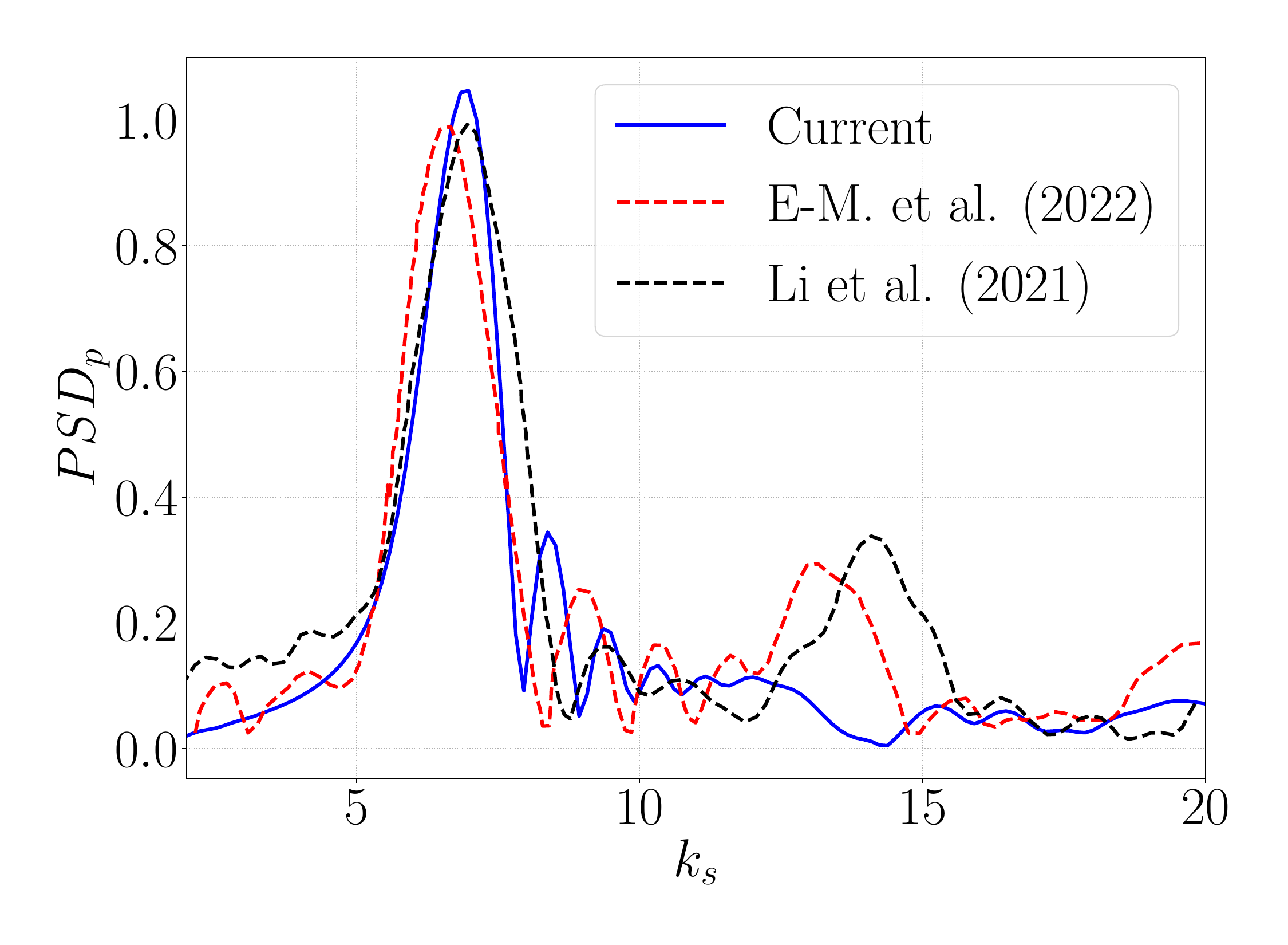}
		\end{subfigure}
		\begin{subfigure}[c]{0.49\linewidth}
			\caption{}\label{FFT_line_fp_Mj_1.45}
			\includegraphics[width=\linewidth,trim=0 0 0 0,clip]{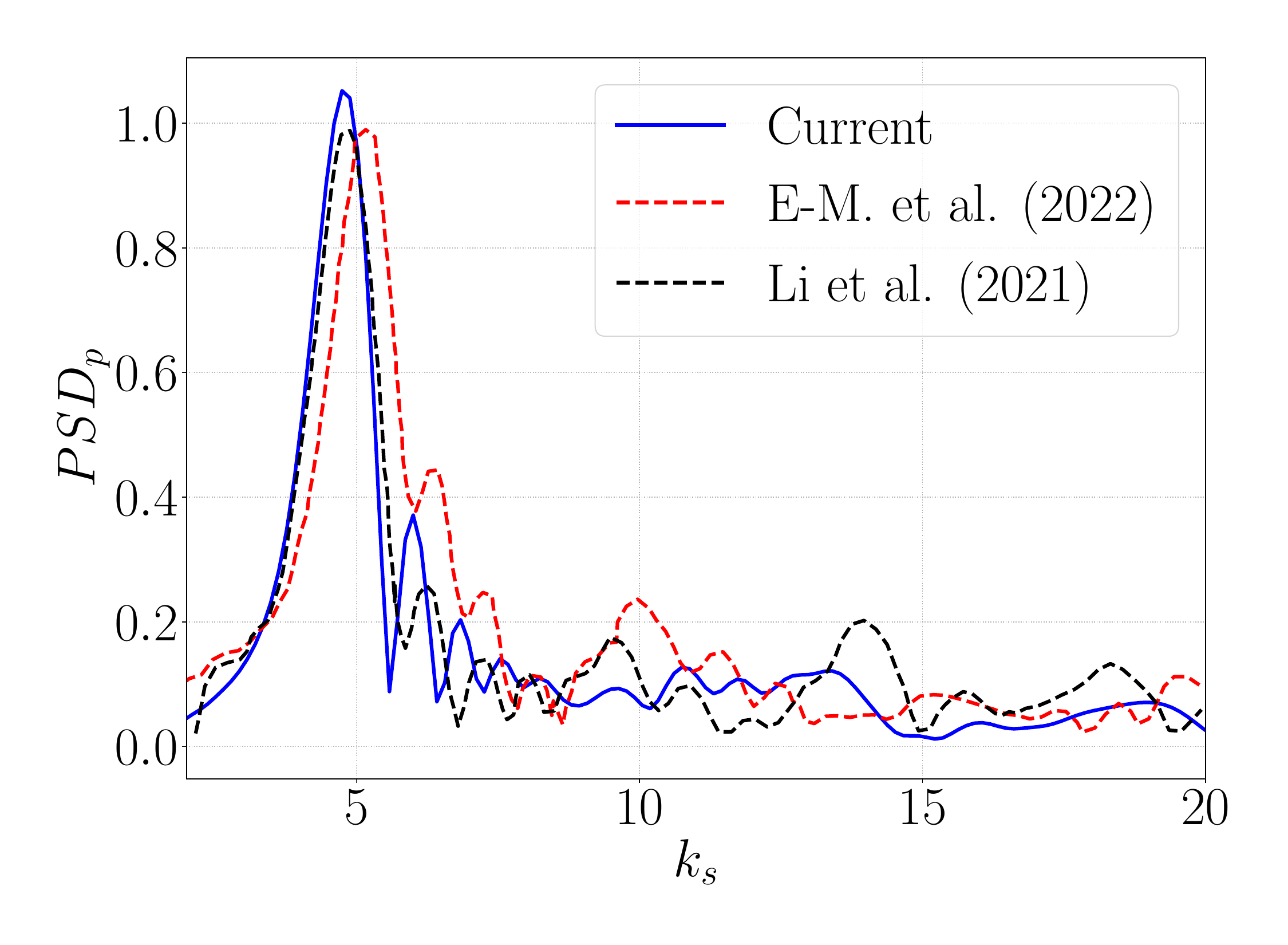}
		\end{subfigure}
		\caption{Power-density spectra of the centre line pressure signal compared with experimental data (\cite{edgington2022unifying}, \cite{li2021screech}). (a) $M_j=1.25$, (b) $M_j=1.45$,
		}\label{FFT_line_fp}
	\end{figure}

	The effectiveness of linear analysis in predicting screech produced by supersonic jets depends on accurately accounting for nonlinear mechanisms through a faithful description of the shock-cell structure in the base flow \citep{EdgingtonMitchell2021}. \fref{FFT_line_fp} shows the power spectral density of the pressure signal along the centreline, with comparisons to experimental data from \citet{edgington2022unifying} and \citet{li2021screech}. The comparison is performed for two levels of under-expansion, $M_j = 1.25$ and $M_j = 1.45$.

	In both cases, the numerical results follow the general behaviour observed in the experimental measurements. The first spectral peak, which corresponds to the shock-cell length, is correctly captured for both values of $M_j$, in terms of both amplitude and wavenumber. A second peak, hereafter denoted $k_{s_2}$, is observed at a higher wavenumber in both the experiments and the URANS fixed-point solutions. The wavenumber and amplitude of this second peak exhibit greater variability in both the experimental data and the URANS simulations, suggesting a stronger dependence on small-scale dynamics and a sensitivity to the turbulence model.

	\begin{figure}
		\centering
		\begin{subfigure}{0.495\linewidth}
			\centering
			\caption{}\label{pressure_line_yz_0_3D_fp_Mj_1.1}
			\includegraphics[width=\linewidth,trim=0 0 0 0,clip]{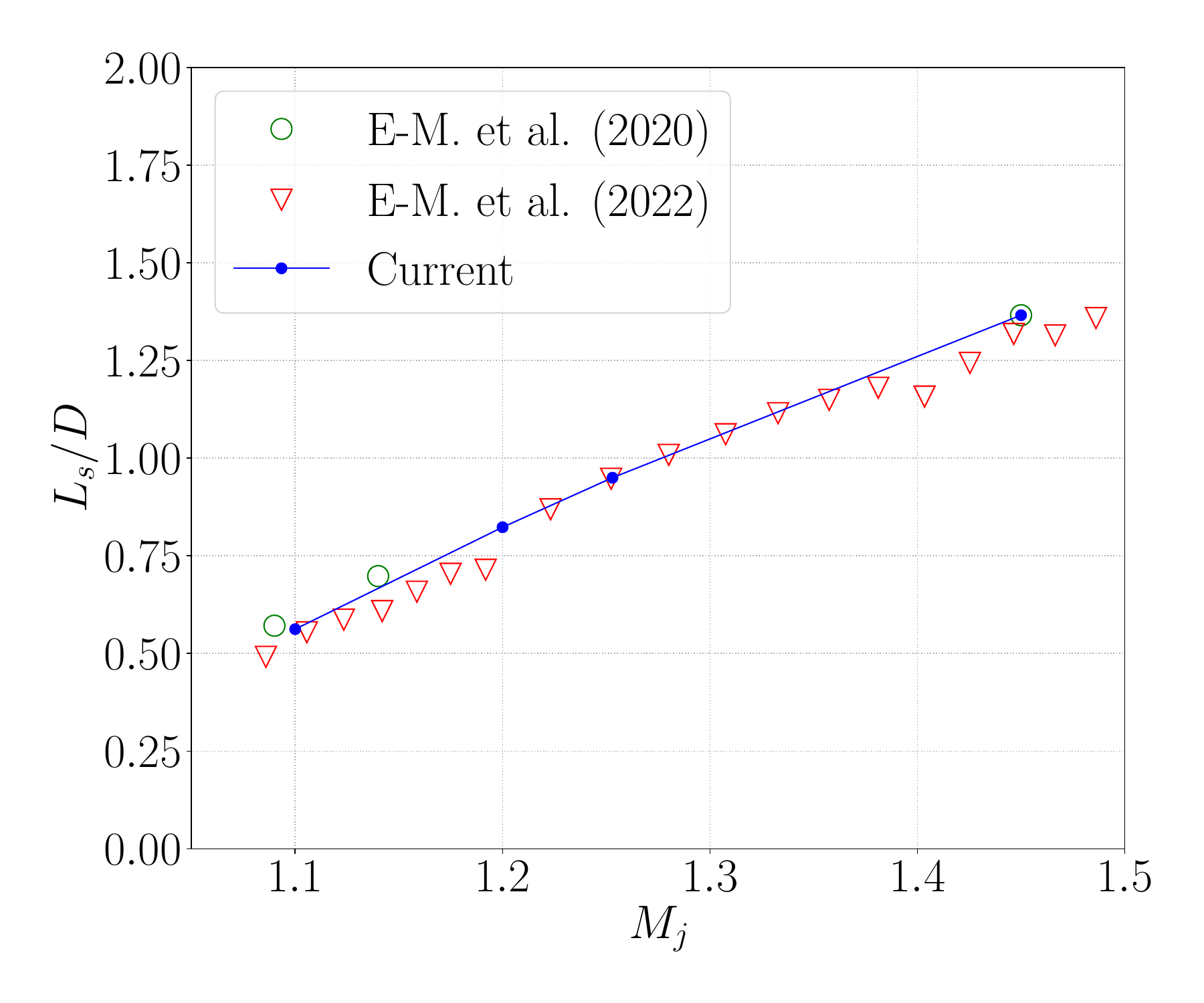}
		\end{subfigure}
		\begin{subfigure}{0.495\linewidth}
			\centering
			\caption{}\label{compare_ks}
			\includegraphics[width=\linewidth,trim=0 0 0 0,clip]{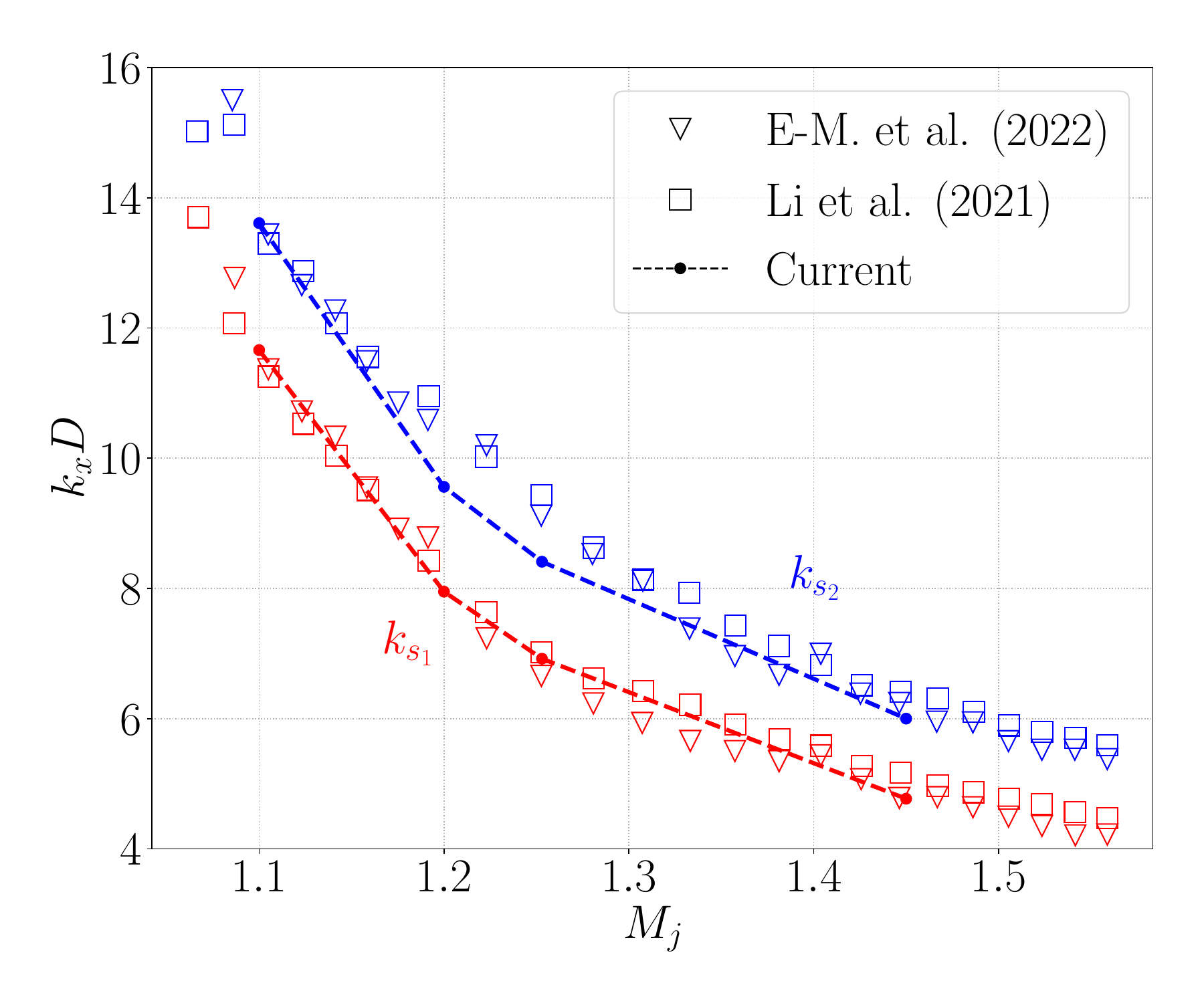}
		\end{subfigure}
		\caption{Baseflow comparison with experimental data from \cite{edgington2022unifying,li2021screech} for a range of jet Mach numbers. (a) Comparison of the shock-cell length calculated from the distance between two pressure peaks.(b) Comparison of the first and second peak on the pressure-signal spectra. $k_{s_1}$ values are shown in red,  $k_{s_2}$ values are shown in blue.
		}\label{shock_spacing}
	\end{figure}

	In \fref{shock_spacing}, the shock-cell length, $k_{s_1}$, and $k_{s_2}$ are compared for all configurations discussed in the paper. The shock-cell length closely follows the experimental data, in agreement with the results shown in \fref{FFT_line_fp}. In \fref{compare_ks}, the two wavenumbers are compared with experimental measurements, showing good agreement for all values of $M_j$ considered for the first wavenumber.

	It is worth noting the variability that can occur in these measurements, even within the experimental data. These results provide strong confidence that the URANS fixed-point solutions correctly capture the key nonlinear features relevant to the development of screech. Global stability analysis of these fixed-point solutions is presented in the following section.

	\section{Global stability analysis} \label{sec:stability}
	\subsection{Numerical parameters for the timestepper stability analysis}
	Global stability analysis is numerically performed, through a time-stepper Krylov-Schur algorithm as described in \secref{sec:computational}.
	\begin{table}
		\caption{Timestepper parameter for the Krylov-Schur algorithm. $M_j$ is the fully expanded Mach number,
			$m$ is the Krylov subspace dimension and $\tau$ is the sampling time.
		}
		\begin{center}
			\begin{tabular}{c|c|c}
				$M_j$   & $m$  &  $\tau$       \\
				\hline
				1.1     & 400  & $0.195 \; t$  \\
				1.2     & 500  & $0.2 \;   t$  \\
				1.25    & 500  & $0.2 \;   t$  \\
				1.45    & 500  & $0.3 \;   t$  \\
				\hline
			\end{tabular}\label{tab_3D_KS}
		\end{center}
	\end{table}
	\tabref{tab_3D_KS} shows the Krylov-Schur parameters used for the calculation. The Krylov dimension $m$ represents the total number of snapshots used to approximate the spectrum of the Jacobian matrix.
	To properly calculate the modes frequency, at least ten periods of the instability must be covered by the linear simulation~\citep{Frantz2023}, which according to our sampling time amounts to a dimension $m \approx 100$ for the Krylov-subspace. Since it is known that screech frequency is dependent on the $M_j$, and it decreases for increasing $M_j$, the Krylov subspace is chosen to be larger for larger $M_j$. Since the linearized solver is initialized with white noise, an initial transient must be ejected before starting the stability calculation, which could take from $100$ to $200$ sampling time depending on the expansion configuration of the jet $M_j$, so that a total Krylov subspace dimension of $m \approx 400-500$ have been chosen.

	\subsection{Modal analysis}

	\begin{figure}
		\centering
		\begin{subfigure}[c]{0.49\linewidth}
			\centering
			\caption{}\label{spectrum_1.1}
			\includegraphics[width=\linewidth,trim=0 0 0 0,clip]{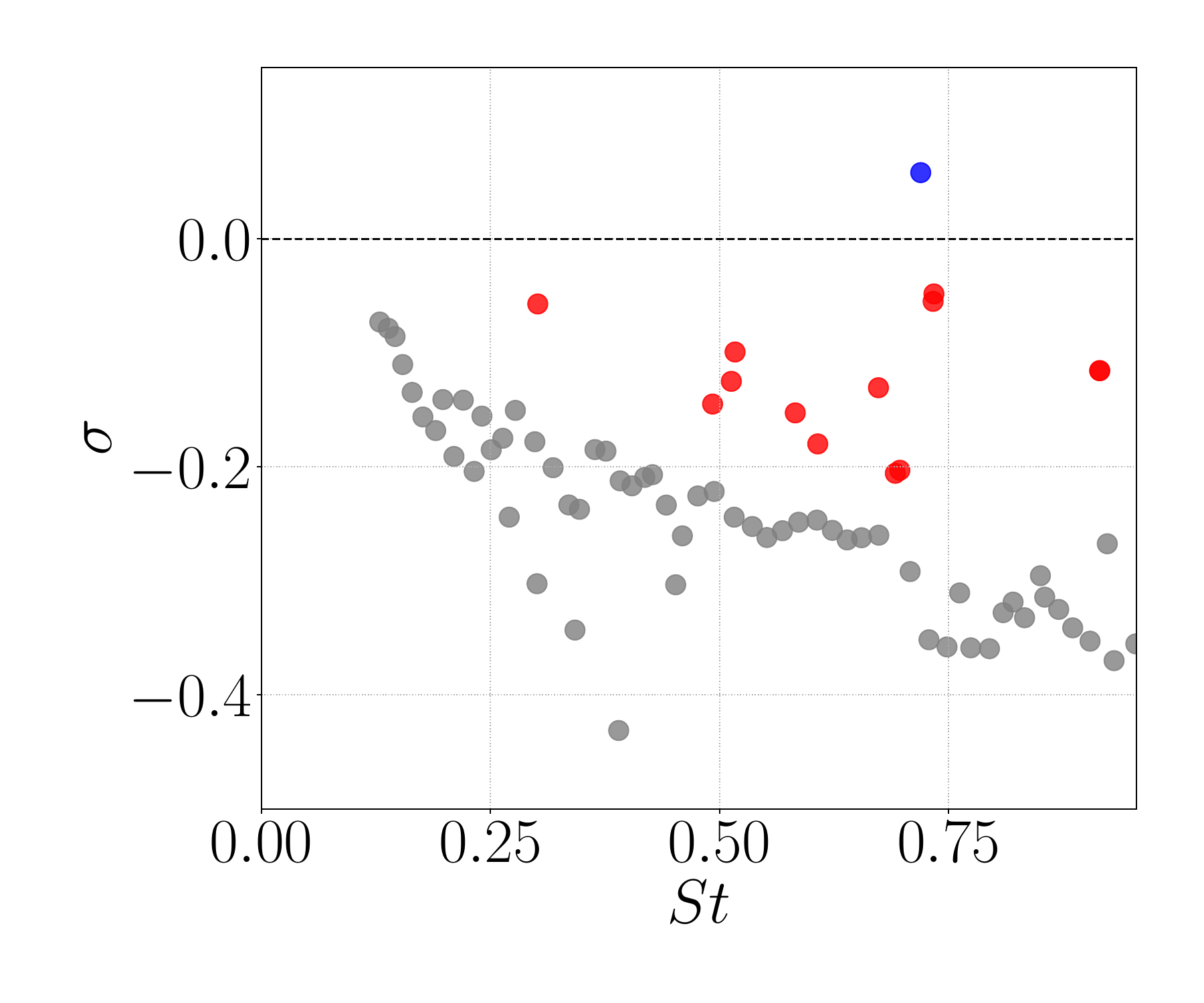}
		\end{subfigure}
		\begin{subfigure}[c]{0.49\linewidth}
			\centering
			\caption{}\label{spectrum_1.2}
			\includegraphics[width=\linewidth,trim=0 0 0 0,clip]{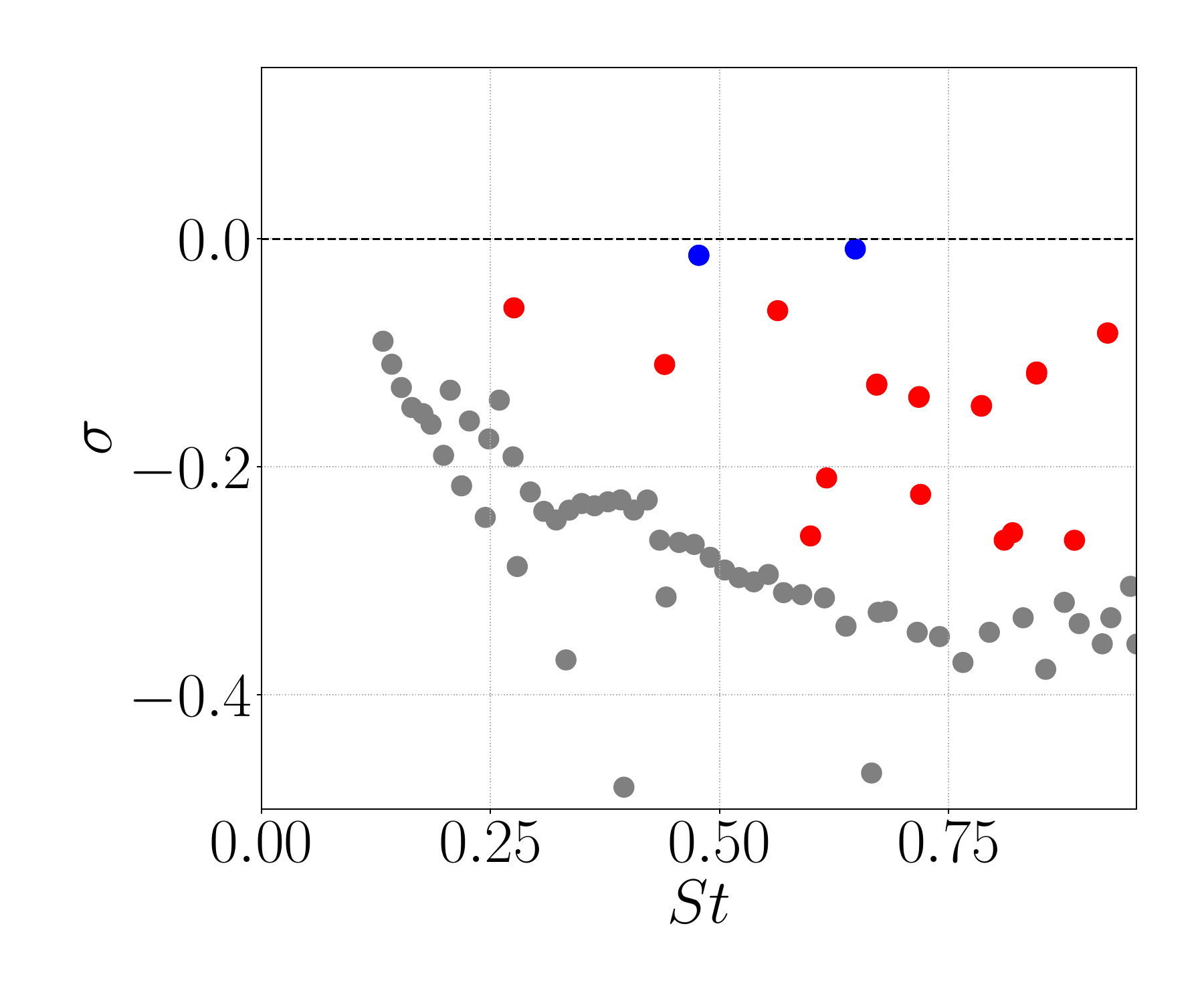}
		\end{subfigure}\\
		\begin{subfigure}[c]{0.49\linewidth}
			\centering
			\caption{}\label{spectrum_1.25}
			\includegraphics[width=\linewidth,trim=0 0 0 0,clip]{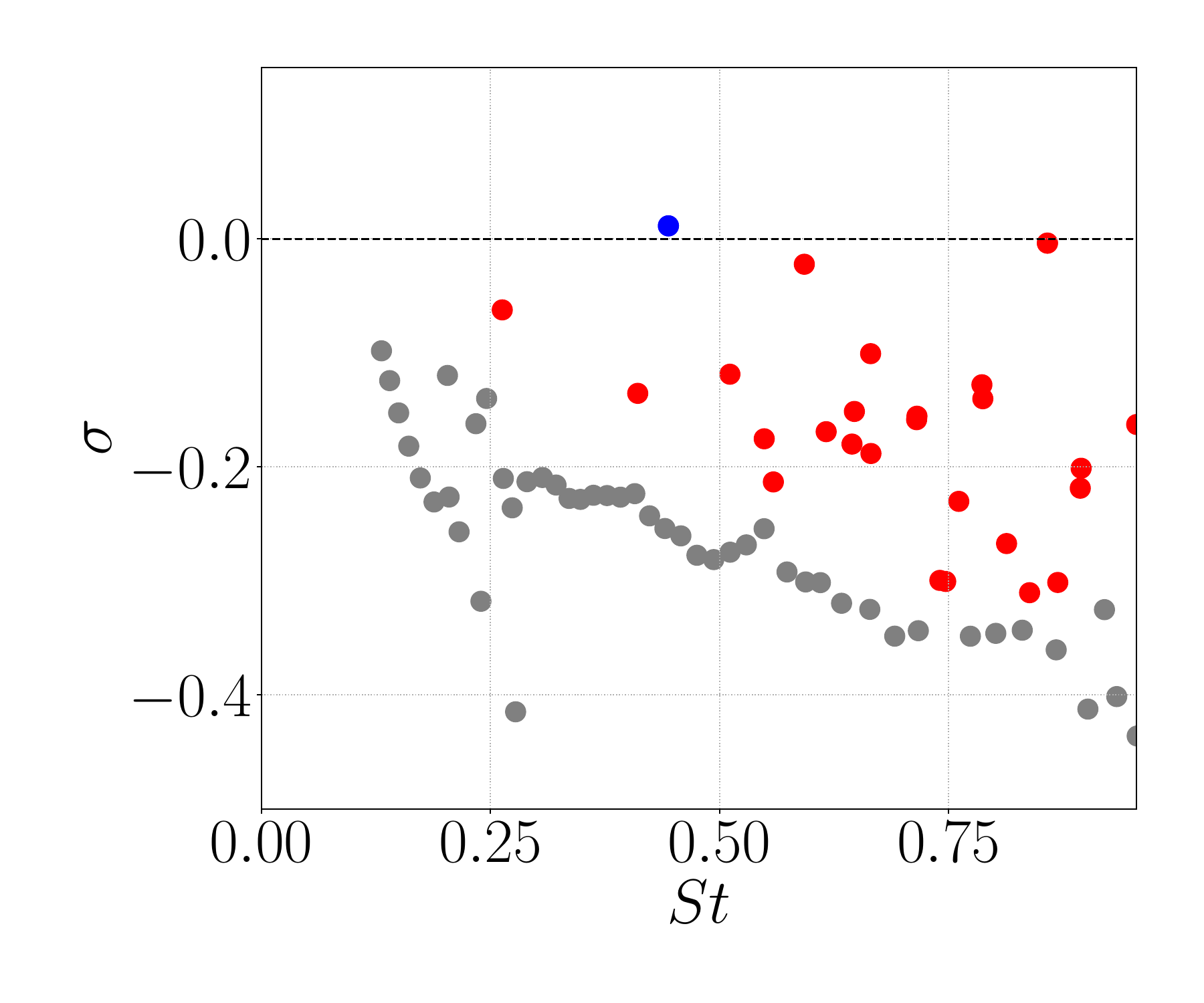}
		\end{subfigure}
		\begin{subfigure}[c]{0.49\linewidth}
			\centering
			\caption{}\label{spectrum_1.45}
			\includegraphics[width=\linewidth,trim=0 0 0 0,clip]{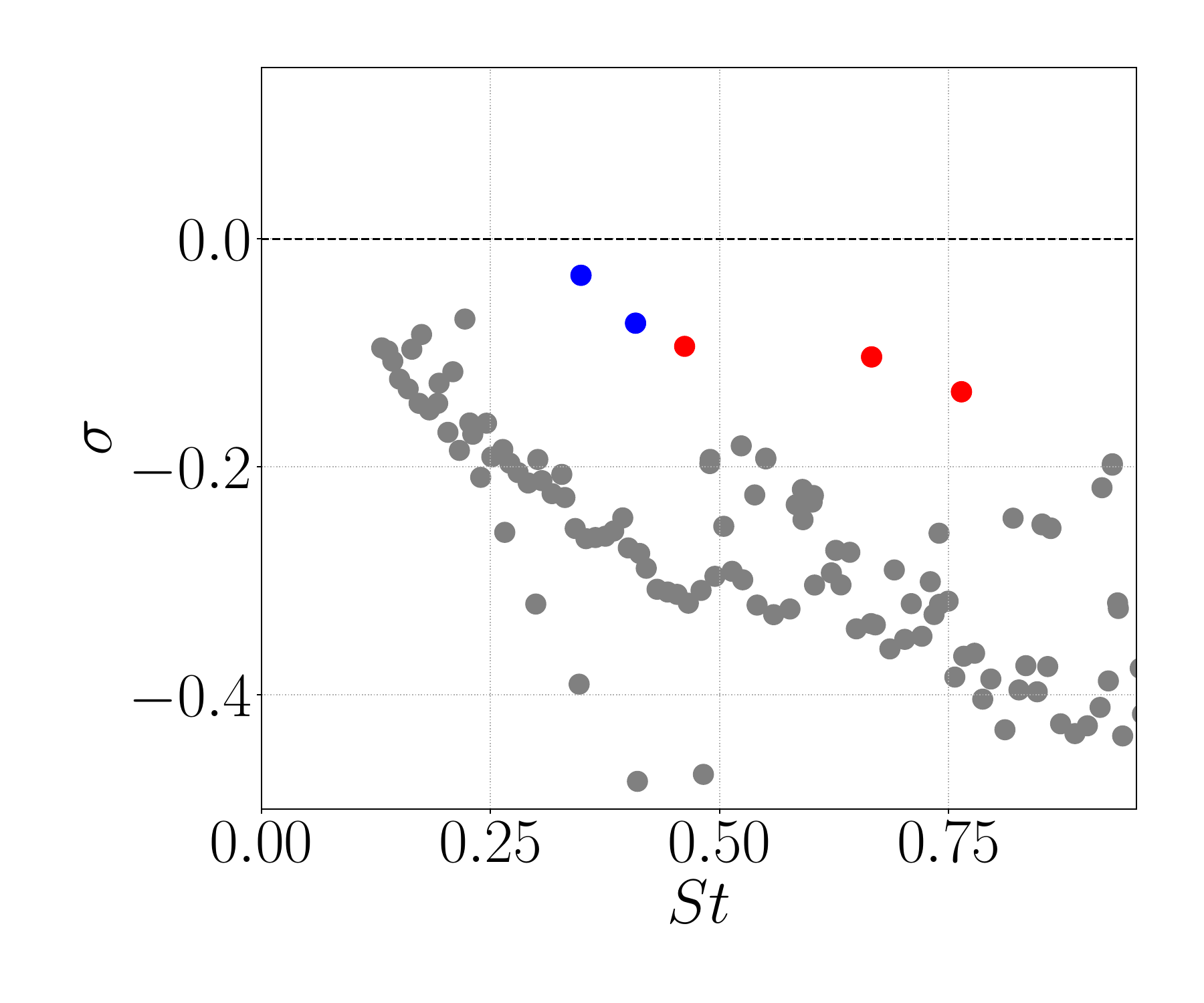}
		\end{subfigure}
		\caption{Spectrum resulting from global stability analysis at different $M_j$. The eigenvalues ($\lambda = \sigma + \omega i$, $St = 2 \pi \omega $) colored in red and blue are the one that numerically converged ($\varepsilon_r < 10^{-8}$), while the ones in grey do not converged ($\varepsilon_r > 10^{-4}$). The eigenvalues chosen for the analysis are highlighted in blue. (a) $M_j=1.1$, (b) $M_j=1.2$, (c) $M_j=1.25$, (d) $M_j=1.45$.
		}\label{stab_3D}
	\end{figure}

	Computing modal instabilities for turbulent flow has a different meaning than computing them for laminar flow. First and foremost, it is essential to distinguish between instabilities that develop spatially and those that are self-sustaining temporally. The former are related to receptivity mechanisms and can be calculated using a resolvent method. The latter correspond to resonances within the flow and are superimposed on the turbulent flow. Only the second type of instability is addressed in this paper.  Interpreting these modes in turbulent flow is more complex, particularly the amplification factor $\sigma$, as it ideally indicates whether the instability is self-sustaining or not. However, in reality, for this amplification factor to be intrinsic, the scales driving it must be sufficiently uncorrelated with the turbulence scales. When this is not entirely the case, the amplification factor then depends on the turbulence dynamics. In a URANS context, it will depend on the turbulence model used. In this case, only the nature of the mode is important; it must be a mode isolated from branches of convective modes and whose spatial envelope is localized. An example of a self-sustaining instability well decoupled from the turbulence is the case of transonic buffeting around an airfoil, where the sign of the amplification factor gives the threshold for the onset of flow buffeting \citep{CGM07,CGM09}.

	In \fref{stab_3D}, the calculated eigenvalues $\lambda$ of the form $\lambda = \sigma + \omega i$ are shown. The spectra exhibit different behaviours depending on the level of under-expansion. For $M_j=1.1$ and $M_j=1.25$, one unstable mode can be found ($\sigma > 0$), while for $M_j=1.2$ and $M_j=1.45$, only stable eigenvalues are found ($\sigma <0$). In each figure, the modes selected for the analysis are highlighted in blue. The criteria for selecting the modes are based on different factors. The most important one is the instability of the mode, which corresponds to the real part of the eigenvalue being positive. In this case, the fixed-point solution is unstable and the linear dynamics represent a self-sustained mechanism that can be related to the screech tone. In the configurations where no unstable modes are found, the nonlinear simulation converges to a stationary state, meaning that no self-sustained behaviour has been observed from the simulations. Yet, even in the stable cases, important physical characteristics found in experiments can be captured from stable modes, mainly the spatial distribution of the eigenfunction and the frequency of the oscillation.

	All spectra present a branch of convective modes (highlighted in grey) that are numerically weakly converged ($\varepsilon_{r} > 10^{-4}$). These modes are not interesting from a temporal stability point of view, since they are linked to convective instabilities present in the shear layer of the jet, that are for the most part spatially amplified. Another family of modes are the ones that are stable but detached from the convective branch (highlighted in red); these modes can provide useful information, since they select a frequency at which the mode detaches from the convective branch to express some oscillatory behaviour in the flow. An a priori way to select these modes is to study the ones that are the most detached from the convective branch, while at the same time doing a posteriori analysis in order to make sure that the modes analysed do not present artificial oscillations or non-physical behaviour.

	An explanation for why some configurations present stable modes and others do not, has to be found in the dissipative aspects of the turbulence model.
	The dissipative nature of the eddy-viscosity modelled via the Boussinesq hypothesis tends to damp perturbations that arise in the shear layer, preventing the feedback loop from being self-sustained in simulations for some $M_j$. This aspect has been analysed via a sensitivity case study presented in \secref{sec:sensitivity}, which looks at two-dimensional planar screeching jets and illustrates the high sensitivity of the real part of the eigenvalue to the choice of turbulence model. It has to be noted, however, that while the growth rate of the mode can change drastically depending on the turbulence modelling, the imaginary part of the eigenvalue, which represents the oscillation frequency, stays approximately the same for the entire range of pressure ratios, regardless of the model.
	This is because the features of the base flow that select the frequency of screech are well captured by the turbulence model in the fixed point solution.
	The modes selected for the analysis, under the mentioned criteria, are the ones highlighted in blue.
	A Strouhal number $St=\omega/2\pi$ can be extracted from the angular frequency of the mode. This frequency is the one at which the mode oscillates in time, giving us an estimate for the frequency of the acoustic waves present in the mode and that can be linked to the screech phenomenon.
	\begin{figure}
		\centering
		\begin{subfigure}[c]{0.49\linewidth}
			\centering
			\caption{}\label{compare_st}
			\includegraphics[width=\linewidth,trim=0 0 0 0,clip]{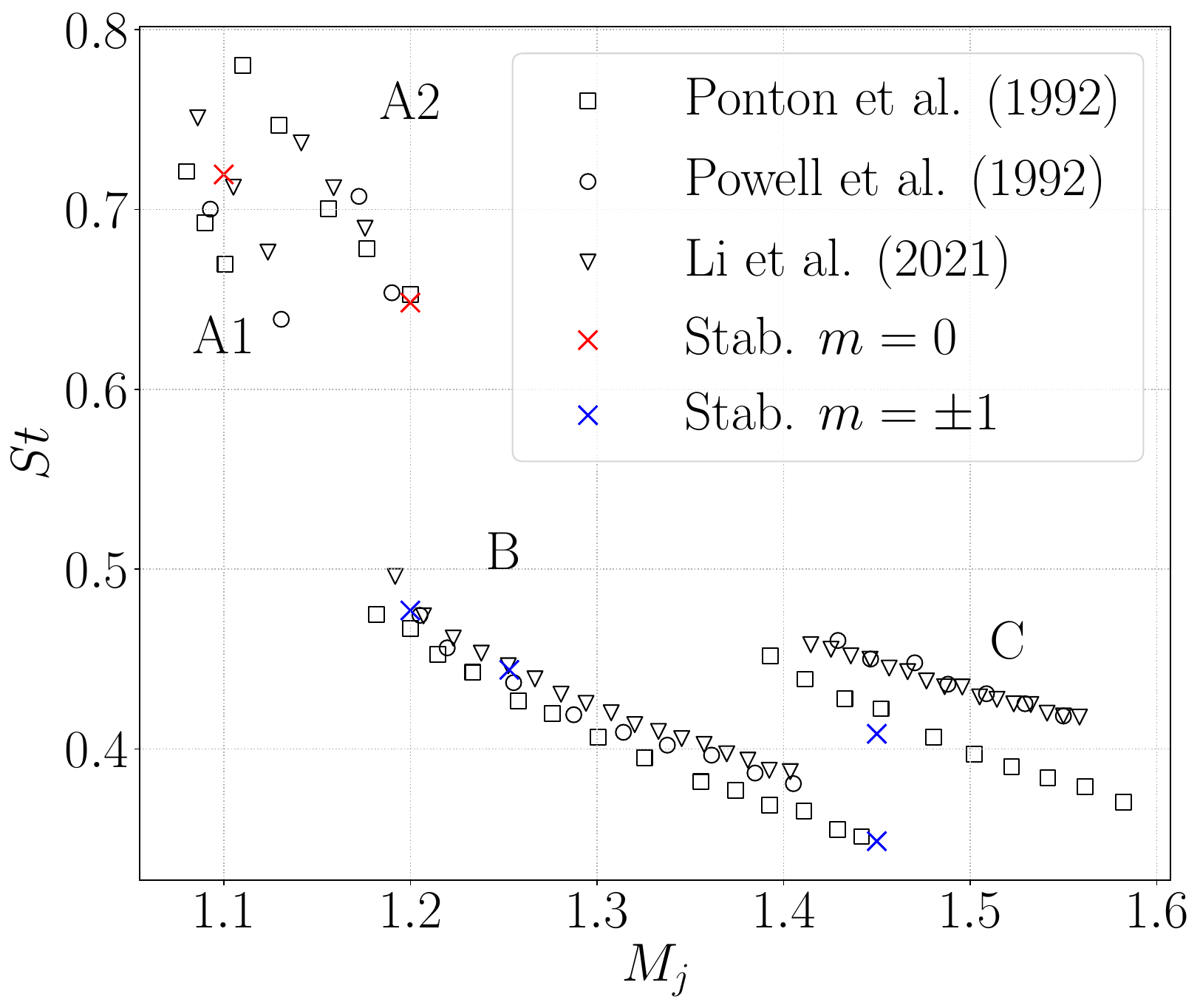}
		\end{subfigure}
		\begin{subfigure}[c]{0.49\linewidth}
			\centering
			\caption{}\label{compare_wavelength}
			\includegraphics[width=\linewidth,trim=0 0 0 0,clip]{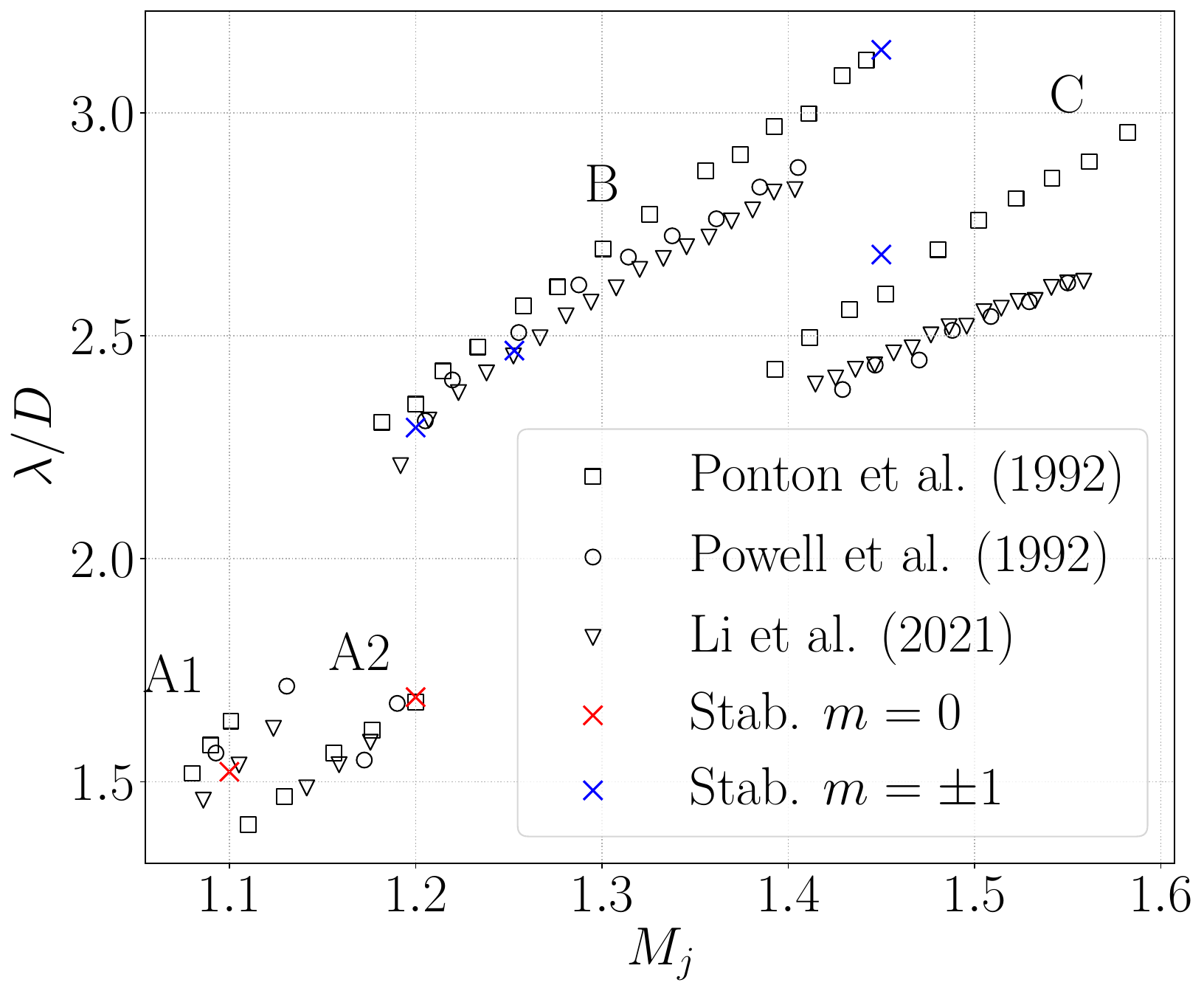}
		\end{subfigure}
		\caption{Comparison between experimental data \citep{Ponton1992,powell1992observations,li2021screech} and global stability analysis for different $M_j$. (a) Strouhal number $St=\omega/2 \pi$, (b) wavelength $\lambda = (2 \pi c_{\infty})/\omega$.
			For the stability calculation, the modes of azimuthal number $m=0$ and $m=\pm 1$ are highlighted in red and blue respectively.
		}\label{compare_expe}
	\end{figure}
	
	\fref{compare_expe} presents a comparison between the Strouhal and wavelength values calculated from the imaginary part $\omega$ of the selected modes at different $M_j$ and measurements coming from various experiments \citep{Ponton1992,powell1992observations,li2021screech}.
	It can be seen from this comparison that, within the variability of the different experiments, the estimated frequencies of the stability modes are in good agreement with the different families of modes measured in experiments, given the fact that the experimental configurations can vary in some parameters (Reynolds number, nozzle lip thickness, ...).
	These families of screech modes are characterised by their azimuthal wavenumber $m$: axisymmetric modes and flapping/helical modes.
	The axisymmetrical modes A1 and A2, which are observed in the low $M_j$ range for $M_j=1.1$ and $M_j=1.2$, are characterised by a streamwise velocity perturbation of toroidal shape, which is linked to an azimuthal wavenumber $m=0$ for the streamwise velocity fluctuation. The experiments and the stability analysis calculations show the exclusive nature of these modes as it has been recently described via local stability analysis in \cite{nogueira2022closure}. The flapping B modes (observed for $M_j=1.2$, $M_j=1.25$ and $M_j=1.45$) and the helical mode C (observed for $M_j=1.45$) have an azimuthal wavenumber of $m= \pm 1$, reflecting a flapping or helical movement of the real part of the streamwise velocity $u_x$ perturbation and of the acoustic wave.

\subsection{Axisymmetric ($m=0$) modes }

\begin{figure}
	\centering
	\begin{subfigure}[c]{0.49\linewidth}
		\centering
		\caption{}\label{modeA1_ux}
		\includegraphics[width=\linewidth,trim=0 0 0 0,clip]{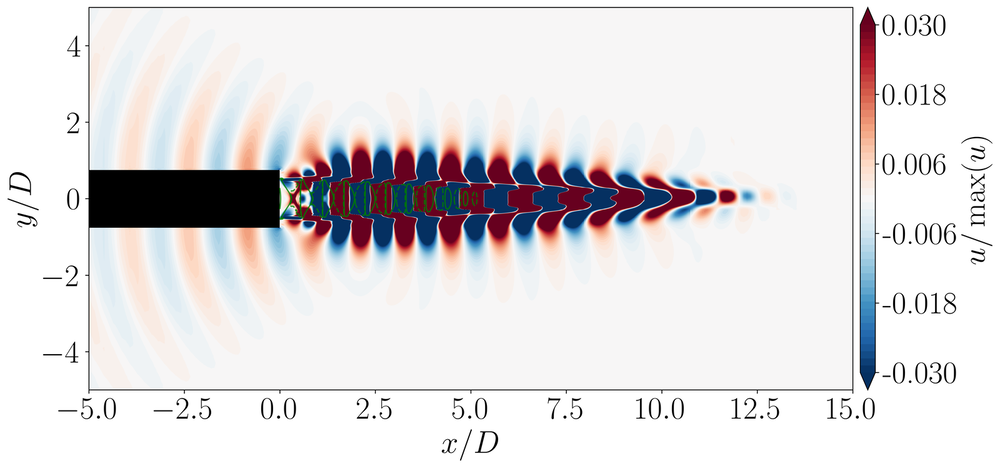}
	\end{subfigure}
	\begin{subfigure}[c]{0.49\linewidth}
		\centering
		\caption{}\label{modeA1_vx}
		\includegraphics[width=\linewidth,trim=0 0 0 0,clip]{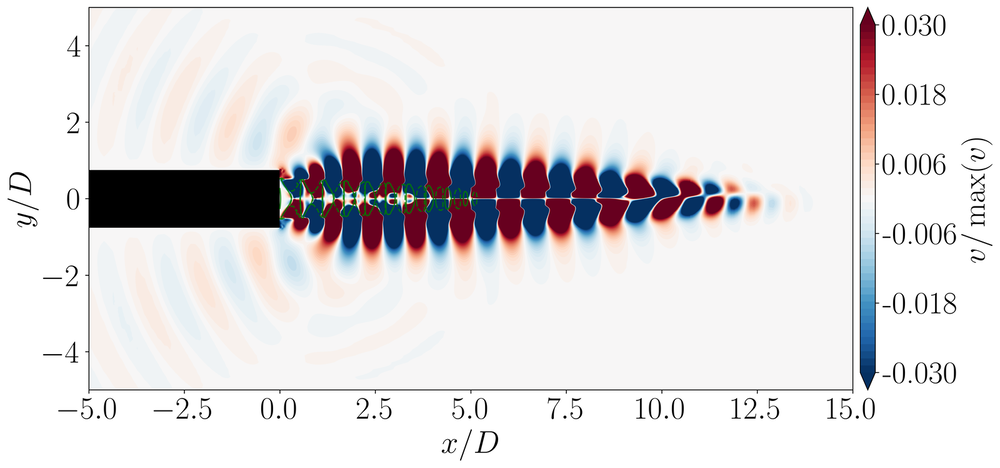}
	\end{subfigure}
	\begin{subfigure}[c]{0.49\linewidth}
		\centering
		\caption{}\label{modeA1_rho}
		\includegraphics[width=\linewidth,trim=0 0 0 0,clip]{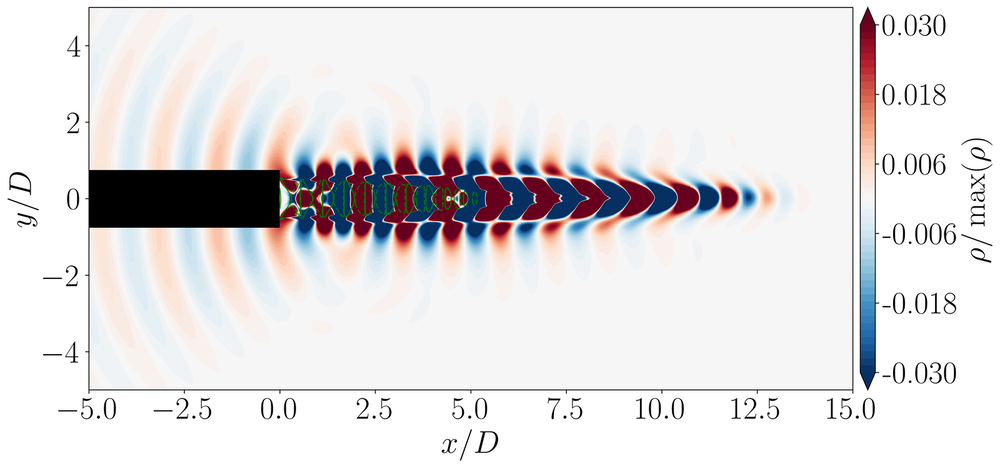}
	\end{subfigure}
	\begin{subfigure}[c]{0.49\linewidth}
		\centering
		\caption{}\label{modeA1_kt}
		\includegraphics[width=\linewidth,trim=0 0 0 0,clip]{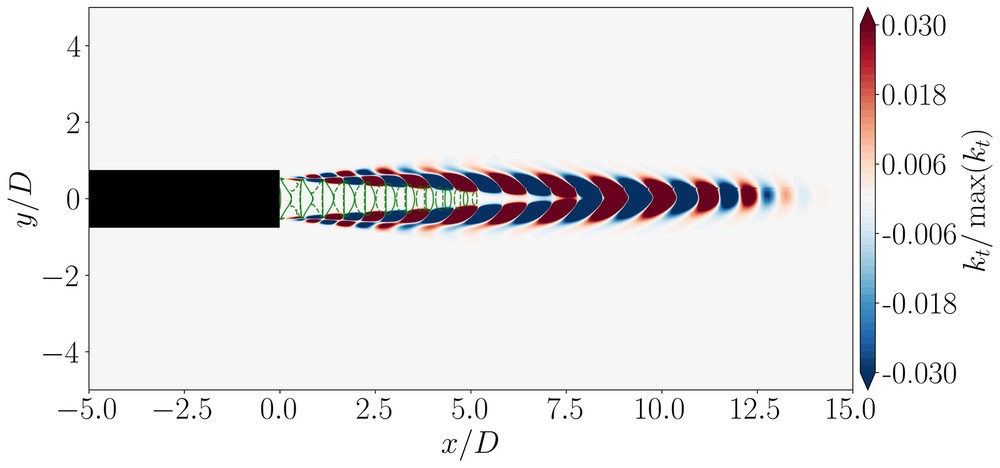}
	\end{subfigure}
	\caption{Spatial distribution of the real part of mode A1 at $M_j=1.1$. (a) Streamwise velocity component $u$, (b) radial velocity component $v$, (c) density $\rho$, (d) turbulent kinetic energy $k_t$. The green contours represent the shock-cell structure of the fixed point solution.}\label{modesA1_xy}
\end{figure}

\begin{figure}
	\centering
	\begin{subfigure}[c]{0.49\linewidth}
		\centering
		\caption{}\label{modeA2_ux}
		\includegraphics[width=\linewidth,trim=0 0 0 0,clip]{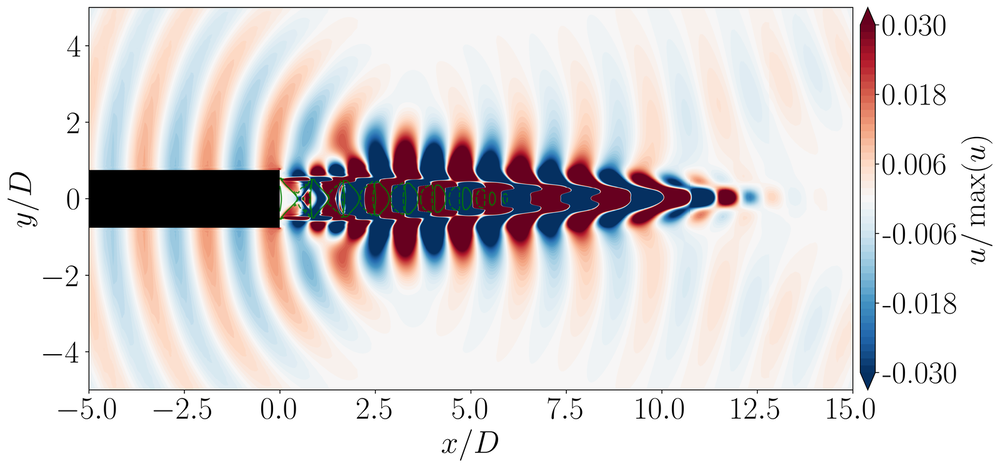}
	\end{subfigure}
	\begin{subfigure}[c]{0.49\linewidth}
		\centering
		\caption{}\label{modeA2_vx}
		\includegraphics[width=\linewidth,trim=0 0 0 0,clip]{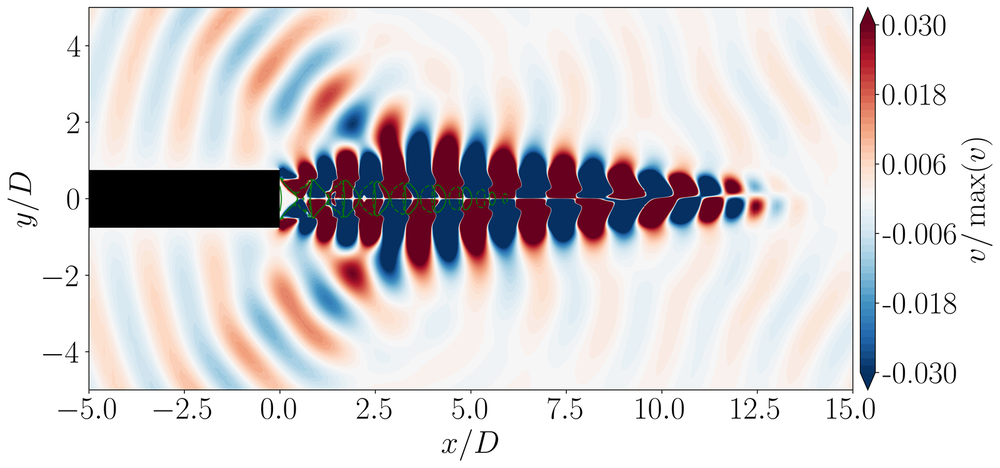}
	\end{subfigure}
	\begin{subfigure}[c]{0.49\linewidth}
		\centering
		\caption{}\label{modeA2_rho}
		\includegraphics[width=\linewidth,trim=0 0 0 0,clip]{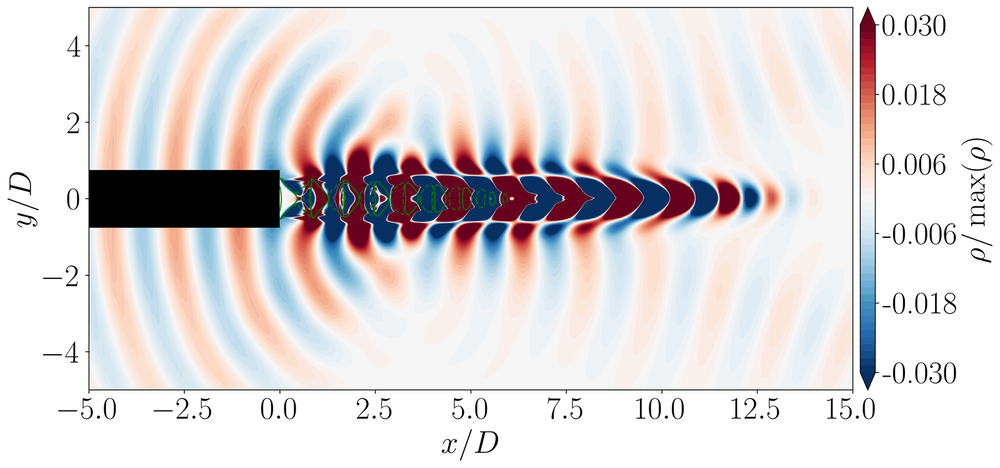}
	\end{subfigure}
	\begin{subfigure}[c]{0.49\linewidth}
		\centering
		\caption{}\label{modeA2_kt}
		\includegraphics[width=\linewidth,trim=0 0 0 0,clip]{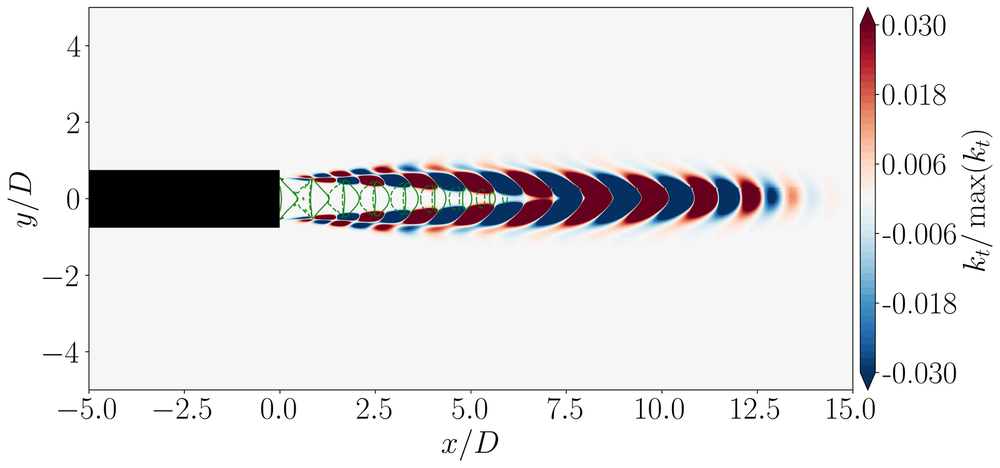}
	\end{subfigure}
	\caption{Spatial distribution of the real part of mode A2 at $M_j=1.2$. (a) Streamwise velocity component $u$, (b) radial velocity component $v$, (c) density $\rho$, (d) turbulent kinetic energy $k_t$. The green contours represent the shock-cell structure of the fixed point solution.}\label{modesA2_xy}
\end{figure}

 The spatial distribution of the axisymmetric modes, observed at $M_j=1.1$ (mode A1) and at $M_j=1.2$ (mode A2) is analysed. The real part of the perturbation fields is shown in \fref{modesA1_xy} and \fref{modesA2_xy}. The green isolines represent the shock-cell structure of the fixed point solution and are identified via the velocity divergence $\nabla \cdot \mathbf{u}$.

For jet operating condition, the streamwise velocity component $u$ is dominated by a  Kelvin--Helmholtz-type structure and represents the downstream-propagating part of the feedback loop. The spatial structure of the wavepacket highlight the shock-cell pattern, showing that the global mode is linked to the quasi-periodic shock structure of the fixed point solution.
The radial velocity component $v$ presents an antisymmetric structure in the $x-y$ plane. The alternating positive and negative regions are located around the jet boundary. This confirms that the perturbation corresponds to a toroidal oscillation of the jet, as expected for an $m=0$ mode. 
Density, streamwise velocity and radial velocity all share the characteristic upstream propagating acoustic waves typical of screech. Both modes A1 and A2 have a downstream propagating acoustic wave recognisable with the Mach wave radiation typical of supersonic jet flows.
Whithin the URANS framework, the turbulent kinetic energy $k_t$ is part of the perturbed variables, so that the turbulence modeling information can be analysed. $k_t$ is mainly inside the jet and in the shear-layer region. Contrary to the other fields, it does not display an acoustic radiation pattern. This is expected since the perturbation of the turbulence model is linked to the vortical part of the instability and to the regions where the shear layer is present, and where the turbulent viscosity is more active. The spatial distribution of $k_t$ follows the Kelvin--Helmholtz wavepacket and is also modulated by the shocks, indicating that the turbulent quantities are involved in the linear dynamics of the screech mode.

Comparing the two axisymmetric modes, A1 at $M_j=1.1$ is more compact and presents a shorter streamwise wavelength as expected. The perturbation reaches its largest amplitude closer to the nozzle and decays over a shorter distance. Mode A2 at $M_j=1.2$ has a larger streamwise wavelength and extends further downstream. This is consistent with the decrease of the screech frequency when the jet Mach number is increased. The acoustic component visible in the density field is also more developed for A2, suggesting a stronger radiation of the mode at this level of underexpansion.

\begin{figure}
	\centering
	\begin{subfigure}[c]{0.49\linewidth}
		\centering
		\caption{}\label{modeA1_crossec_u}
		\includegraphics[width=\linewidth,trim=0 0 0 0,clip]{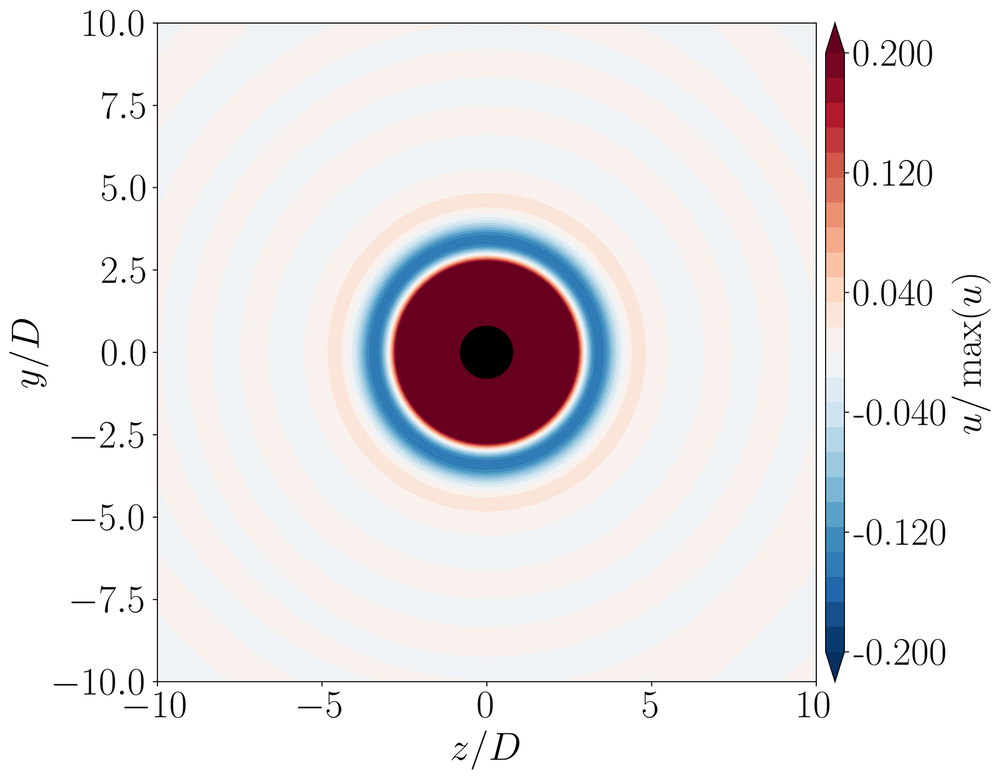}
	\end{subfigure}
	\begin{subfigure}[c]{0.49\linewidth}
		\centering
		\caption{}\label{modeA2_crossec_u}
		\includegraphics[width=\linewidth,trim=0 0 0 0,clip]{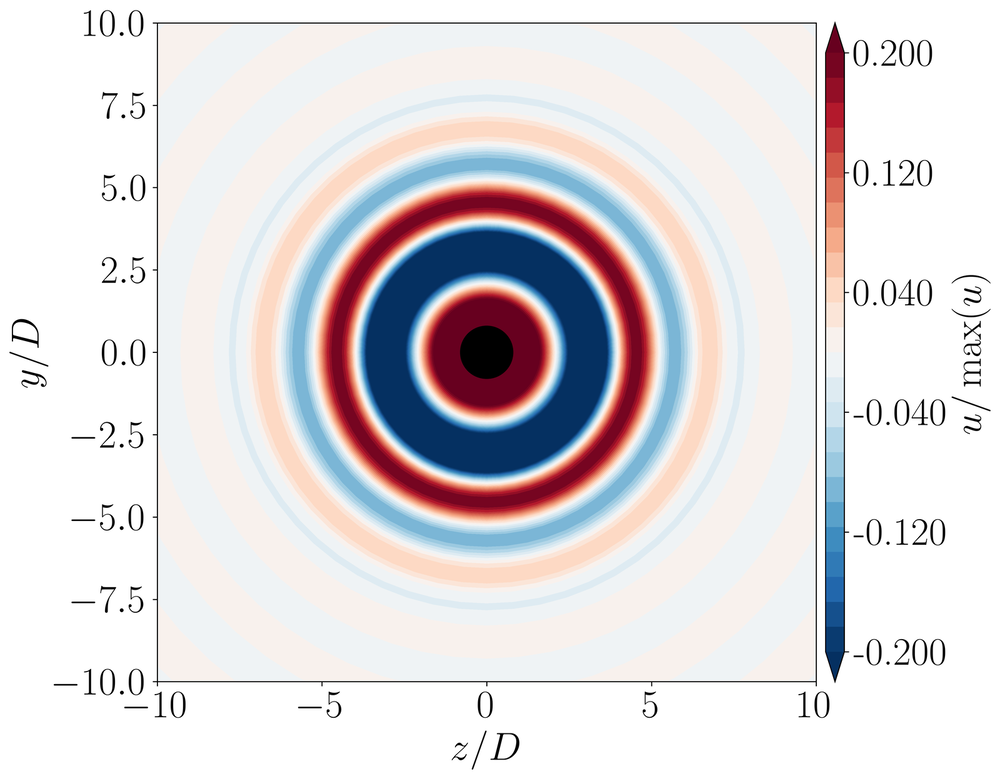}
	\end{subfigure}
	\caption{Cross-sectional distribution of the streamwise velocity component $u$ for the axisymmetric modes. Plane taken at $x = -0.5D$ (a) Mode A1 at $M_j=1.1$, (b) mode A2 at $M_j=1.2$.}\label{modesA_crossec}
\end{figure}

 \fref{modesA_crossec} shows the cross-sections $y-z$ plane taken at $x/D=-0.5$, clearly reveal the toroidal structure of the acoustic field that propagates upstream.

\subsection{Helical and flapping nature of the \texorpdfstring{$m=\pm 1$}{m=±1} modes}\label{sec:helical_flapping}

In this section the $m=\pm1$ modes are analysed. These modes are observed for different levels of underexpansion, namely at $M_j=1.2$ (mode B), $M_j=1.25$ (mode B) and $M_j=1.45$ (mode B and mode C). Since the base flow is axisymmetric, the linearised operator is invariant with respect to azimuthal rotations. 
Therefore, the perturbation can be decomposed into Fourier modes in the azimuthal direction. In this formalism, the modes $m = \pm 1$ correspond to two simple eigenvalues, being respectively left and right helical modes. In a fully 3-D formulation (no azimuthal Fourier decomposition), these two modes $m = \pm 1$ become an eigenvalue of multiplicity two.

This characteristic is recovered numerically. For example, for each case in which B or C modes are identified, two discrete eigenvalues appear in the spectrum with almost identical temporal growth rates $\sigma$ and angular frequencies $\omega$, as reported in \tabref{tab_3D_m1}. The small differences between the eigenvalues can be attributed to accumulation of numerical errors when calculating the fixed point or when solving the linear timestepper.

\begin{table}
	\caption{Growth rate $\sigma$ and angular frequency $\omega$ for the paired $m = \pm 1$ screech modes at different $M_j$.}
	\begin{center}
		\begin{tabular}{c|c|c|c}
			Mode   & $M_j$  &  $\sigma$   & $\omega$  \\
			\hline
			B     & 1.20   &  $-$0.01443 & 2.999761  \\
			B*     & 1.20   &  $-$0.01454 & 2.999790  \\
			\hline
			B     & 1.25   &  $+$0.01130 & 2.79102   \\
			B*     & 1.25   &  $+$0.01133 & 2.79104   \\
			\hline
			B     & 1.45   &  $-$0.03204 & 2.1932    \\
			B*     & 1.45   &  $-$0.03205 & 2.1935    \\
			\hline
			C     & 1.45   &  $-$0.07400 & 2.56569   \\
			C*     & 1.45   &  $-$0.07405 & 2.56555   \\
			\hline
		\end{tabular}\label{tab_3D_m1}
	\end{center}
\end{table}
Because the sub-eigenspace associated with this eigenvalue of multiplicity 2 is two-dimensional, any linear combination of the two eigenvectors is also an eigenvector. The numerical solver therefore recovers one basis of this eigenspace, which can depend on the numerical method, on the initial condition used in the Krylov procedure, or on small symmetry-breaking errors introduced by the discretisation. 
The resulting mode may appear as a helical mode, corresponding to a single rotating wave with azimuthal wavenumber $m=+1$ or $m=-1$, or as a flapping mode, produced by the superposition of the two counter-rotating components. 

\begin{figure}
	\centering
	\begin{subfigure}[c]{0.32\linewidth}
		\centering
		\caption{}\label{B1_azi_1.25}
		\includegraphics[width=\linewidth,trim=0 0 0 0,clip]{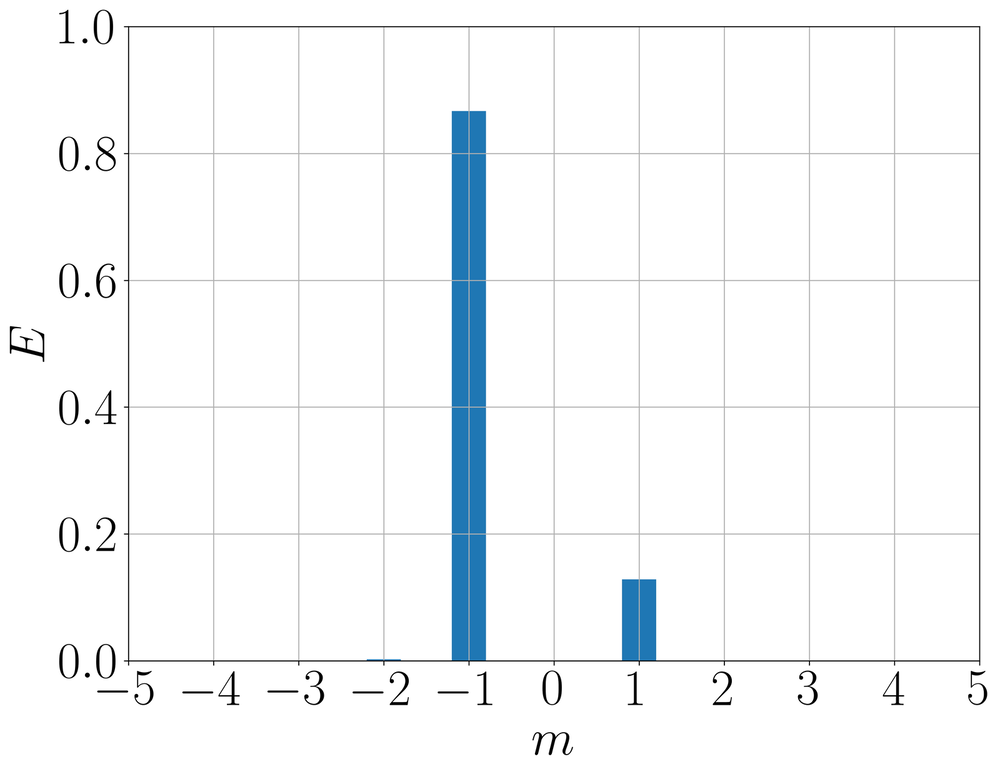}
	\end{subfigure}
	\begin{subfigure}[c]{0.32\linewidth}
		\centering
		\caption{}\label{B1_azi_1.45}
		\includegraphics[width=\linewidth,trim=0 0 0 0,clip]{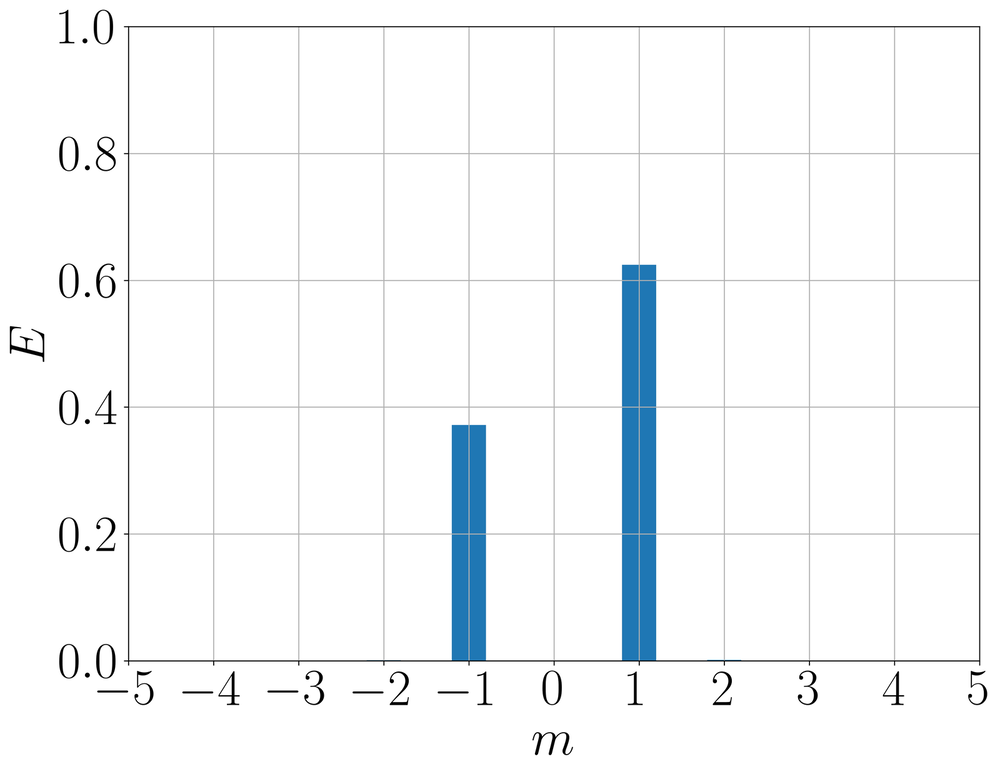}
	\end{subfigure}
	\begin{subfigure}[c]{0.32\linewidth}
		\centering
		\caption{}\label{C1_azi_1.45}
		\includegraphics[width=\linewidth,trim=0 0 0 0,clip]{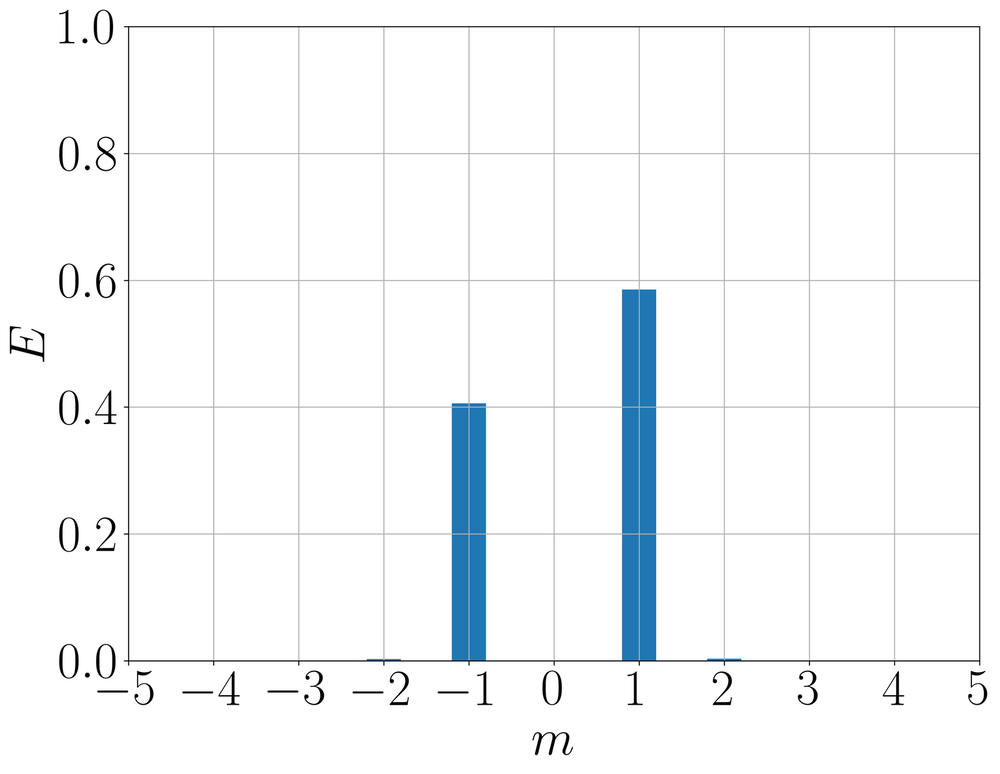}
	\end{subfigure}
	\caption{Azimuthal Fourier decomposition of the streamwise velocity component for the degenerate $m=\pm1$ modes. (a) Mode B at $M_j=1.25$, (b) mode B at $M_j=1.45$, (c) mode C at $M_j=1.45$.}\label{dec_azi}
\end{figure}

\begin{figure}
	\centering
	\begin{subfigure}[c]{0.32\linewidth}
		\centering
		\caption{}\label{azi_p_B145}
		\includegraphics[width=\linewidth,trim=0 0 0 0,clip]{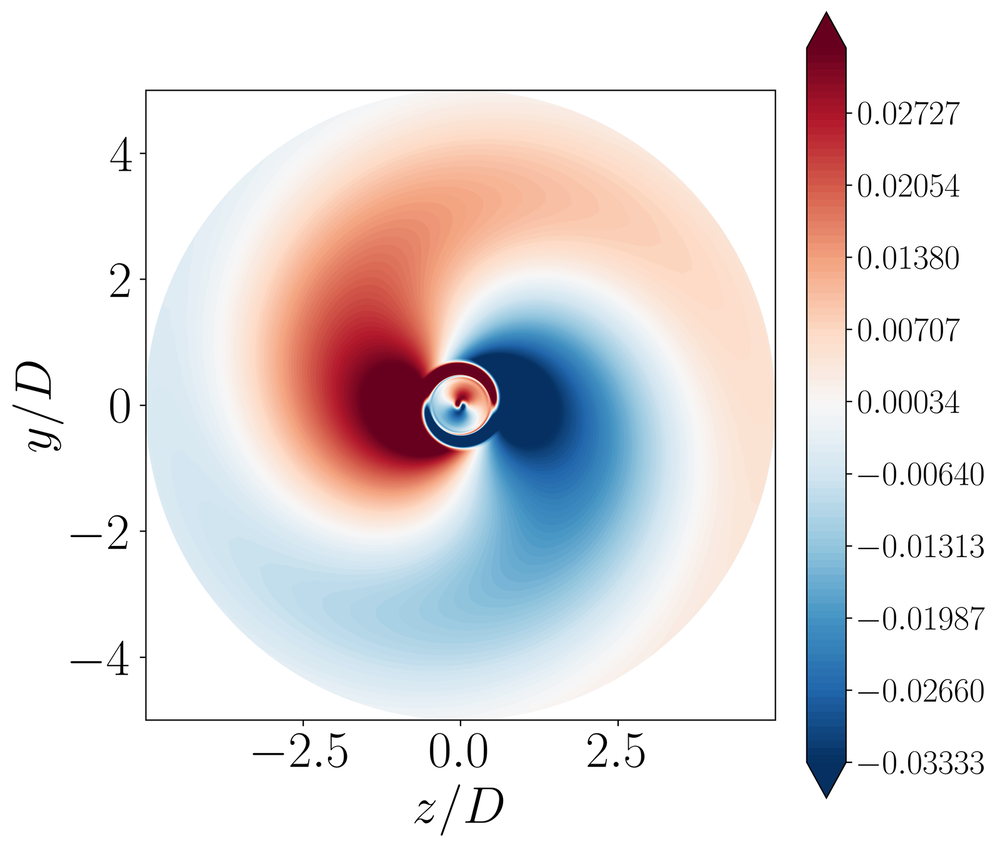}
	\end{subfigure}
	\begin{subfigure}[c]{0.32\linewidth}
		\centering
		\caption{}\label{azi_minus_B145}
		\includegraphics[width=\linewidth,trim=0 0 0 0,clip]{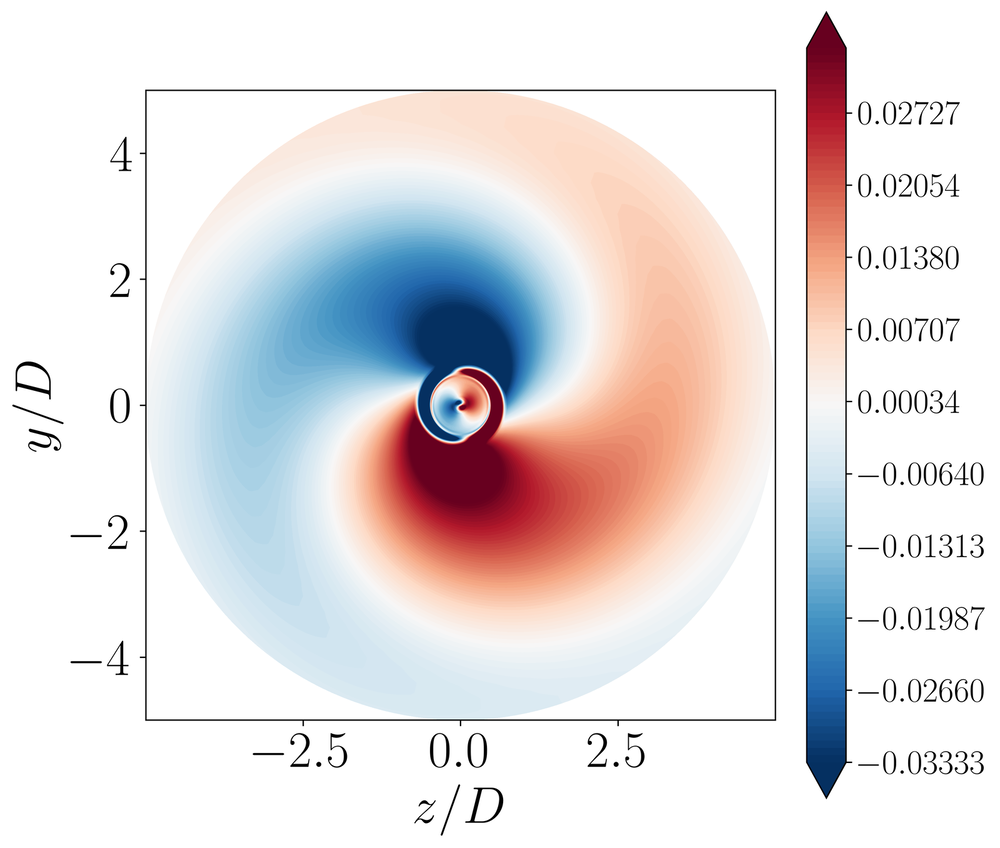}
	\end{subfigure}
	\begin{subfigure}[c]{0.32\linewidth}
		\centering
		\caption{}\label{azi_flap_B145}
		\includegraphics[width=\linewidth,trim=0 0 0 0,clip]{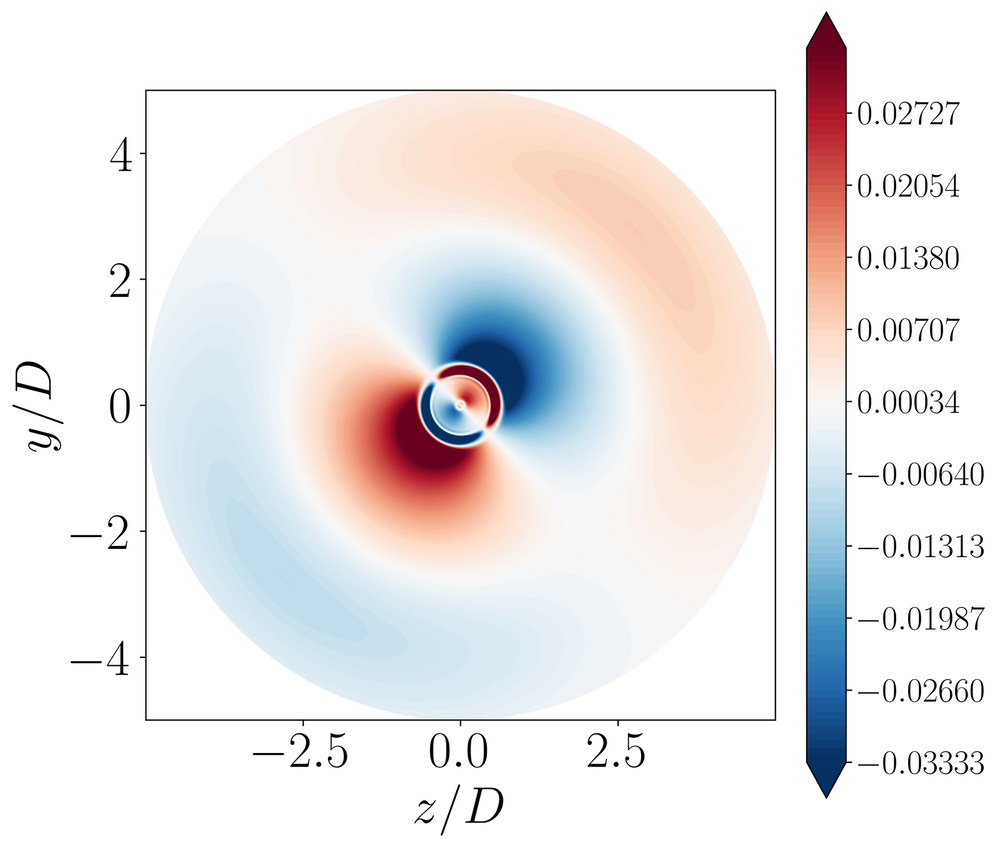}
	\end{subfigure}
	\caption{Azimuthal decomposition of mode B at $M_j=1.45$ at the cross-sectional plane $x/D=1$. (a) $m=+1$ component $u_{+1}$, (b) $m=-1$ component $u_{-1}$ (c) superposition $u_{+1}+u_{-1}$, which recovers a flapping pattern.}\label{azi_B145}
\end{figure}

However, when the basic flow is axisymmetric, each calculated eigenvalue has a unique azimuthal wavenumber value $m$, which is well verified in \fref{dec_azi}.
Indeed, the azimuthal Fourier decomposition of the mode B in \fref{dec_azi} shows that the energy of the computed eigenvectors is distributed over the two azimuthal wavenumbers $m=1$ and $m=-1$.
The repartition between the two components depends on the particular eigenvector returned by the solver, but the sum of the two eigenfunctions derived from the double eigenvalues is independent of the chosen basis.
In order to analyze these modes more simply, we have chosen to represent them in the orthonormal basis compatible with an approach where the azimuthal direction is decomposed into a Fourier series.
For example, by separating the two azimuthal Fourier components, a flapping structure can be reconstructed from the superposition of the two helical waves. This is shown in \fref{azi_B145} for mode B at $M_j=1.45$. Each component individually corresponds to a helical perturbation rotating in opposite directions, and their sum recovers a flapping pattern.

The selection of a purely helical or flapping structure in experiments is therefore expected to involve initial condition, boundary condition and small symmetry-breaking mechanisms. In particular, small asymmetries in the nozzle geometry or in the inflow conditions may favour one linear combination of the eigenspace. The selection of a flapping or helical structure cannot be determined from the linear axisymmetric eigenproblem alone. A receptivity mechanism must be specified in order to determine the weight of each mode in the dynamics.

For example, the B mode is generally observed as a flapping mode in experimental facilities, whereas the C mode has been reported over less consistent ranges of $M_j$ \citep{andre2011experimental,mercier2017experimental}. This suggests that the observed flapping or helical character is not fixed by the linear eigenspace alone, but also by the receptivity, and weak imperfections and at the end the nonlinearities of the experimental configuration.

In this sense, the flapping B mode can be interpreted as a robust eigenvector direction in the $m=\pm 1$ two-dimensional subspace, while the direction of the helical perturbation associated with mode C appears more sensitive to the details of the configuration. The present linear analysis does not by itself predict which combination will be selected in the saturated regime, but it shows that both helical and flapping perturbations belong to the same sub-eigenspace. 

In the following the real part of the perturbation is shown in the $x$--$y$ plane.

\begin{figure}
	\centering
	\begin{subfigure}[c]{0.49\linewidth}
		\centering
		\caption{}\label{modeB1_1.2_ux}
		\includegraphics[width=\linewidth,trim=0 0 0 0,clip]{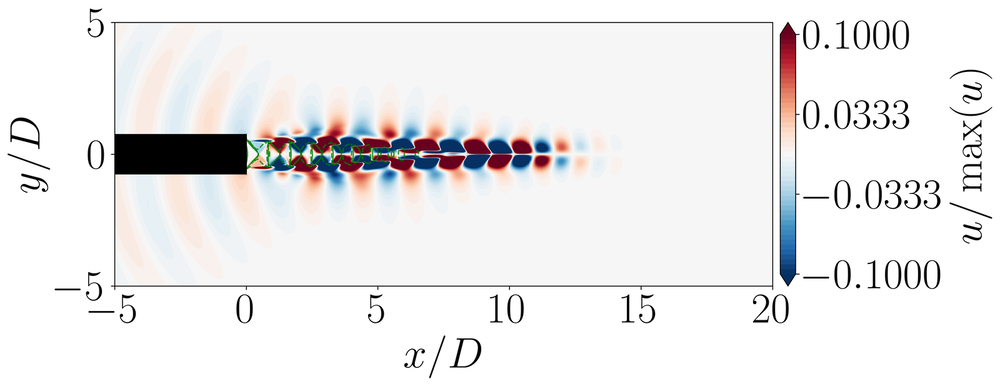}
	\end{subfigure}
	\begin{subfigure}[c]{0.49\linewidth}
		\centering
		\caption{}\label{modeB1_1.25_ux}
		\includegraphics[width=\linewidth,trim=0 0 0 0,clip]{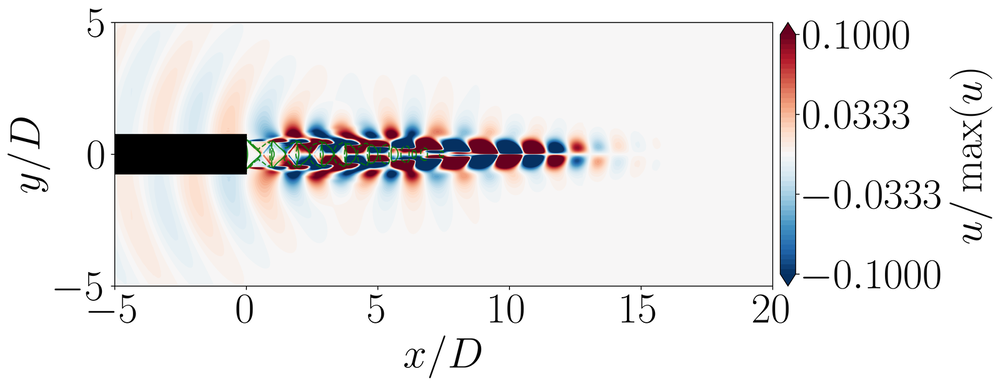}
	\end{subfigure}
	\begin{subfigure}[c]{0.49\linewidth}
		\centering
		\caption{}\label{modeB1_1.2_rho}
		\includegraphics[width=\linewidth,trim=0 0 0 0,clip]{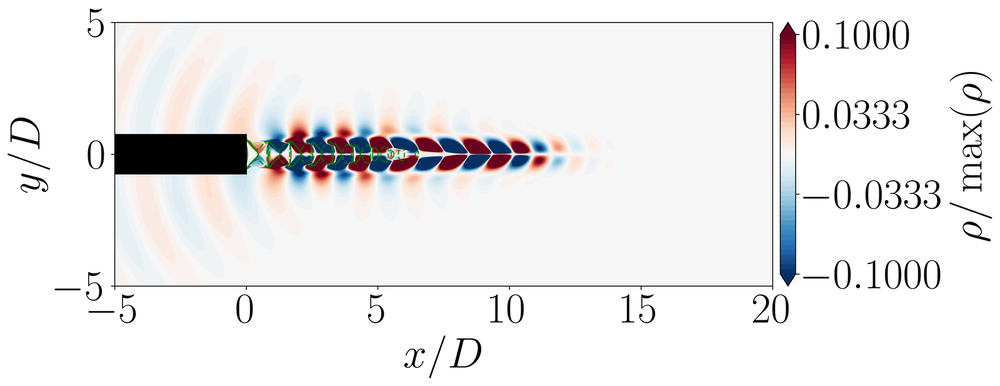}
	\end{subfigure}
	\begin{subfigure}[c]{0.49\linewidth}
		\centering
		\caption{}\label{modeB1_1.25_rho}
		\includegraphics[width=\linewidth,trim=0 0 0 0,clip]{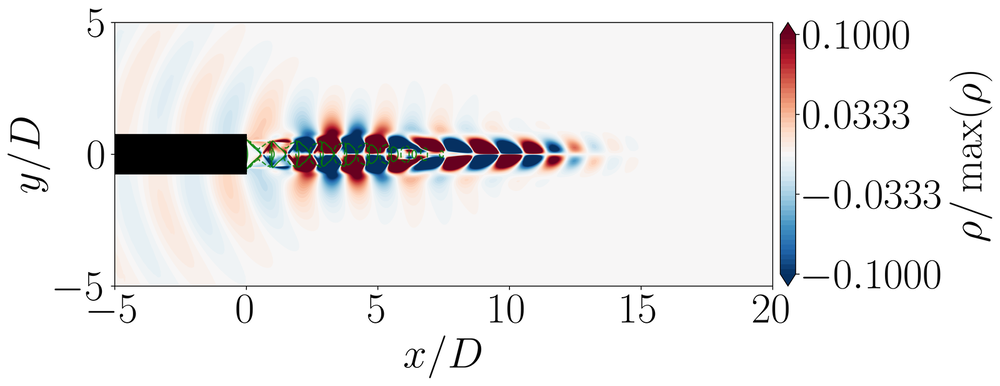}
	\end{subfigure}
	\begin{subfigure}[c]{0.49\linewidth}
		\centering
		\caption{}\label{modeB1_1.2_kt}
		\includegraphics[width=\linewidth,trim=0 0 0 0,clip]{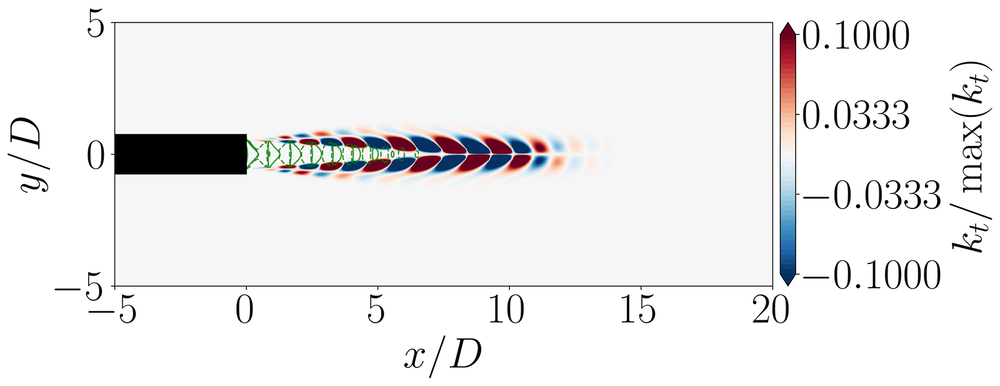}
	\end{subfigure}
	\begin{subfigure}[c]{0.49\linewidth}
		\centering
		\caption{}\label{modeB1_1.25_kt}
		\includegraphics[width=\linewidth,trim=0 0 0 0,clip]{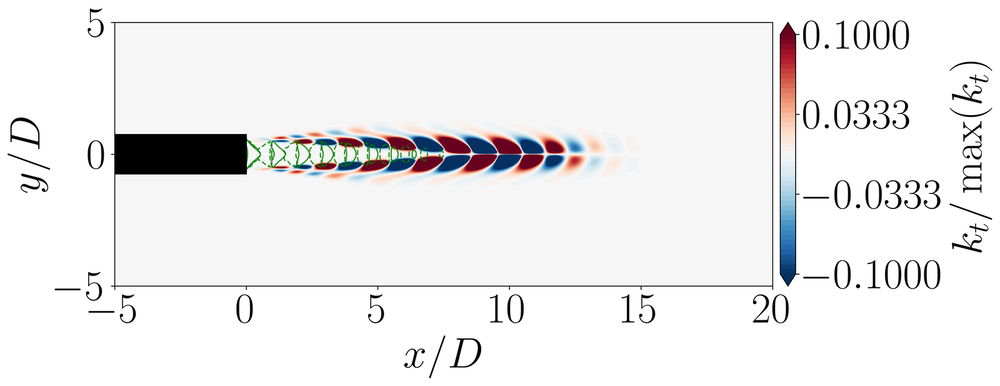}
	\end{subfigure}
	\caption{Spatial distribution of the real part of mode B for the weakly underexpanded cases. (a,c,e) $M_j=1.2$, (b,d,f) $M_j=1.25$. (a,b) Streamwise velocity component $u$, (c,d) density $\rho$, (e,f) turbulent kinetic energy $k_t$. The green contours represent the shock-cell structure of the fixed point solution.}\label{modesB1_weak_spatial}
\end{figure}

\begin{figure}
	\centering
	\begin{subfigure}[c]{0.49\linewidth}
		\centering
		\caption{}\label{modeB1_1.45_ux}
		\includegraphics[width=\linewidth,trim=0 0 0 0,clip]{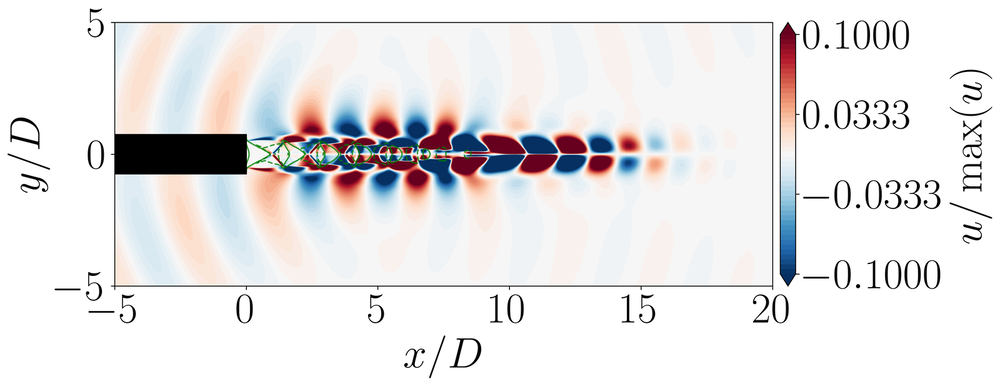}
	\end{subfigure}
	\begin{subfigure}[c]{0.49\linewidth}
		\centering
		\caption{}\label{modeC1_1.45_ux}
		\includegraphics[width=\linewidth,trim=0 0 0 0,clip]{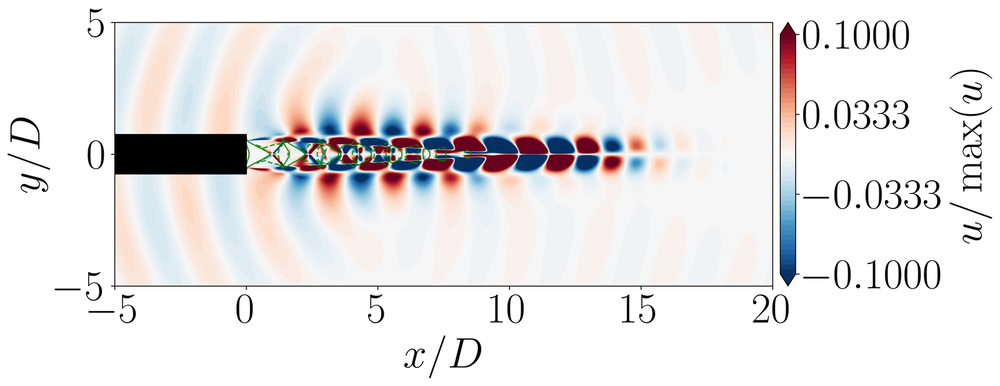}
	\end{subfigure}
	\begin{subfigure}[c]{0.49\linewidth}
		\centering
		\caption{}\label{modeB1_1.45_rho}
		\includegraphics[width=\linewidth,trim=0 0 0 0,clip]{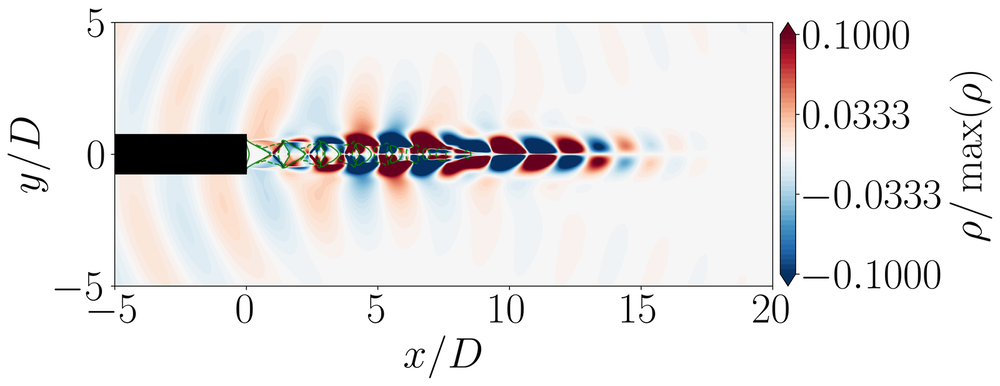}
	\end{subfigure}
	\begin{subfigure}[c]{0.49\linewidth}
		\centering
		\caption{}\label{modeC1_1.45_rho}
		\includegraphics[width=\linewidth,trim=0 0 0 0,clip]{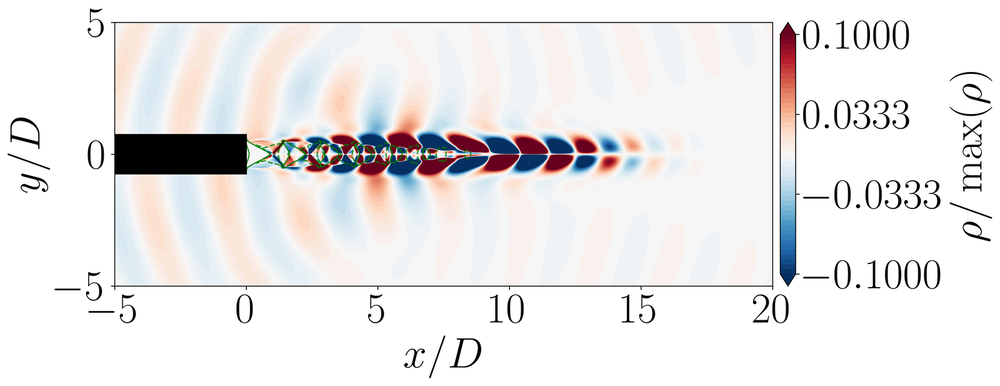}
	\end{subfigure}
	\begin{subfigure}[c]{0.49\linewidth}
		\centering
		\caption{}\label{modeB1_1.45_kt}
		\includegraphics[width=\linewidth,trim=0 0 0 0,clip]{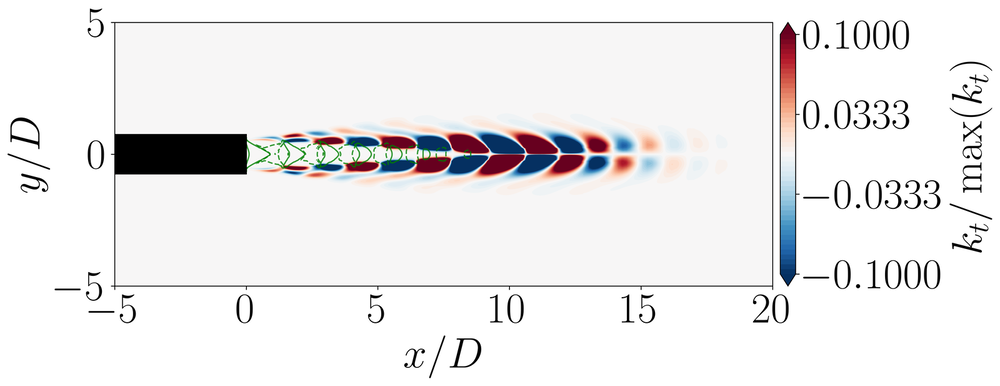}
	\end{subfigure}
	\begin{subfigure}[c]{0.49\linewidth}
		\centering
		\caption{}\label{modeC1_1.45_kt}
		\includegraphics[width=\linewidth,trim=0 0 0 0,clip]{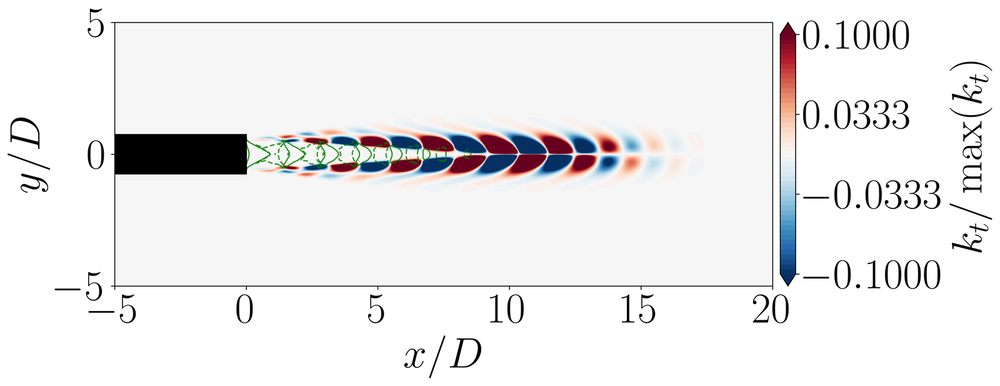}
	\end{subfigure}
	\caption{Spatial distribution of the real part of the $m=1$ modes at $M_j=1.45$. (a,c,e) Mode B, (b,d,f) mode C. (a,b) Streamwise velocity component $u$, (c,d) density $\rho$, (e,f) turbulent kinetic energy $k_t$. The green contours represent the shock-cell structure of the fixed point solution.}\label{modesBC_1.45_spatial}
\end{figure}

The spatial distribution of the $m=\pm1$ modes is similar to that of the axisymmetric modes. In all cases, the mode is dominated by the Kelvin-Helmholtz-type instability in the shear layer, while the upstream acoustic typical of screech is present.

The main difference with the axisymmetric case lies in the azimuthal organisation. In the present $x$--$y$ cuts, the pure $m=+1$ component appears as an antisymmetric perturbation with respect to the jet axis.

\fref{modesB1_weak_spatial} shows that the B mode keeps the same general structure between $M_j=1.2$ and $M_j=1.25$. Increasing the Mach number mainly increases the streamwise wavelenght of the wavepacket, since the the shock length also increses, showing the modulation of the mode with respect to the shock-cell structure.

The comparison between B and C at $M_j=1.45$ is shown in \fref{modesBC_1.45_spatial}. Both modes contain the same elements observed at lower Mach number, but they correspond to different resonant interactions with the base flow, as will be further analysed in the following section. Mode B has a longer streamwise wavelength, consistent with its lower frequency. This difference is visible in all components and in the relative position of the perturbation with respect to the shock-cell pattern.

\section{Waves and energy decomposition of the resonance loop}\label{sec:waves}
The modes presented in \secref{sec:stability} exhibit acoustic content and contain important information about the self-sustained mechanism governing screech. To extract the different waves involved in the instability mechanism, the modes are analysed in Fourier space, transformed along the streamwise direction $x$. This approach has recently been adopted in the works of \citet{EdgingtonMitchell2021,nogueira2022closure,edgington2022unifying}, where it was successfully used to characterise the waves that form the resonance loop and their interactions.
\subsection{Comparison with experimental data of the wave content}
\begin{figure}
	\centering
	\begin{minipage}[c]{0.9\linewidth} 
		\centering
		\begin{subfigure}[c]{0.49\linewidth}
			\caption{}\label{contour_fft_u_Mj_1.1}
			\includegraphics[width=\linewidth,trim=0 0 0 0,clip]{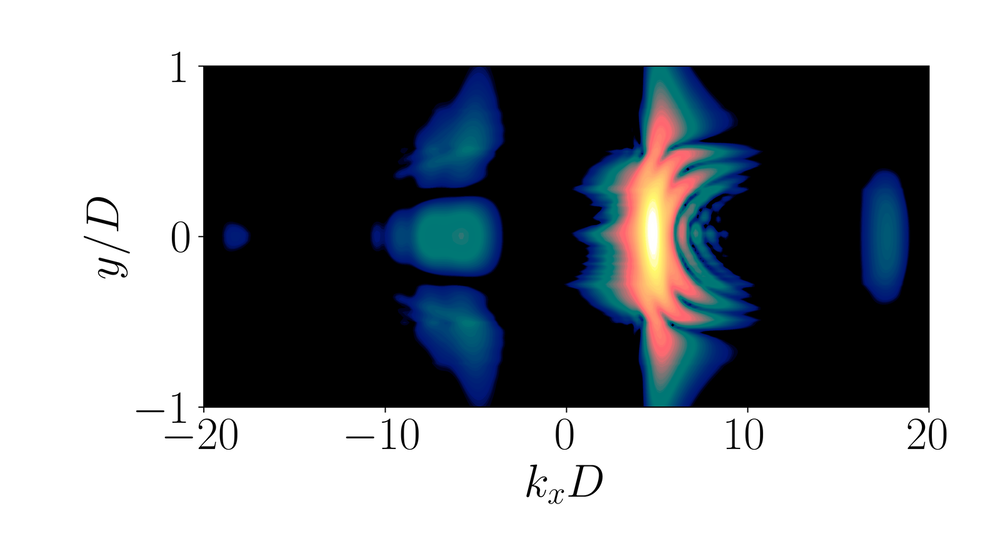}
		\end{subfigure}
		\begin{subfigure}[c]{0.49\linewidth}
			\caption{}\label{contour_expe_fft_u_Mj_1.1}
			\includegraphics[width=\linewidth,trim=0 0 0 0,clip]{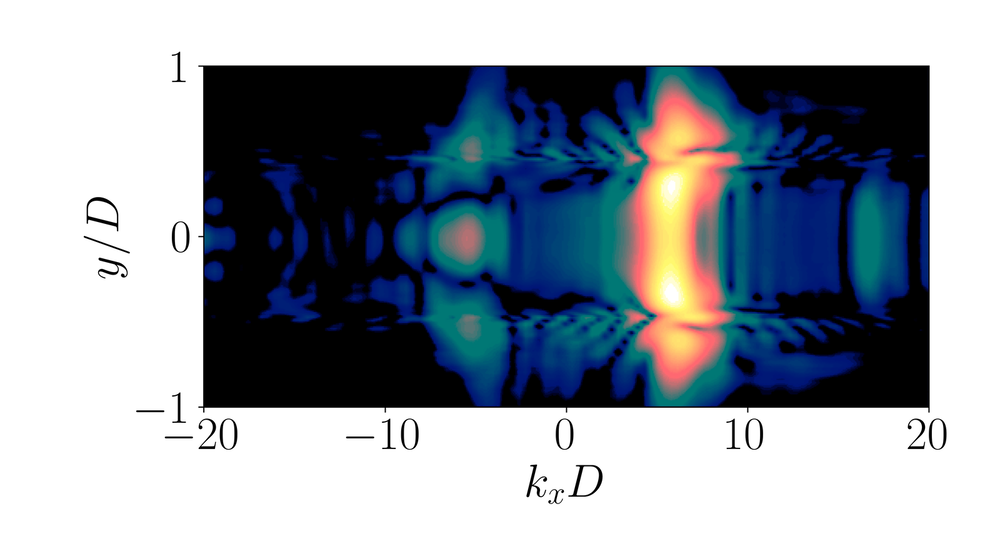}
		\end{subfigure}\\
		\begin{subfigure}[c]{0.49\linewidth}
			\caption{}\label{contour_fft_v_Mj_1.1}
			\includegraphics[width=\linewidth,trim=0 0 0 0,clip]{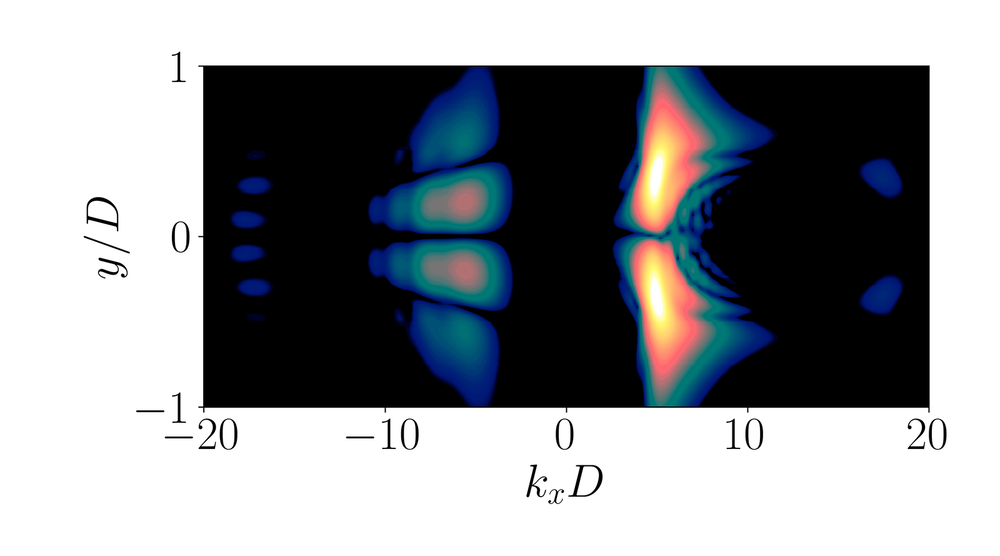}
		\end{subfigure}
		\begin{subfigure}[c]{0.49\linewidth}
			\caption{}\label{contour_expe_fft_v_Mj_1.1}
			\includegraphics[width=\linewidth,trim=0 0 0 0,clip]{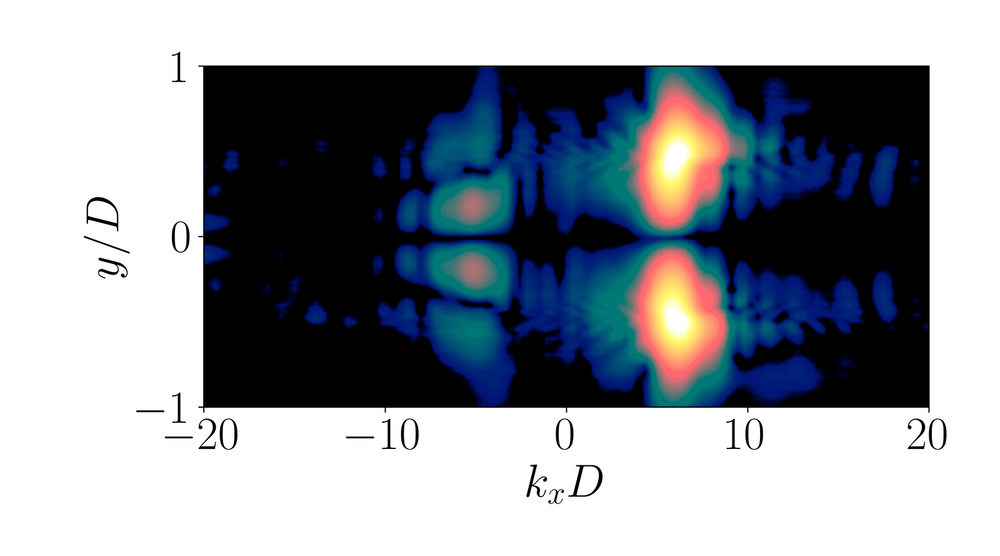}
		\end{subfigure}
	\end{minipage}
	\begin{minipage}[c]{0.09\linewidth}
				\centering
		\vspace{-0.1cm}
		\begin{subfigure}[c]{\linewidth}
			\includegraphics[width=\linewidth,trim=0 0 0 0,clip]{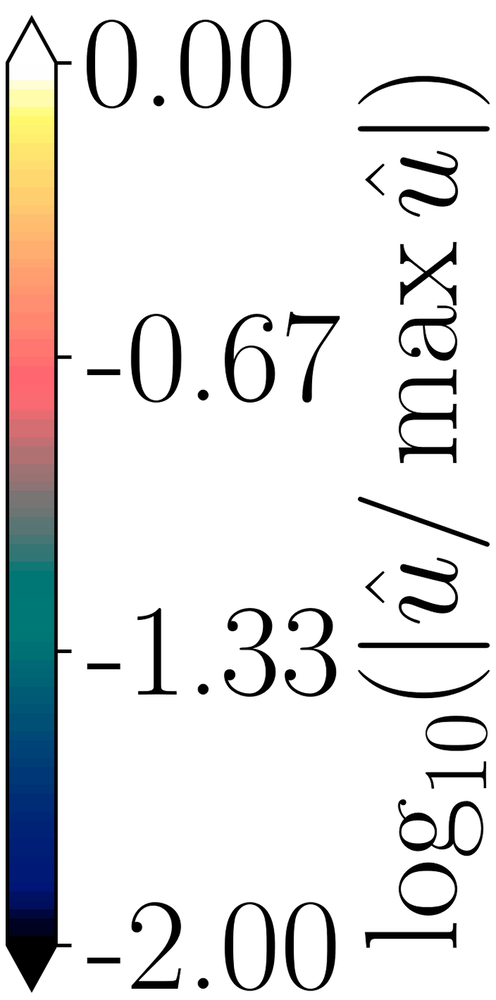}
		\end{subfigure}\\
		\begin{subfigure}[c]{\linewidth}
			\vspace{1.2cm}
			\includegraphics[width=\linewidth,trim=0 0 0 0,clip]{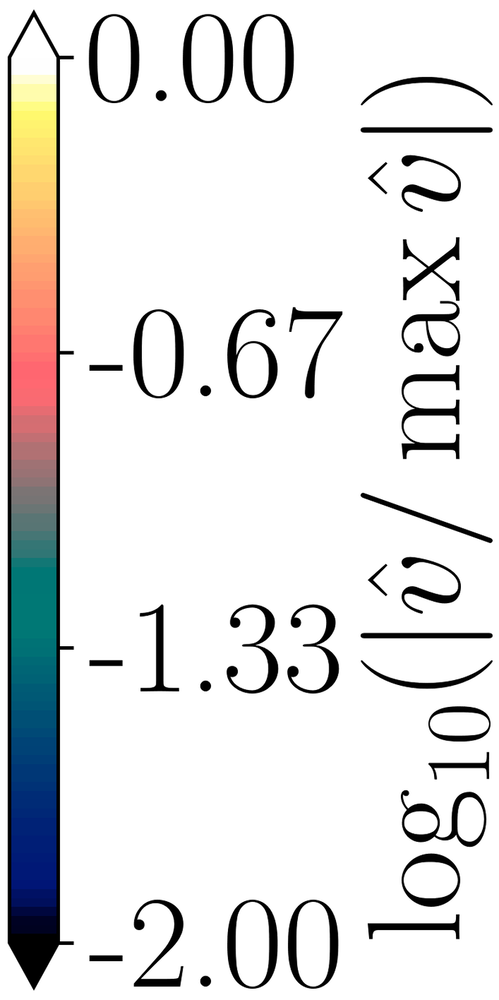}
		\end{subfigure}
	\end{minipage}
		\caption{Power spectral density of the velocity components compared to POD modes from experiments \citep{EdgingtonMitchell2021} for mode A1 at $M_j=1.1$, (experiments are at $M_j=1.09$). (a,b) Streamwise velocity component, (c,d) normal velocity component, (a,c) global stability analysis results, (b,d) experimental POD data.
		}\label{contour_fft_Mj_1.1}
	\end{figure}
A Fast Fourier Transform (FFT) of the streamwise and normal velocity components, $u$ and $v$, of the complex modes is performed in the streamwise direction on the two-dimensional plane defined at $z/D = 0$.

\fref{contour_fft_Mj_1.1} presents a comparison between the power spectral density of modes obtained from a POD of experimental data \citep{EdgingtonMitchell2021} and from global stability analysis for the $M_j = 1.1$ configuration. The contour plots show the intensity of the streamwise Fourier transform as a function of the streamwise wavenumber $k_x$ and the normal coordinate $y$. For both the streamwise and normal velocity components, three distinct structures are observed both in stability and experimental data.

The most energetic structure appears at approximately $k_x D \approx 5$ and corresponds to the spectral signature of a Kelvin--Helmholtz-like instability propagating downstream. The two other structures, located at higher wavenumber ($k_x D \approx 17$) and at negative wavenumber ($k_x D \approx -5.7$), result from the interaction between the downstream-propagating Kelvin--Helmholtz wave and the shock-cell structure.

These waves arise from the interaction between the Kelvin--Helmholtz wave and the shock-cell structure: one wave corresponds to the sum of the dominant shock-cell spectral peak $k_{s_1}$ and the Kelvin--Helmholtz wavenumber $k^+_{kh}$, while the second corresponds to their difference. Since, for all configurations considered, the wavenumber $k_{s_1}$ is larger than $k^+_{kh}$, the wave resulting from the difference,
\[
k^- =  k^+_{kh} - k_{s_1} 
\]
has a negative wavenumber.

The sign of the wavenumber corresponds to the sign of the phase velocity of the wave. Note that, for all cases studied in this work, the phase and group velocities share the same sign \citep{EdgingtonMitchell2021}. Negative wavenumbers therefore indicate upstream-propagating waves. The structure with negative phase velocity corresponds to the guided jet mode (GJM, or $k^-$), which has been the focus of numerous recent studies and has been identified as the wave responsible for closing the screech feedback loop and for several other jet-related acoustic phenomena, as discussed in \secref{sec:introduction}.

Comparison with experimental data shows that the three structures extracted from the stability analysis agree well with those identified in experimental POD modes \citep{EdgingtonMitchell2021}, in terms of spatial structure, amplitude, and wavenumber. Both the streamwise $u$ and normal $v$ velocity components exhibit good agreement with the experimental results.

\begin{figure}
	\centering
	\begin{minipage}[c]{0.9\linewidth} 
		\centering
		\begin{subfigure}[c]{0.49\linewidth}
			\caption{}\label{contour_fft_u_Mj_1.45}
			\includegraphics[width=\linewidth,trim=0 0 0 0,clip]{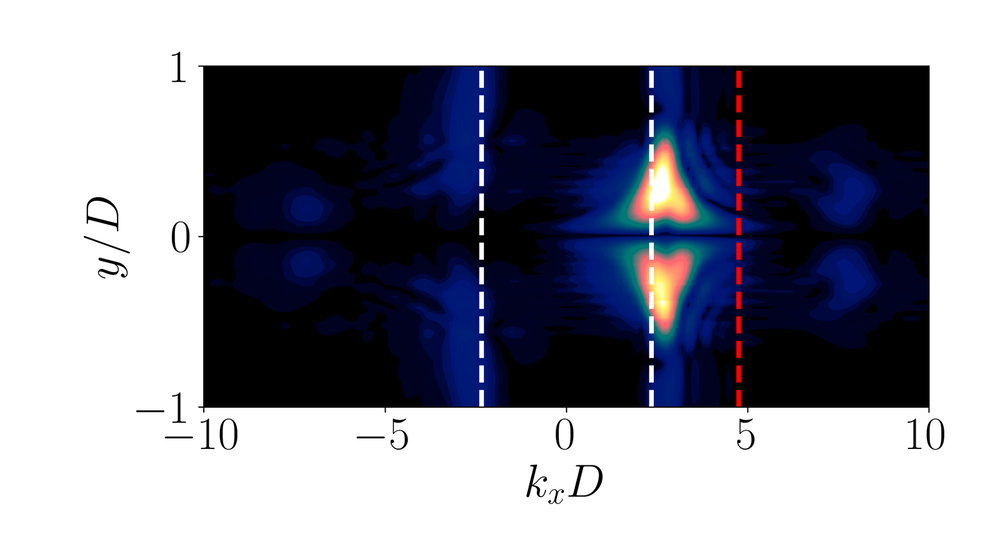}
			\end{subfigure}
		\begin{subfigure}[c]{0.49\linewidth}
			\caption{}\label{contour_expe_fft_u_Mj_1.45}
			\includegraphics[width=\linewidth,trim=0 0 0 0,clip]{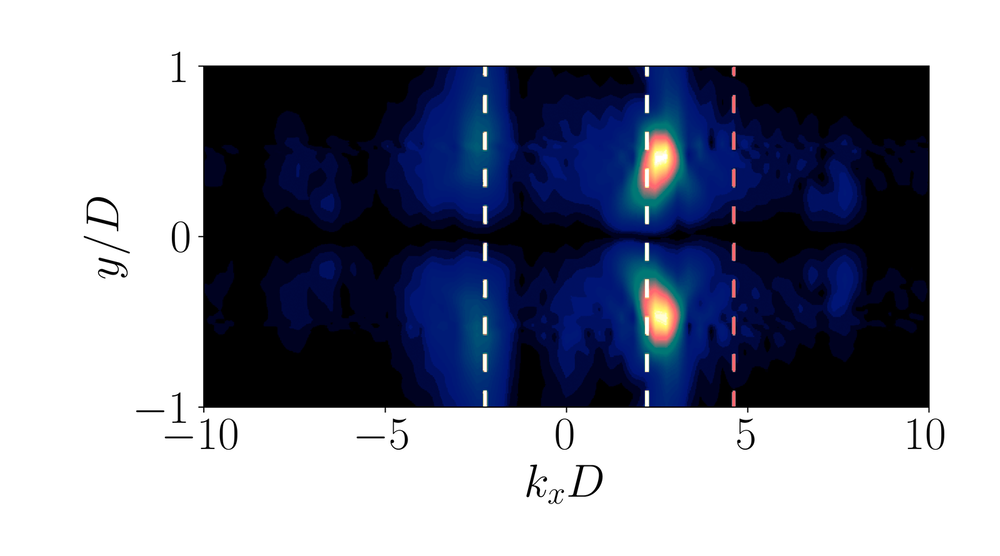}
		\end{subfigure}\\
		\begin{subfigure}[c]{0.49\linewidth}
			\caption{}\label{contour_fft_v_Mj_1.45}
			\includegraphics[width=\linewidth,trim=0 0 0 0,clip]{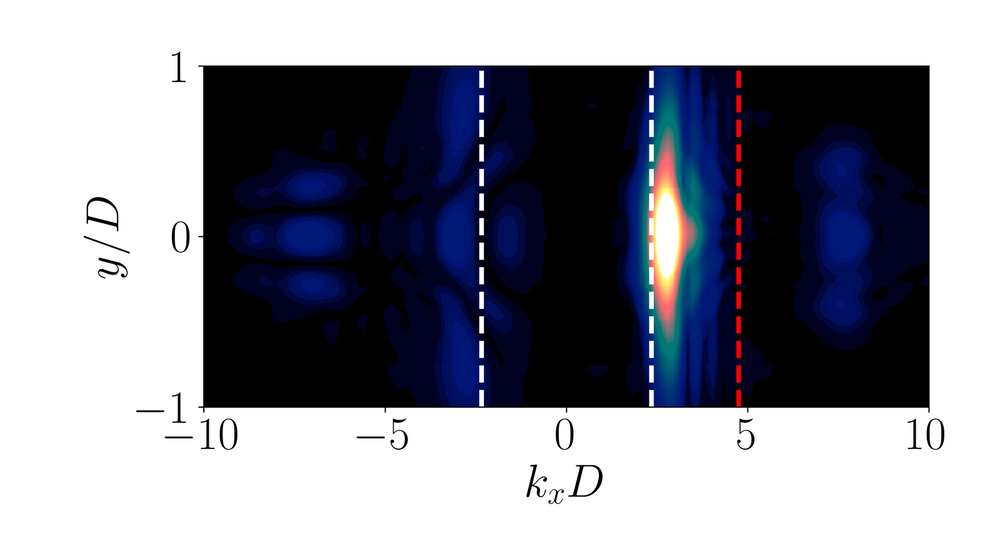}
		\end{subfigure}
		\begin{subfigure}[c]{0.49\linewidth}
			\caption{}\label{contour_expe_fft_v_Mj_1.45}
			\includegraphics[width=\linewidth,trim=0 0 0 0,clip]{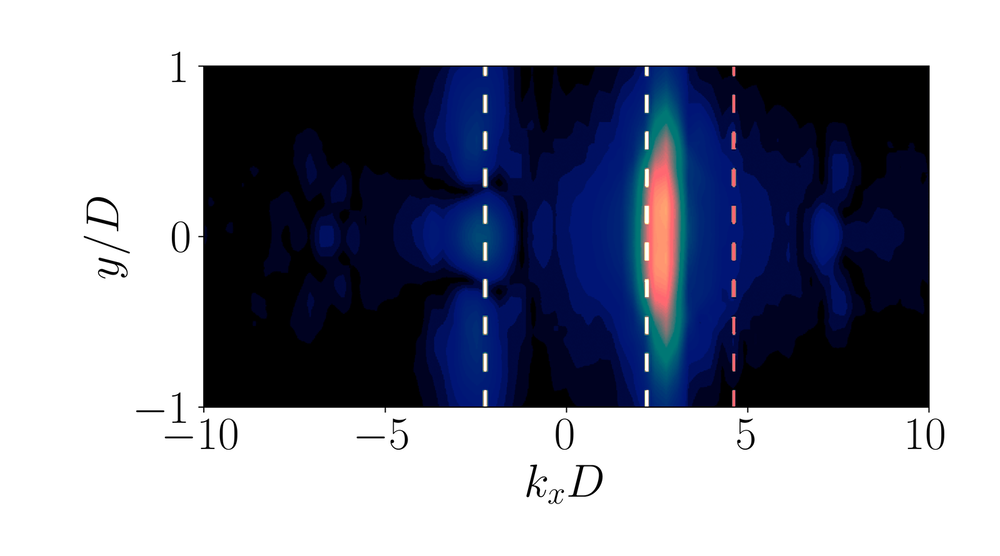}
		\end{subfigure}
	\end{minipage}
	\begin{minipage}[c]{0.09\linewidth}
		\centering
		\begin{subfigure}[c]{\linewidth}
			\includegraphics[width=\linewidth,trim=0 0 0 0,clip]{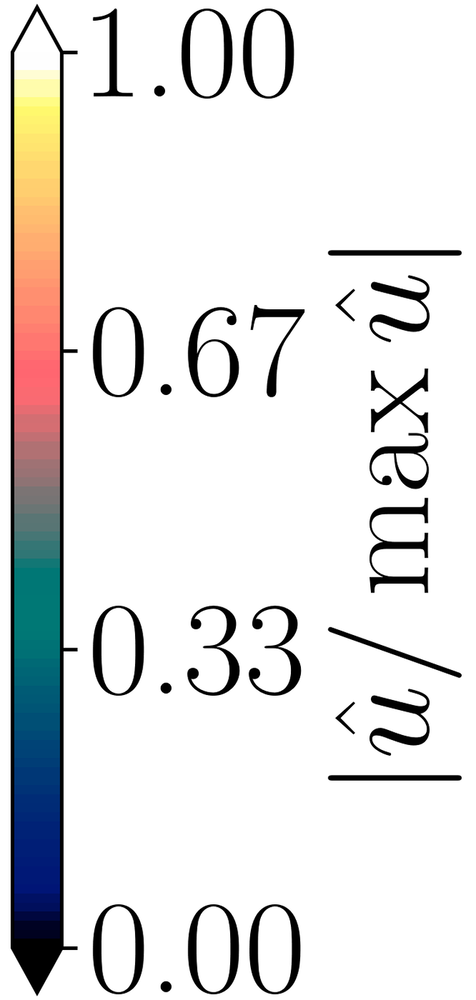}
		\end{subfigure}\\
		\begin{subfigure}[c]{\linewidth}
			\vspace{1.1cm}
			\includegraphics[width=\linewidth,trim=0 0 0 0,clip]{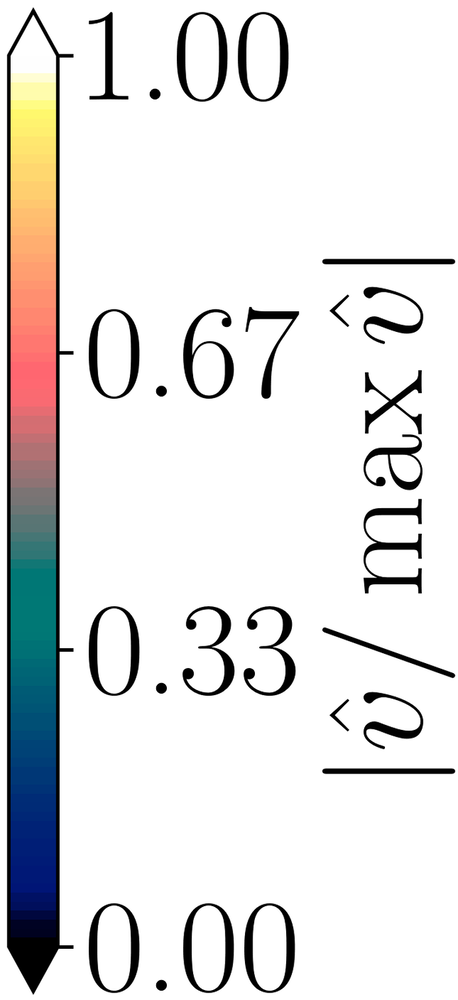}
		\end{subfigure}
	\end{minipage}
		\caption{Power spectral density of the velocity components compared to POD modes from experiments \citep{EdgingtonMitchell2021} for mode C at $M_j=1.45$. (a,b) Streamwise velocity component, (c,d) normal velocity component, (a,c) global stability analysis results, (b,d) experimental POD data. The white dashed lines represent the wavenumber associated with ambient sound speed $c_{\infty}$, the red dashed line is $k_{s_1}$.
		}\label{contour_fft_Mj_1.45}
	\end{figure}
	\fref{contour_fft_Mj_1.45} shows a comparison of the power spectral density (PSD) between stability analysis and experimental data for mode C at $M_j = 1.45$. For this configuration, the three structures appear with a different spatial distribution compared to those shown in \fref{contour_fft_Mj_1.1} at $M_j = 1.1$, since mode C at $M_j = 1.45$ is antisymmetric and not axially symmetric in the streamwise velocity component in the 2-D plane.

	 The plot clearly shows that the shock-cell wavenumber $k_{s_1} \approx 5$ is larger than the Kelvin--Helmholtz wavenumber $k^+_{kh} \approx 3$. Once again, the interaction given by the difference $k^+_{kh} - k_{s_1}$ yields a negative wavenumber, which is associated with the GJM.

	The expected three structures are observed in the PSD data at both positive and negative wavenumbers. All three exhibit lower wavenumbers compared to the $M_j = 1.1$ case, since at this level of underexpansion all waves, together with the shock spacing, have larger wavelengths. The three structures agree well with the experimental data in terms of amplitude, wavenumber, and spatial structure, reinforcing the idea that modes computed using URANS modelling correctly capture the resonance behaviour of screech.

	In addition, this analysis demonstrates that stability analysis and linear modes can accurately predict screech behaviour even in cases of strong underexpansion, where shocks are stronger and a linear methodology might not be expected \emph{a priori} to be applicable.

	The ability to compute the base flow while accurately capturing strong shocks is a key advantage of the URANS framework. Indeed, numerical shock-capturing schemes can resolve steep gradients in high-Mach-number flows, providing precise estimate on the effects of the shock-cell structure on the instability at high levels of under-expansion. It should be noted, however, that shock-capturing schemes are, in most cases, highly non-differentiable and often require automatic differentiation in order be used whithin a linear framework.

\begin{figure}
	\centering
	\begin{minipage}[c]{0.9\linewidth} 
		\centering
		\begin{subfigure}[c]{0.49\linewidth}
			\caption{}\label{contour_K_kh_u_Mj_1.1}
			\includegraphics[width=\linewidth,trim=0 0 0 0,clip]{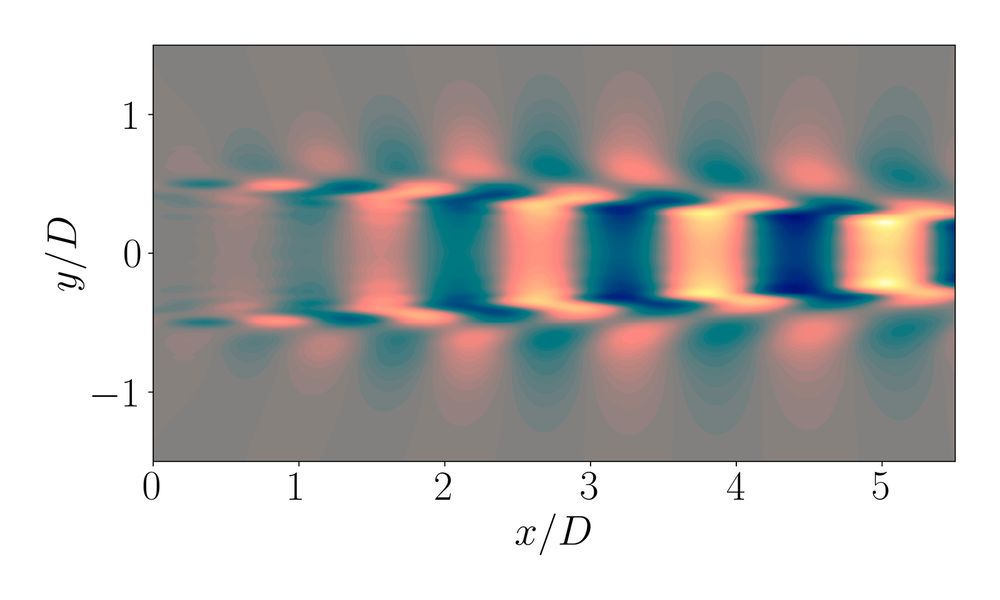}
		\end{subfigure}
		\begin{subfigure}[c]{0.49\linewidth}
			\caption{}\label{contour_expe_K_kh_u_Mj_1.1}
			\includegraphics[width=\linewidth,trim=0 0 0 0,clip]{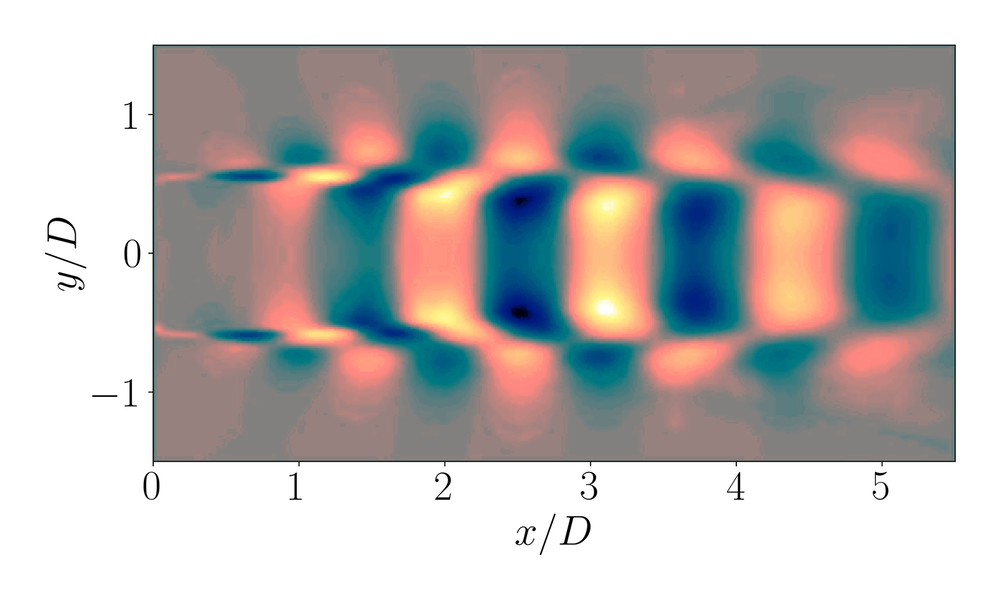}
		\end{subfigure}
		\begin{subfigure}[c]{0.49\linewidth}
			\caption{}\label{contour_K-_u_Mj_1.1}
			\includegraphics[width=\linewidth,trim=0 0 0 0,clip]{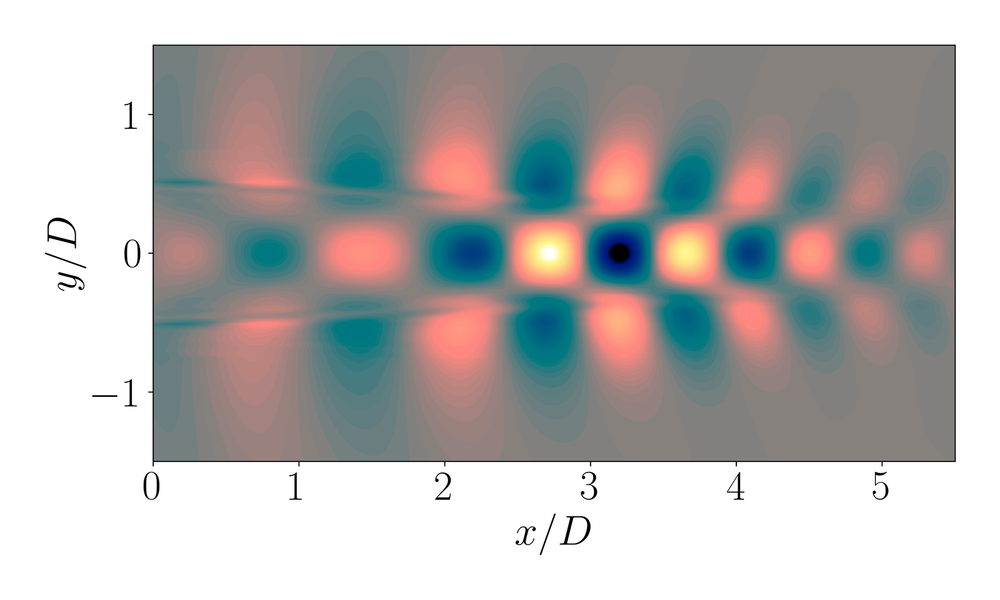}
		\end{subfigure}
		\begin{subfigure}[c]{0.49\linewidth}
			\caption{}\label{contour_expe_K-_u_Mj_1.1}
			\includegraphics[width=\linewidth,trim=0 0 0 0,clip]{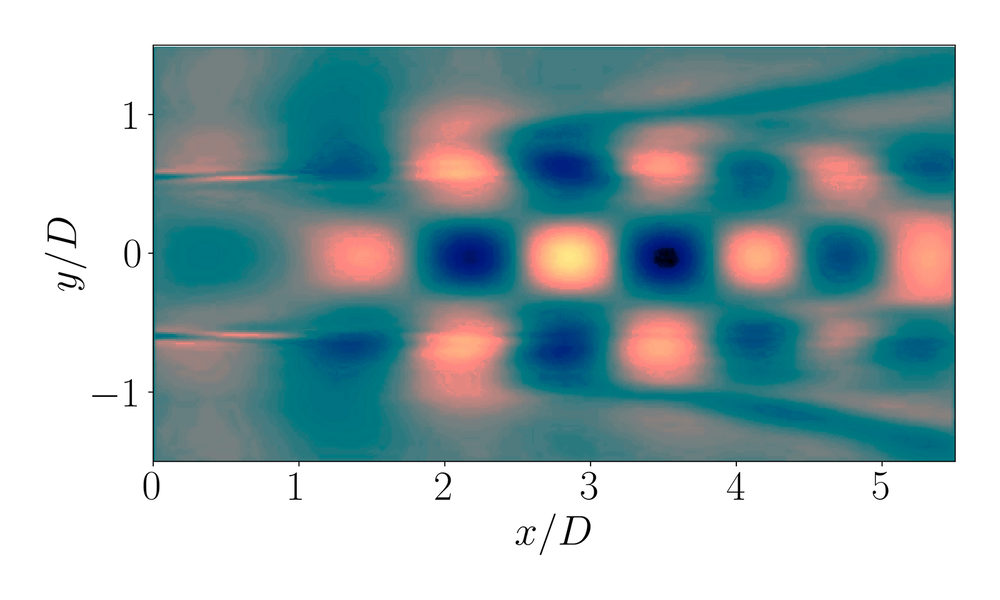}
		\end{subfigure}
	\end{minipage}
	\begin{minipage}[c]{0.09\linewidth}
		\centering
		\includegraphics[width=\linewidth]{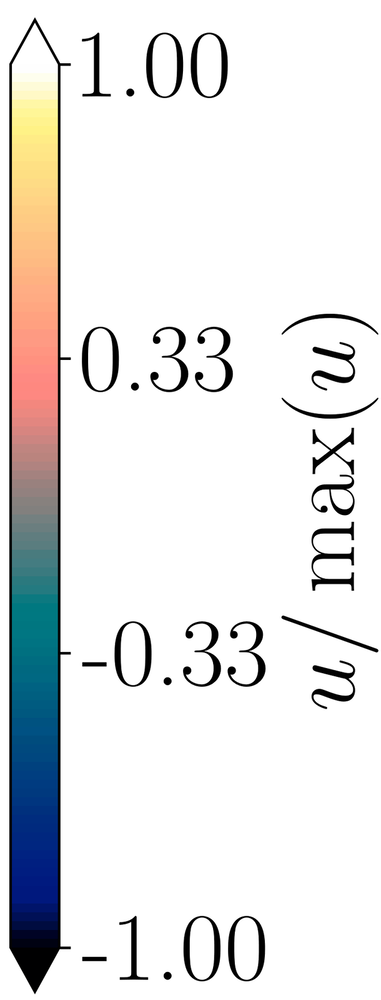}
	\end{minipage}
		\caption{Spatial distribution of the extracted waves from the streamwise velocity $u$ compared to POD modes from experiments \citep{EdgingtonMitchell2019} for mode A1 at $M_j=1.1$. (a,c) Global stability analysis, (b,d) POD data, (a,b) $k^+_{kh}$ wave (c,d) $k^-$ wave
}\label{contour_K_Mj_1.1}
	\end{figure}

	To identify the different wave components of the mode, the spatial distribution of the wave structures is extracted by filtering the streamwise spectral data using a cosine-tapered filter with a window size of $k_w D = 2\pi$, centred on the maximum PSD level of each structure. Once a given structure is isolated, an inverse transform is applied and the spatial distribution of the wave is reconstructed in physical space, following the methodology of \citet{EdgingtonMitchell2021}.

	\fref{contour_K_Mj_1.1} shows a comparison of the spatial distributions of the KH and GJM waves obtained from stability analysis (\fref{contour_K_kh_u_Mj_1.1} and \fref{contour_K-_u_Mj_1.1}) with those extracted from POD of experimental data \citep{EdgingtonMitchell2021} (\fref{contour_expe_K_kh_u_Mj_1.1} and \fref{contour_expe_K-_u_Mj_1.1}). The real part of the streamwise velocity component is shown. Both the KH and GJM waves show qualitative agreement with the experimental data. The GJM only shows minor discrepancies in the subsonic region, where the experimental data appear noisier than those obtained from stability analysis.

	\begin{figure}
		\begin{subfigure}[c]{0.49\linewidth}
			\caption{}\label{contour_k-radial_u_Mj_1.1}
			\includegraphics[width=\linewidth,trim=0 0 0 0,clip]{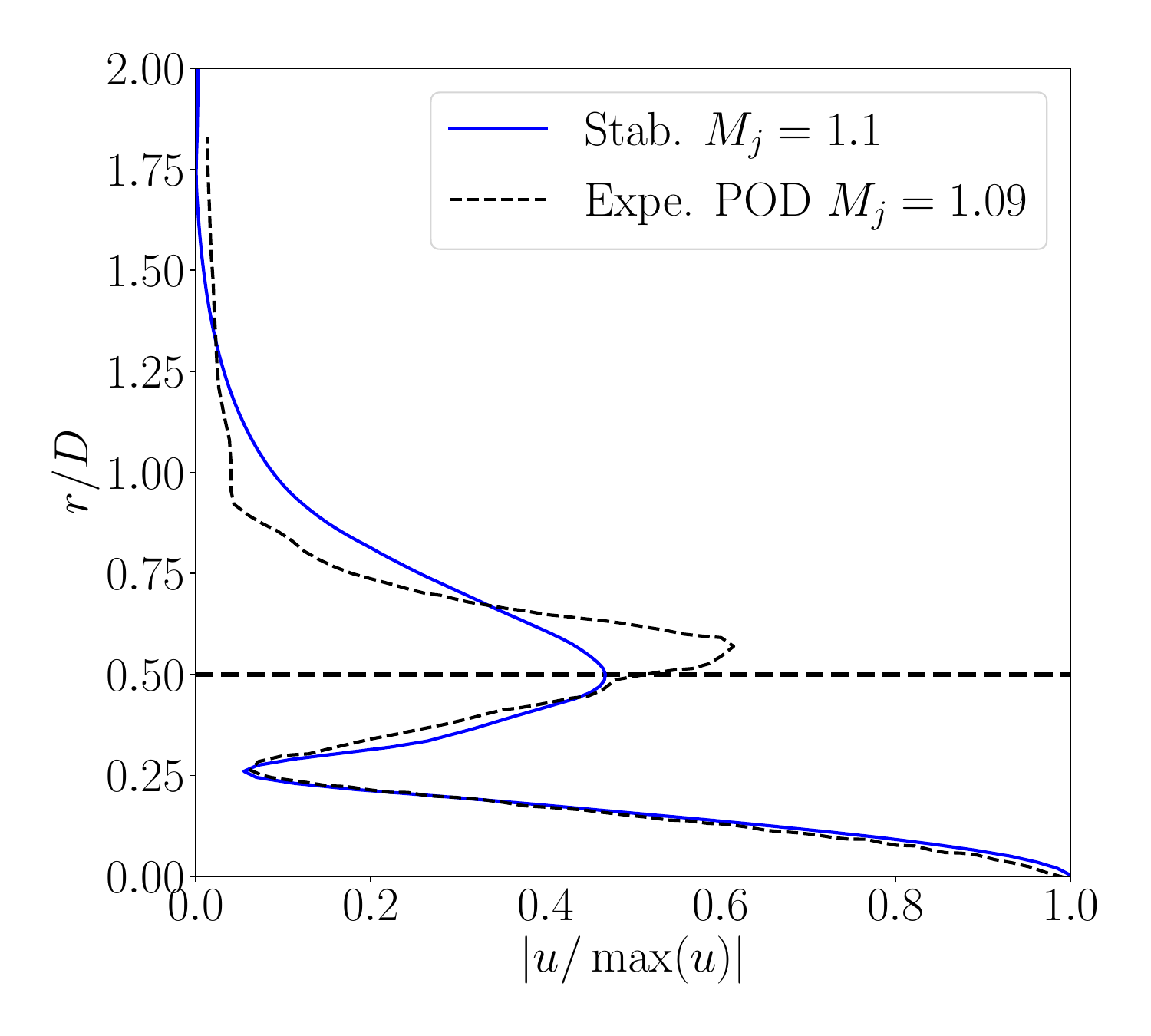}
		\end{subfigure}
		\begin{subfigure}[c]{0.49\linewidth}
			\caption{}\label{contour_kkh_radial_u_Mj_1.1}
			\includegraphics[width=\linewidth,trim=0 0 0 0,clip]{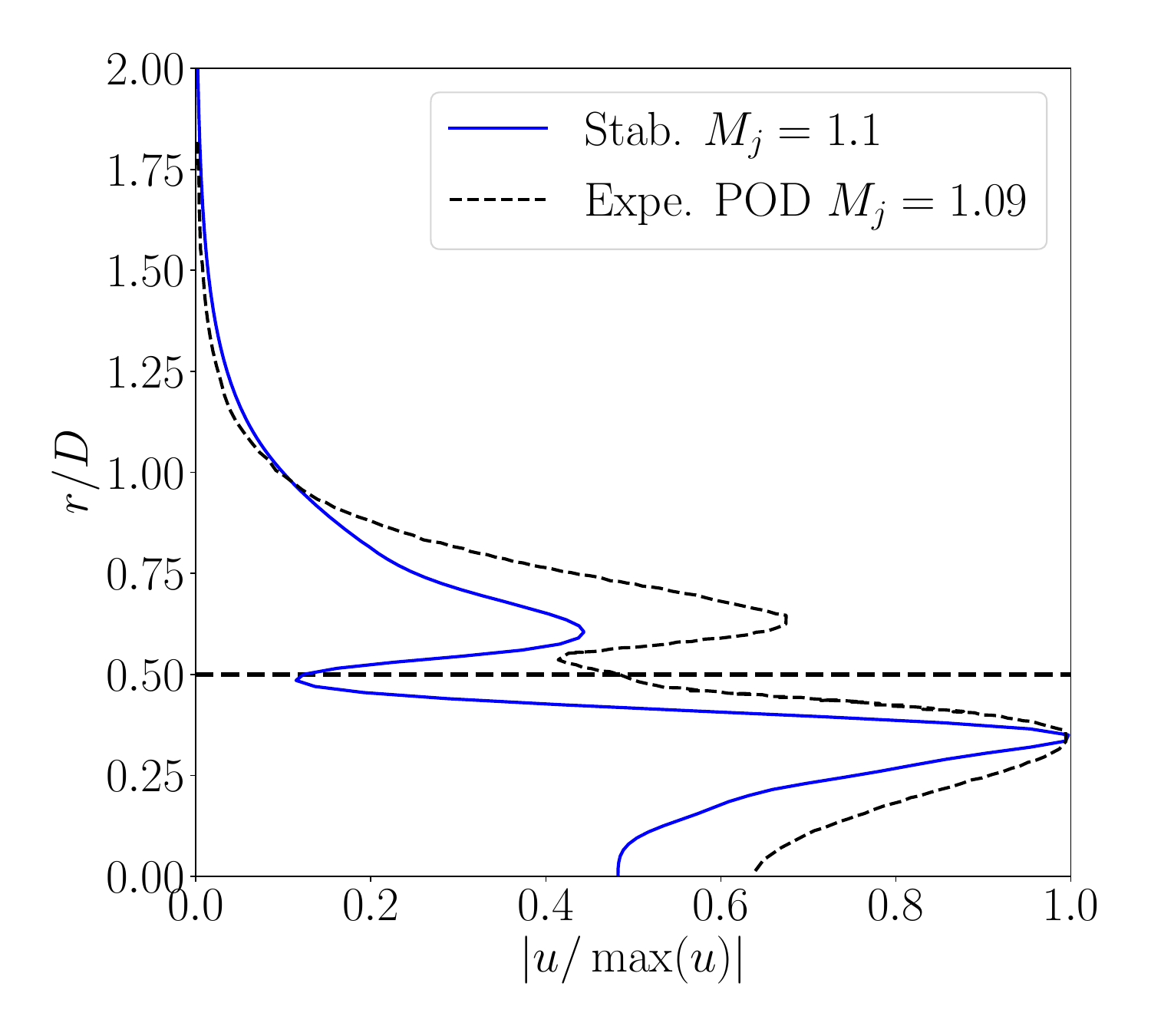}
		\end{subfigure}
		\caption{Radial structure of the real part of streamwise velocity component for mode A1 at $M_j=1.1$ at $x/D=2.75$. (a) $k^-$ wave,  (b) $k^+_{kh}$ wave. Comparison with experimental data from \cite{EdgingtonMitchell2021}. }\label{contour_radial_Mj_1.1}
	\end{figure}

It has been extensively demonstrated \citep{mancinelli2021complex,bogey2017feedback,nogueira2024guided,Tam_Hu_1989} that the GJM propagates upstream of the jet, even in supersonic configurations.
By analysing the radial structure of the GJM (\fref{contour_k-radial_u_Mj_1.1}), its support is seen primarily in the supersonic region of the jet. However, for the wave to propagate upstream, it must also have positive support in the subsonic region of the jet \citep{Tam_Hu_1989}. Comparison of the GJM radial structure with POD data from \cite{EdgingtonMitchell2021} shows good agreement across almost the entire support. For comparison, \fref{contour_kkh_radial_u_Mj_1.1} shows the radial structure of the KH wave. The main difference is that, while the GJM grows and reaches a peak amplitude at the shear layer, the KH mode is minimal at the jet boundary, with its support extending through both the subsonic and supersonic regions of the flow.

\subsection{The staging phenomenon}
As previously mentioned, the screech tone exhibits a staging behaviour in terms of acoustic frequency. As shown in \fref{compare_expe}, for certain levels of underexpansion, the screech tone undergoes a jump in frequency and symmetry for a small change in $M_j$, transitioning from a toroidal acoustic wave to a flapping/helical wave.

Recent studies \citep{nogueira2022closure,edgington2022unifying} have shown that this change in frequency is associated with the interaction between the downstream-propagating KH wave and the shock-cell structure of the jet. The shock-cell structure exhibits a quasi-periodic pattern in the streamwise direction, with two dominant modes appearing in the Fourier spectra at $k_{s_1}$ and $k_{s_2}$, respectively (\fref{FFT_line_fp}). It is argued that, for modes A1 and B, the KH wave interacts primarily with the first shock-cell mode, whereas for modes A2 and C, it interacts with the second. Upstream-propagating waves with distinct spatio-temporal wavenumbers are then generated, explaining the jumps in screech frequency observed in the experiments, depending on which interaction exhibits the dominant growth rate.

\begin{figure}
	\centering
	\begin{subfigure}[c]{0.49\linewidth}
		\caption{}\label{contour_staging_u_Mj_1.1}
		\includegraphics[width=\linewidth,trim=0 0 0 0,clip]{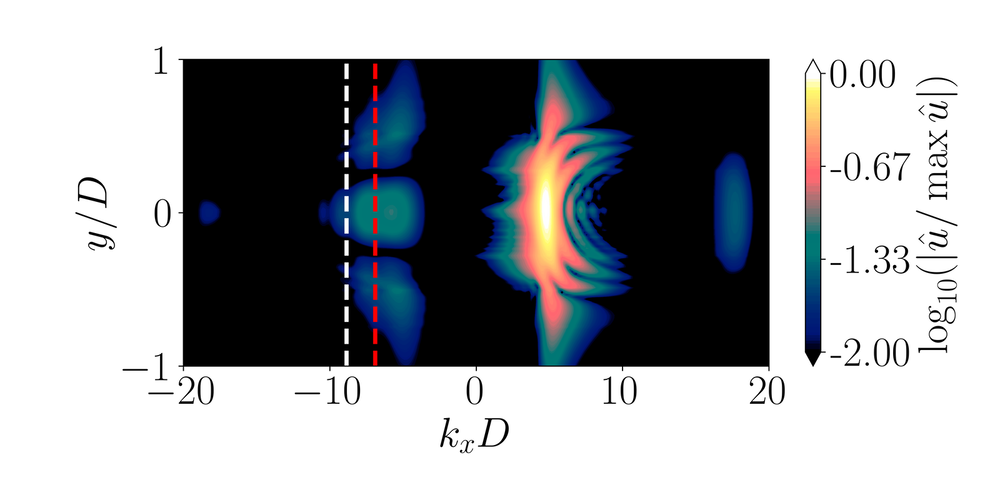}
	\end{subfigure}
	\begin{subfigure}[c]{0.49\linewidth}
		\caption{}\label{contour_staging_v_Mj_1.1}
		\includegraphics[width=\linewidth,trim=0 0 0 0,clip]{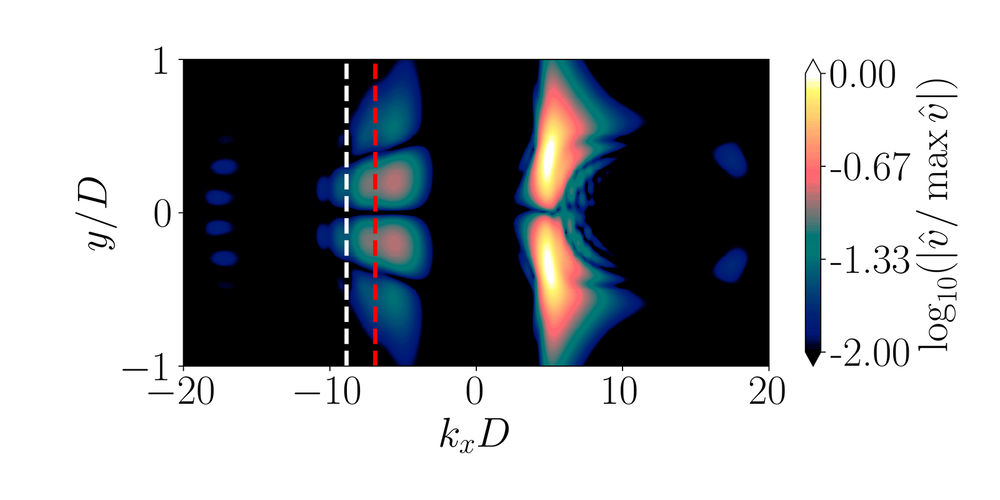}
	\end{subfigure}
	\caption{ Power spectral density of the velocity components for mode A1 at $M_j=1.1$. (a) Streamwise velocity component, (b) normal velocity component. The red and white dashed lines represent the wave numbers $k^+_{kh} - k_{s_1}$ and $k^+_{kh} - k_{s_2}$, respectively.
	}\label{contour_staging_Mj_1.1}
\end{figure}
\begin{figure}
	\centering
	\begin{subfigure}[c]{0.49\linewidth}
		\caption{}\label{contour_staging_Mj_B_1.45}
		\includegraphics[width=\linewidth,trim=0 0 0 0,clip]{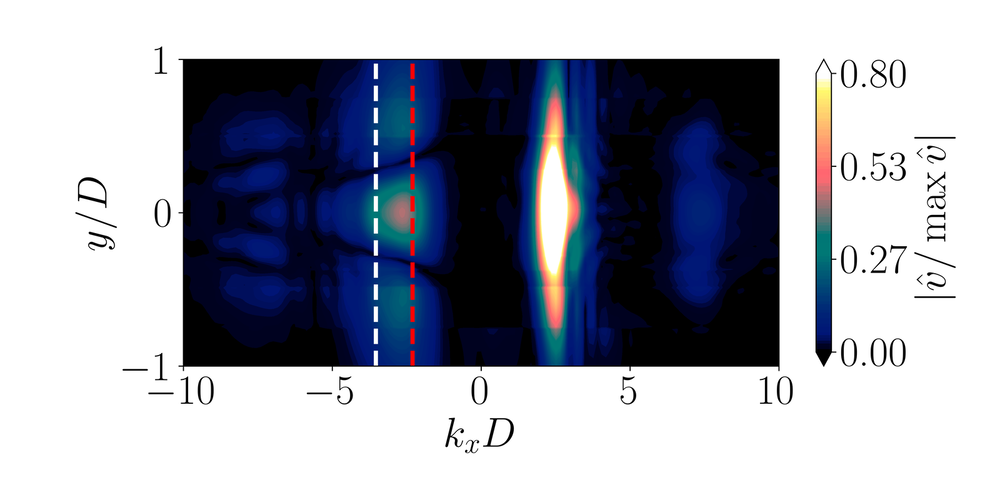}
	\end{subfigure}
	\begin{subfigure}[c]{0.49\linewidth}
		\caption{}\label{contour_staging_Mj_C_1.45}
		\includegraphics[width=\linewidth,trim=0 0 0 0,clip]{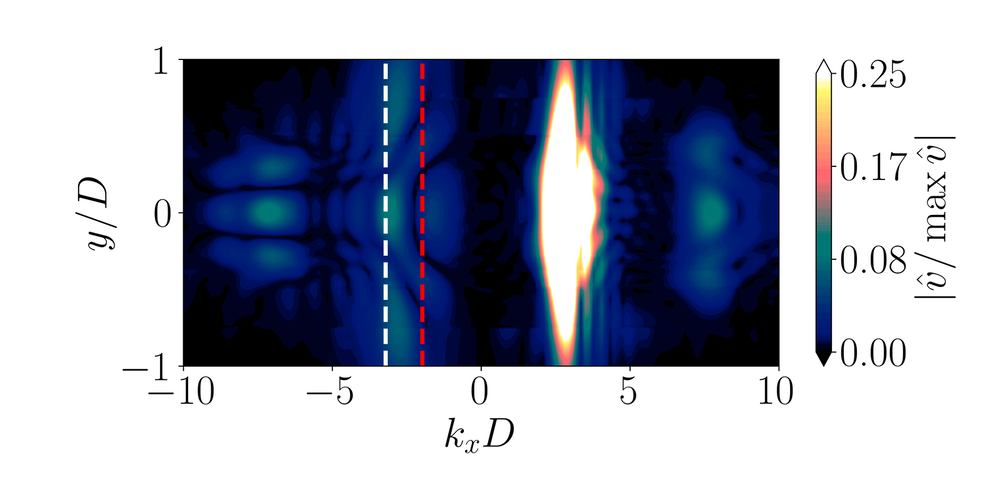}
	\end{subfigure}
	\caption{ Power spectral density transform of the streamwise velocity components for modes B and C at $M_j=1.45$. (a) Mode B, (b) mode C. The red and white dashed lines represent the wave numbers $k^+_{kh} - k_{s_1}$ and $k^+_{kh} - k_{s_2}$, respectively.
 		}\label{contour_staging_Mj1.45}
\end{figure}
\fref{contour_staging_Mj_1.1} shows the streamwise Fourier transform of the global modes corresponding to mode A1 at $M_j = 1.1$, for both velocity components, while \fref{contour_staging_Mj1.45} shows the same Fourier transform for the two global modes observed at $M_j = 1.45$, namely modes B and C (here only the normal velocity component is considered).

Each figure contains the three structures described previously. As mentioned earlier, for modes A1 and B, the GJM wavenumber results from the difference between the KH wavenumber and $k_{s_1}$, i.e., $k^+_{kh} - k_{s_1}$, whereas for modes A2 and C, it results from $k^+_{kh} - k_{s_2}$. To verify this, \fref{contour_staging_u_Mj_1.1} shows the wavenumbers of these differences, with the red line representing $k^+_{kh} - k_{s_1}$ and the white line representing $k^+_{kh} - k_{s_2}$. For modes A1 and B (\fref{contour_staging_u_Mj_1.1} and \fref{contour_staging_Mj_B_1.45}), the GJM wavenumber coincides with the red line, confirming that it arises from the $k^+_{kh} - k_{s_1}$ interaction. For mode C (\fref{contour_staging_Mj_C_1.45}), the GJM wavenumber corresponds to the white line, indicating it results from the $k^+_{kh} - k_{s_2}$ interaction.
\begin{table}
	\caption{Measured wavenumber of the spectral peak for modes A1, B, and C. 
			$k_{s_1}$ and $k_{s_2}$ are the wavenumbers associated with the optimal and suboptimal peaks of the shock-cell spectrum, respectively. 
			$k^+_{kh}$ is the Kelvin--Helmholtz wavenumber, and $k^{-}$ is the GJM wavenumber. 
			The relative error $\epsilon = |k^{-} - (k^+_{kh} - k_{s_1})|/|k^{-}|$ is given in parentheses as a range 
			$[\epsilon_{\min}, \epsilon_{\max}]$ obtained by propagating the uncertainty 
			on all wavenumbers.}
	\begin{center}
		\begin{tabular}{c|c|c|c|c|}
			Mode & $M_j$   & $k^{-}$ & $k^+_{kh} - k_{s_1}$ & $k^+_{kh} - k_{s_2}$ \\
			\hline
			A1 & $1.1$ & $-5.7 \pm 0.070$ & $-6.91 \pm 0.070\ (21\%)$ & $-8.86 \pm 0.070 \ (55\%\text{--}56\%)$ \\
			\hline
			B  & $1.45$ & $-2.37 \pm 0.070$ & $-2.31 \pm 0.070\ (2.5\%\text{--}2.6\%)$ & $-1.978 \pm 0.070 \ (20\%\text{--}21\%)$ \\
			C  & $1.45$ & $-2.79 \pm 0.070$ & $-3.54 \pm 0.070\ (20\%\text{--}21\%)$  & $-3.208 \pm 0.070 \ (12\%\text{--}13\%)$ \\
			\hline
		\end{tabular}\label{tab_k_waves}
	\end{center}
\end{table}
\tabref{tab_k_waves} summarises the wavenumbers associated with the different interactions with the shock spectra. The values are obtained by taking the maximum of the peaks from the wavenumber decomposition. It should be noted that, due to the change in symmetry, the radial velocity component was used to calculate the wavenumber for the $M_j = 1.45$ case.

The table also highlights the relative error between the wavenumber estimated via the interaction (e.g., $k^+_{kh} - k_{s_1}$) and the measured GJM wavenumber $k^{-}$. From these errors, it is evident that for modes A1 and B, the smaller error is associated with the $k_{s_1}$ interaction, whereas for mode C, the interaction with $k_{s_2}$ yields a lower error.

\section{Helmholtz decomposition and wave-component energy budget}\label{sec:helmholtz_waves}

To separate the vortical and dilatational contributions of the perturbation, a Helmholtz decomposition is applied to the velocity components of the global modes. The perturbation velocity is decomposed as
$$
\mathbf{u}=\mathbf{u}_A+\mathbf{u}_{\phi},
$$
where $\mathbf{u}_A$ is the solenoidal component of the velocity, associated with the vortical structures of the mode, and $\mathbf{u}_{\phi}$ is the irrotational component, which contains the compressible contribution, including the acoustic field and the shock-related part of the perturbation (for more details see \secref{app:helmholtz}).

\begin{figure}
	\centering
	\begin{subfigure}[c]{0.49\linewidth}
		\centering
		\caption{}\label{fig_u_helm_A1_Mj11}
		\includegraphics[width=\linewidth,trim=0 100 0 150,clip]{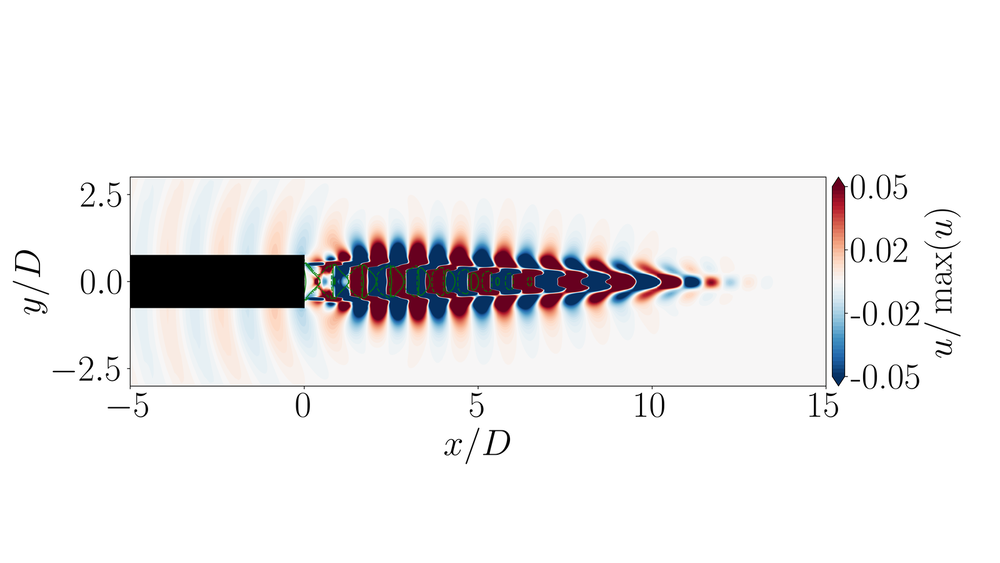}
	\end{subfigure}
	\begin{subfigure}[c]{0.49\linewidth}
		\centering
		\caption{}\label{fig_u_helm_B1_Mj145}
		\includegraphics[width=\linewidth,trim=0 100 0 150,clip]{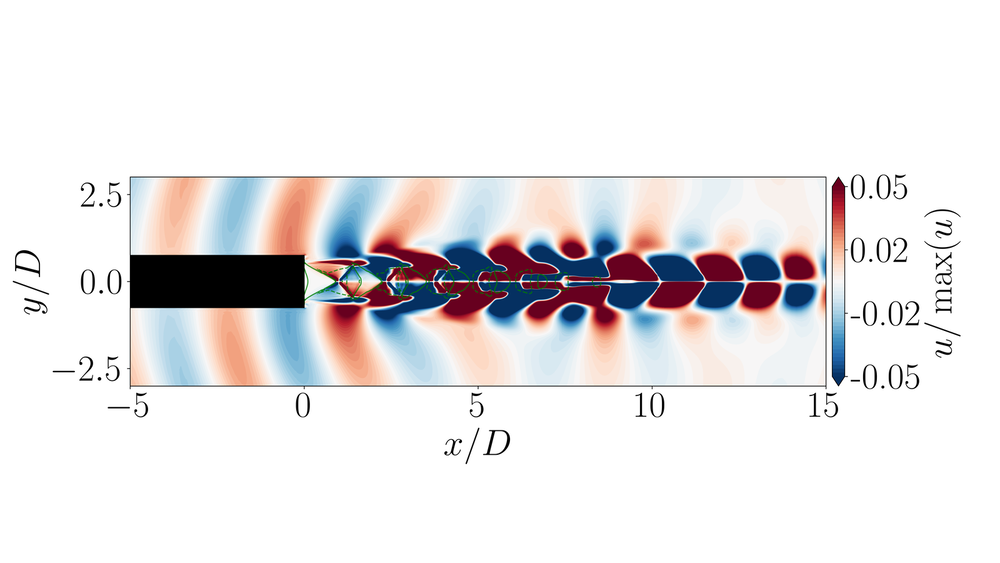}
	\end{subfigure}
	\begin{subfigure}[c]{0.49\linewidth}
		\centering
		\caption{}\label{fig_phi_helm_A1_Mj11}
		\includegraphics[width=\linewidth,trim=0 100 0 150,clip]{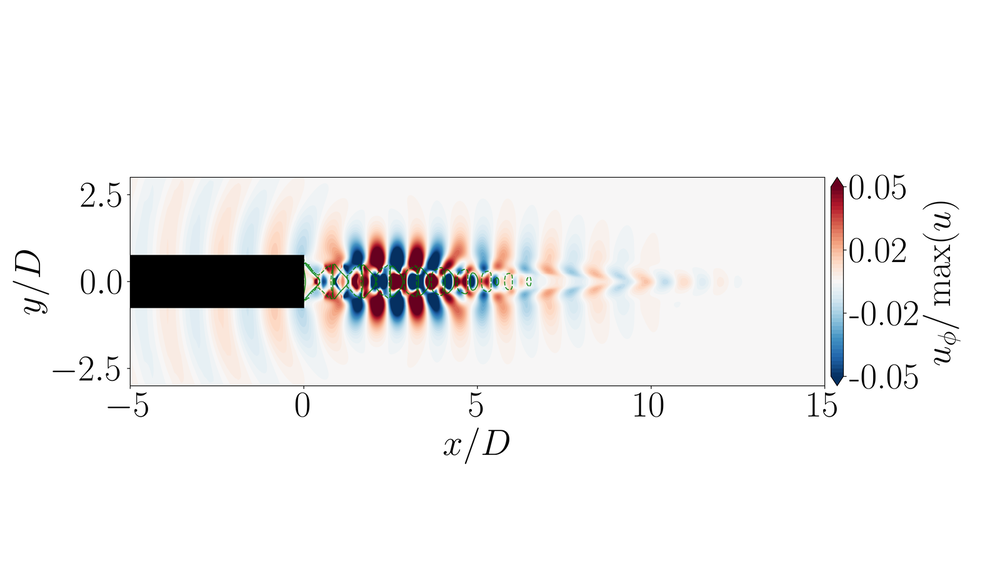}
	\end{subfigure}
	\begin{subfigure}[c]{0.49\linewidth}
		\centering
		\caption{}\label{fig_phi_helm_B1_Mj145}
		\includegraphics[width=\linewidth,trim=0 100 0 150,clip]{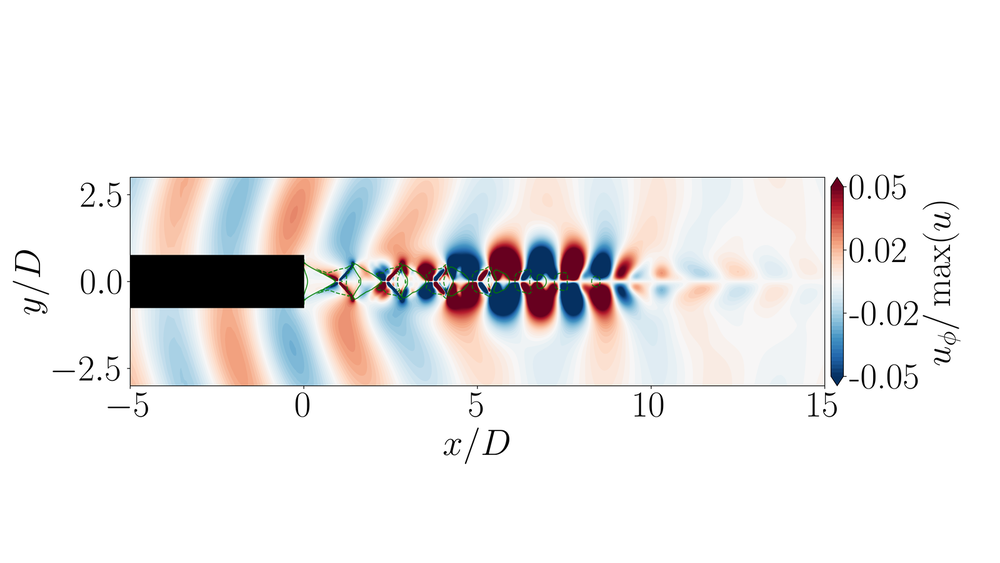}
	\end{subfigure}
	\begin{subfigure}[c]{0.49\linewidth}
		\centering
		\caption{}\label{fig_A_helm_A1_Mj11}
		\includegraphics[width=\linewidth,trim=0 100 0 150,clip]{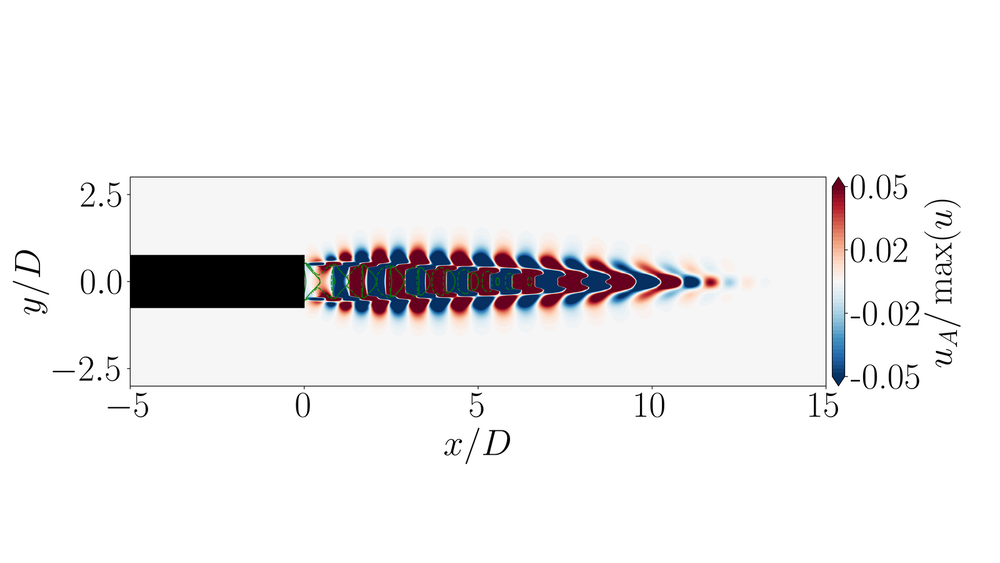}
	\end{subfigure}
	\begin{subfigure}[c]{0.49\linewidth}
		\centering
		\caption{}\label{fig_A_helm_B1_Mj145}
		\includegraphics[width=\linewidth,trim=0 100 0 150,clip]{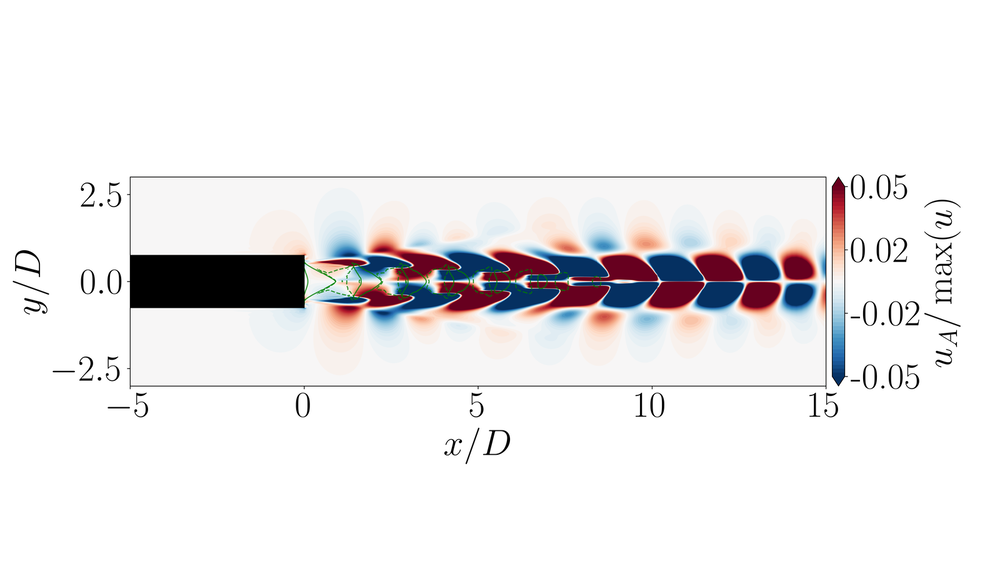}
	\end{subfigure}
	\caption{Helmholtz decomposition of the global modes. (a,c,e) mode A1 at $M_j=1.1$; (b,d,f) mode B at $M_j=1.45$. (a,b) Streamwise velocity perturbation $u$. (c,d) Irrotational streamwise component $u_{\phi}$. (e,f) Solenoidal streamwise component $u_A$. The green contours represent the shock-cell structure of the fixed-point solution.}
	\label{fig_helmholtz_global_modes}
\end{figure}

\fref{fig_helmholtz_global_modes} compares the full streamwise velocity perturbation with its dilatational and solenoidal contributions for mode A1 at $M_j=1.1$ and mode B at $M_j=1.45$. For both operating conditions, the solenoidal component $\mathbf{u}_A$ reproduces most of the spatial structure of the full velocity perturbation. It is mainly concentrated in the jet shear layer and remains confined to the jet region, with only a weak signature in the acoustic field. This component represents the vortical part of the mode and is therefore associated with the downstream Kelvin--Helmholtz wavepacket, which carries most of the perturbation energy.

The dilatational component $\mathbf{u}_{\phi}$ has a different spatial organisation. It contains the compressible part of the perturbation, including both the acoustic field and the part of the mode linked to the shock-cell structure. The modulation by the shock cells is clearly visible in this component, showing that the dilatational field is strongly affected by the fixed-point shock pattern. At the same time, an upstream-propagating acoustic perturbation can also be observed. This contribution is the part of the global mode that is directly related to the acoustic field measured in screech experiments.

This decomposition is useful because it separates the main physical ingredients of the screech feedback loop. The KH wavepacket is mainly contained in the solenoidal field, since it is a vortical and downstream-propagating structure. On the contrary, the $k^-$ wave and the upstream acoustic radiation are contained in the dilatational field, because they are acoustic contributions. In particular, the $k^-$ wave corresponds to the upstream-propagating neutral acoustic wave present in the jet core. The Helmholtz decomposition therefore provides a way to distinguish the downstream vortical part of the loop from the upstream acoustic part, not only from a wavenumber and propagative point of view, but also from an hydrodynamic and acoustic point of view.

\begin{figure}
	\centering
	\begin{subfigure}[c]{0.325\linewidth}
		\caption{}\label{fig_spectrum_phi_A1}
		\includegraphics[width=\linewidth,trim=0 0 0 0,clip]{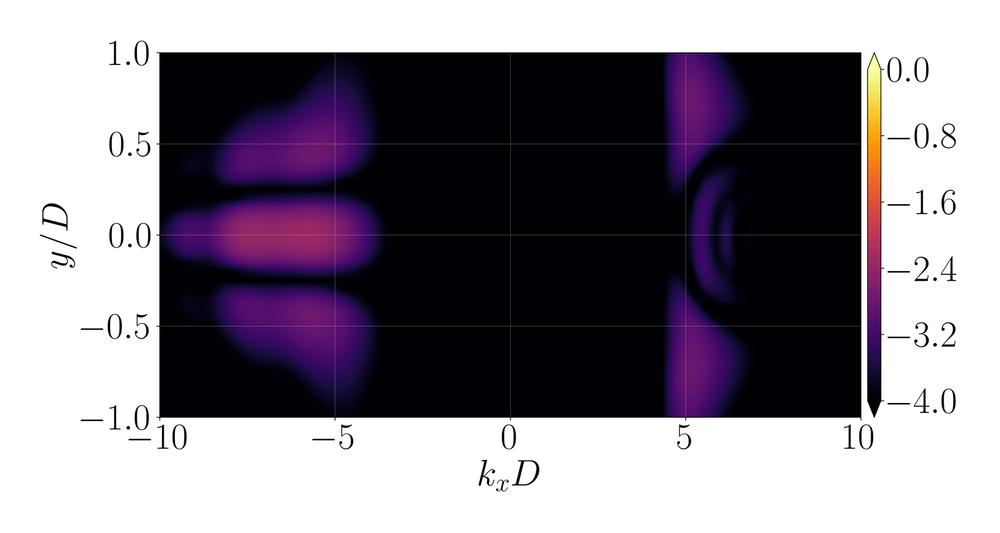}
	\end{subfigure}
	\begin{subfigure}[c]{0.325\linewidth}
		\caption{}\label{fig_spectrum_phi_B1}
		\includegraphics[width=\linewidth,trim=0 0 0 0,clip]{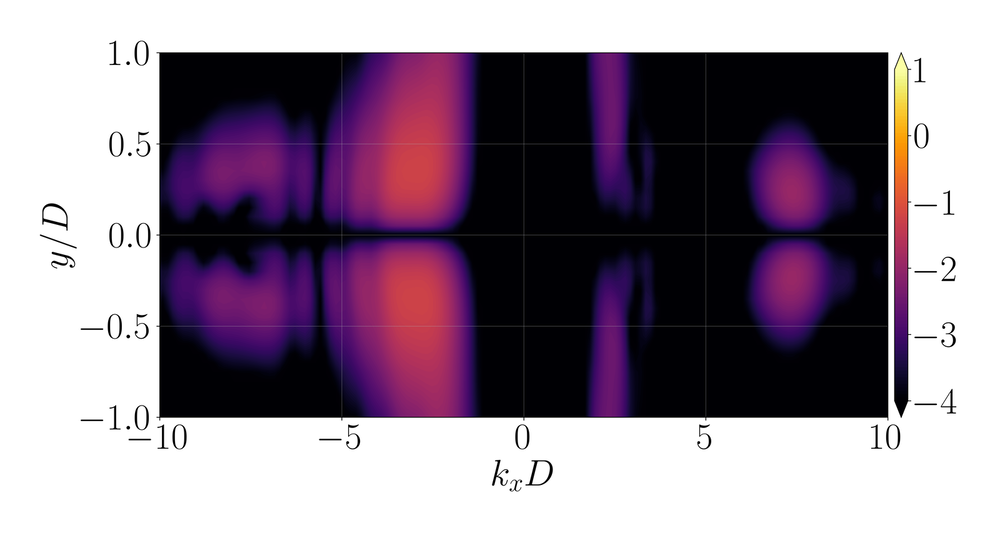}
	\end{subfigure}
	\begin{subfigure}[c]{0.325\linewidth}
		\caption{}\label{fig_spectrum_phi_C1}
		\includegraphics[width=\linewidth,trim=0 0 0 0,clip]{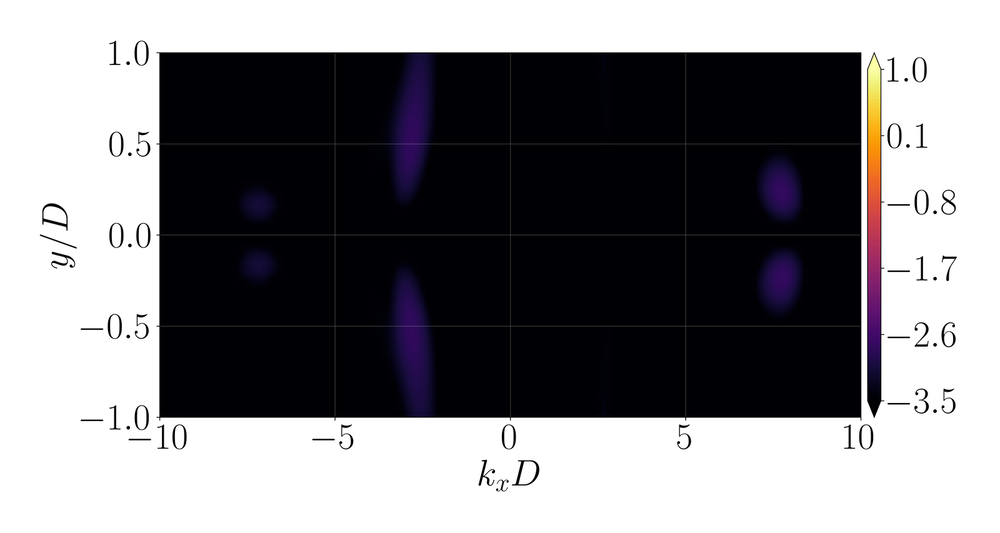}
	\end{subfigure}
	\begin{subfigure}[c]{0.325\linewidth}
		\caption{}\label{fig_spectrum_A_A1}
		\includegraphics[width=\linewidth,trim=0 0 0 0,clip]{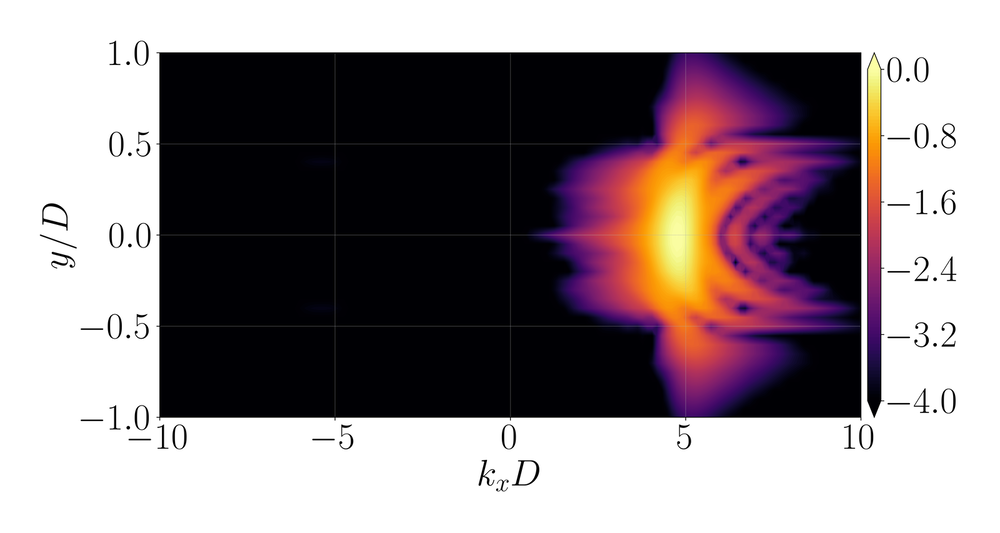}
	\end{subfigure}
	\begin{subfigure}[c]{0.325\linewidth}
		\caption{}\label{fig_spectrum_A_B1}
		\includegraphics[width=\linewidth,trim=0 0 0 0,clip]{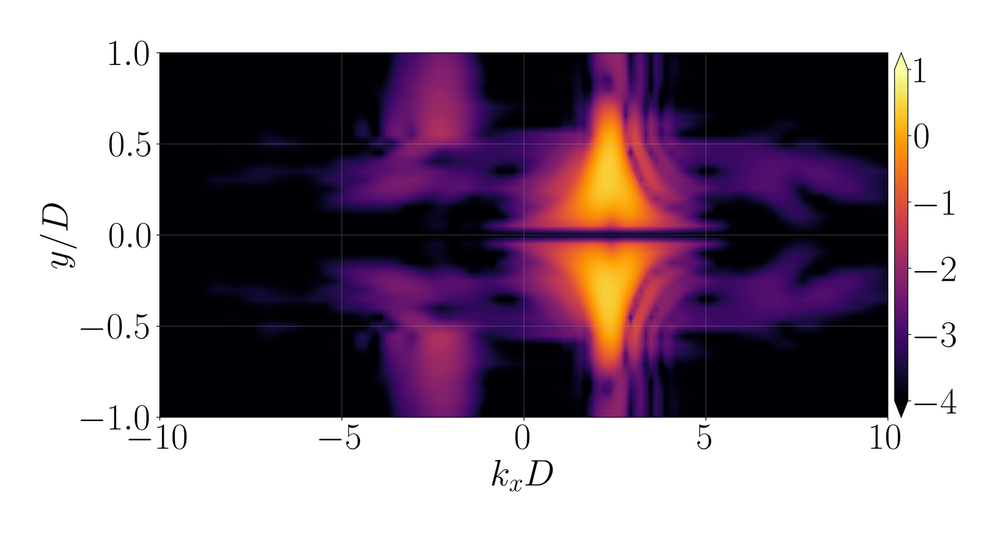}
	\end{subfigure}
	\begin{subfigure}[c]{0.325\linewidth}
		\caption{}\label{fig_spectrum_A_C1}
		\includegraphics[width=\linewidth,trim=0 0 0 0,clip]{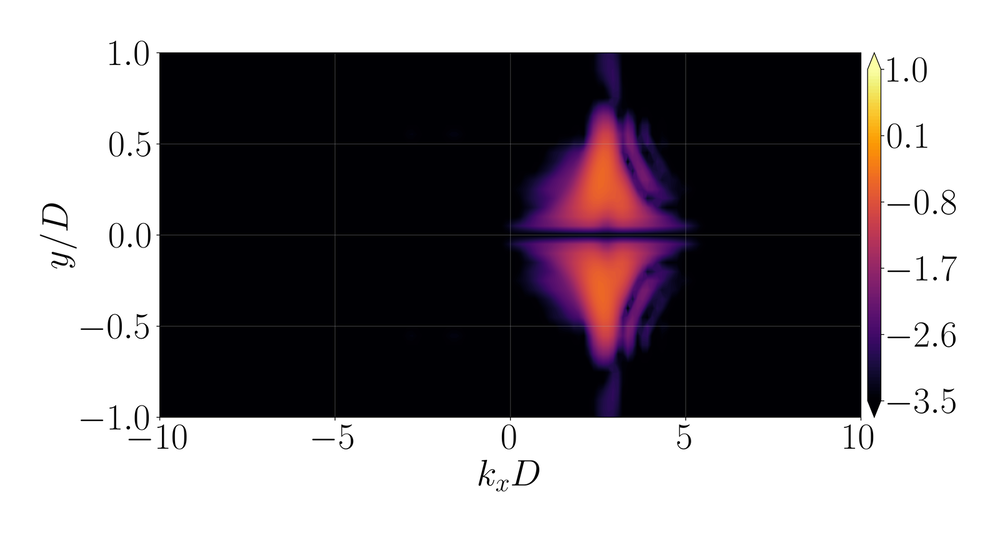}
	\end{subfigure}
	\begin{subfigure}[c]{0.325\linewidth}
		\caption{}\label{fig_spectrum_u_A1}
		\includegraphics[width=\linewidth,trim=0 0 0 0,clip]{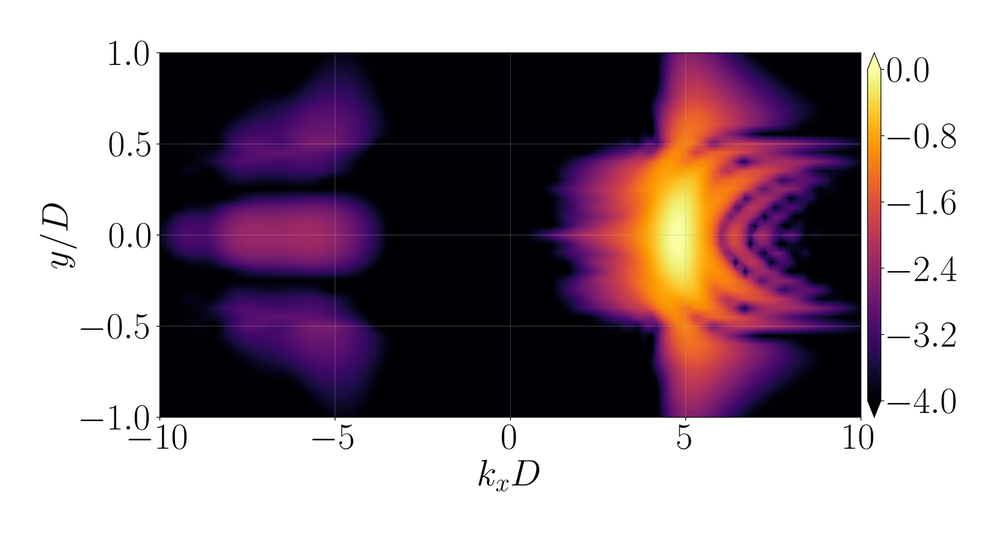}
	\end{subfigure}
	\begin{subfigure}[c]{0.325\linewidth}
		\caption{}\label{fig_spectrum_u_B1}
		\includegraphics[width=\linewidth,trim=0 0 0 0,clip]{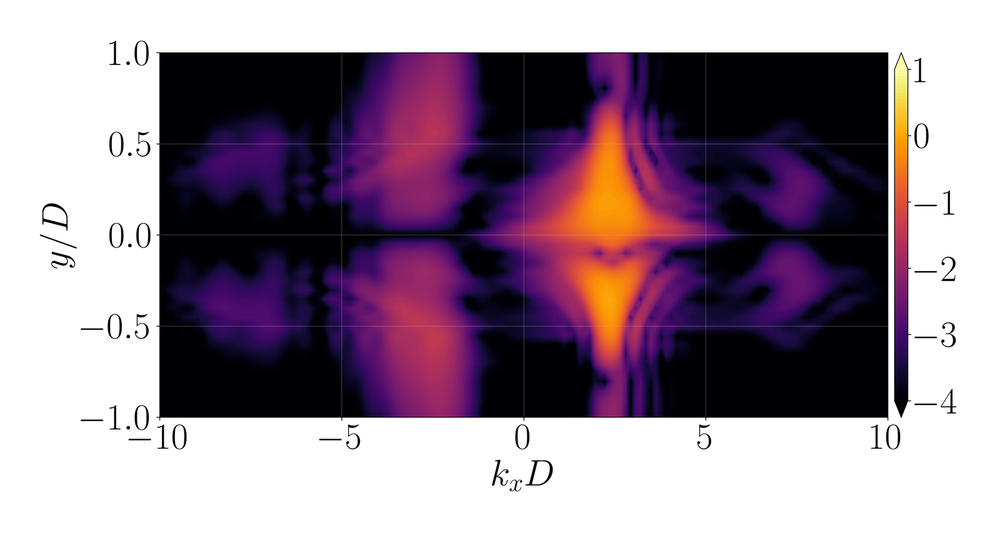}
	\end{subfigure}
	\begin{subfigure}[c]{0.325\linewidth}
		\caption{}\label{fig_spectrum_u_C1}
		\includegraphics[width=\linewidth,trim=0 0 0 0,clip]{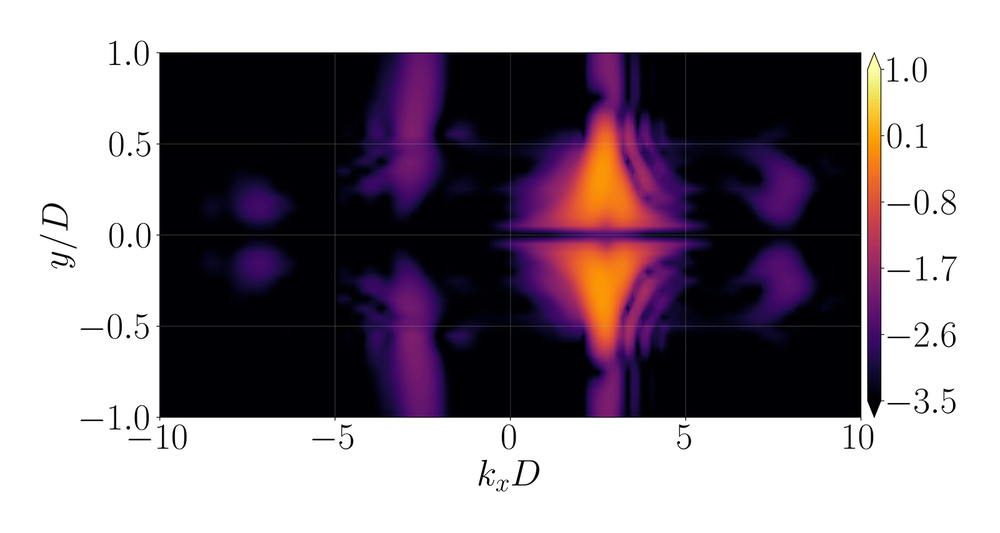}
	\end{subfigure}
	\caption{Streamwise-wavenumber spectra of the decomposed fields. Columns correspond to mode A1 at $M_j=1.1$, mode B at $M_j=1.45$ and mode C at $M_j=1.45$. First row: dilatational component $u_{\phi}$. Second row: solenoidal component $u_{A}$. Third row: full velocity component $u$.}\label{fig_wave_component_spectra}
\end{figure}

This separation is also clear when looking at the streamwise-wavenumber spectra in \fref{fig_wave_component_spectra}. For mode A1, the solenoidal component contains mainly the positive-wavenumber branch, which is associated with the downstream-propagating KH wave. On the contrary, the dilatational component contains a negative-wavenumber contribution inside the jet core, associated with the $k^-$ wave, together with a positive-wavenumber contribution located farther from the core, associated with the acoustic radiation.

The same behaviour is also observed for the B and C modes. The spectrum of the full velocity perturbation is recovered by the combination of the dilatational and solenoidal contributions. The solenoidal field mainly contains the downstream KH wave, while the dilatational field contains most of the upstream-propagating acoustic content. This confirms the acoustic nature of the $k^-$ component and shows that the global stability analysis directly retrieves the waves that compose the screech feedback loop.

As in the previous section, the spatial distribution of the different wavepackets can be reconstructed by filtering the positive and negative streamwise wavenumbers. Here, however, the filtering is applied after the Helmholtz decomposition. The KH wave is therefore extracted from the vortical component, while the $k^-$ wave is extracted from the dilatational component. This choice preserves the physical nature of the extracted structures and avoids mixing vortical and acoustic contributions in the reconstructed wavepackets.

\begin{figure}
	\centering
	\begin{subfigure}[c]{0.49\linewidth}
		\caption{}\label{fig_Akh_A1_Mj11}
		\includegraphics[width=\linewidth,trim=0 100 0 150,clip]{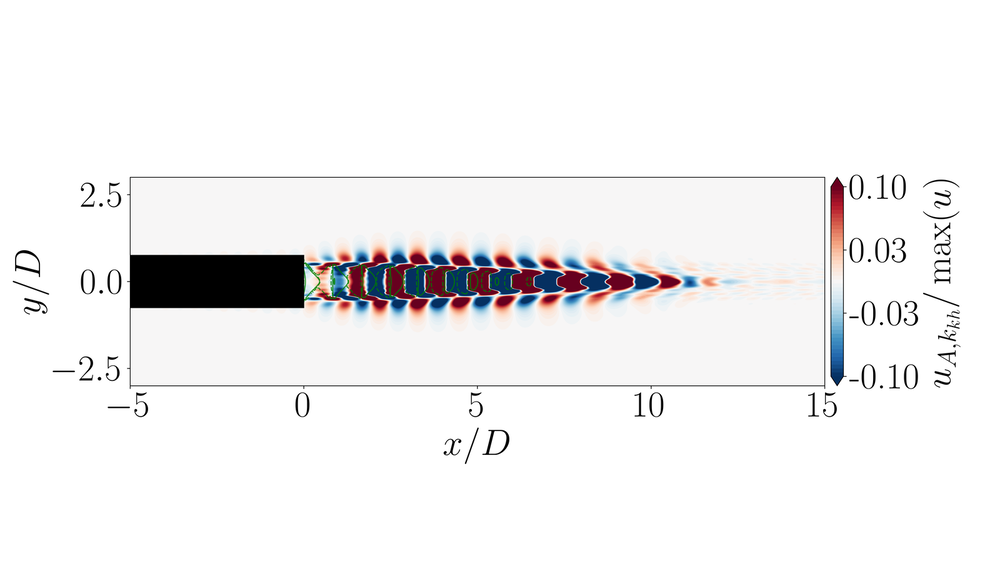}
	\end{subfigure}
	\begin{subfigure}[c]{0.49\linewidth}
		\caption{}\label{fig_phikm_A1_Mj11}
		\includegraphics[width=\linewidth,trim=0 100 0 150,clip]{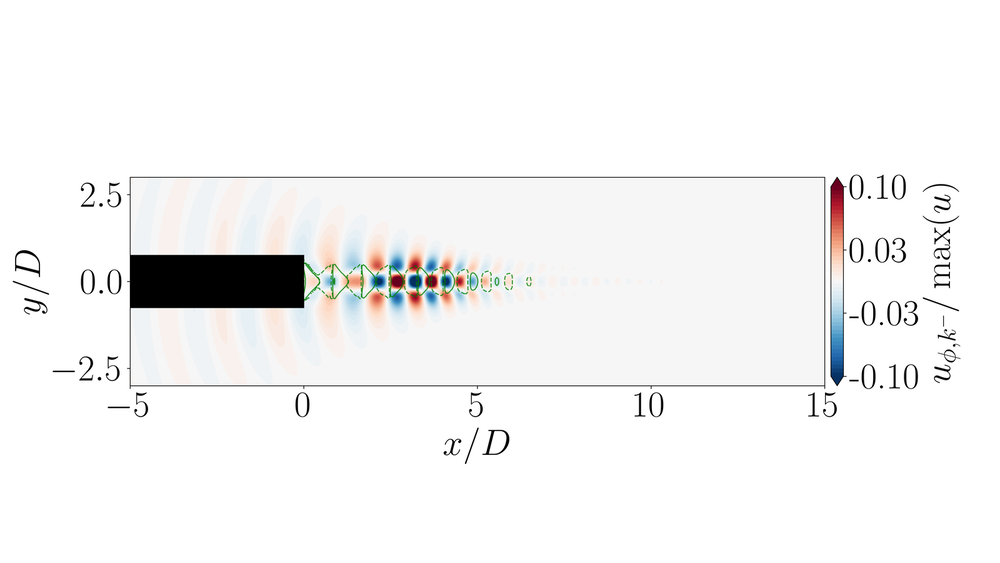}
	\end{subfigure}
	\begin{subfigure}[c]{0.49\linewidth}
		\caption{}\label{fig_Akh_B1_Mj145}
		\includegraphics[width=\linewidth,trim=0 100 0 150,clip]{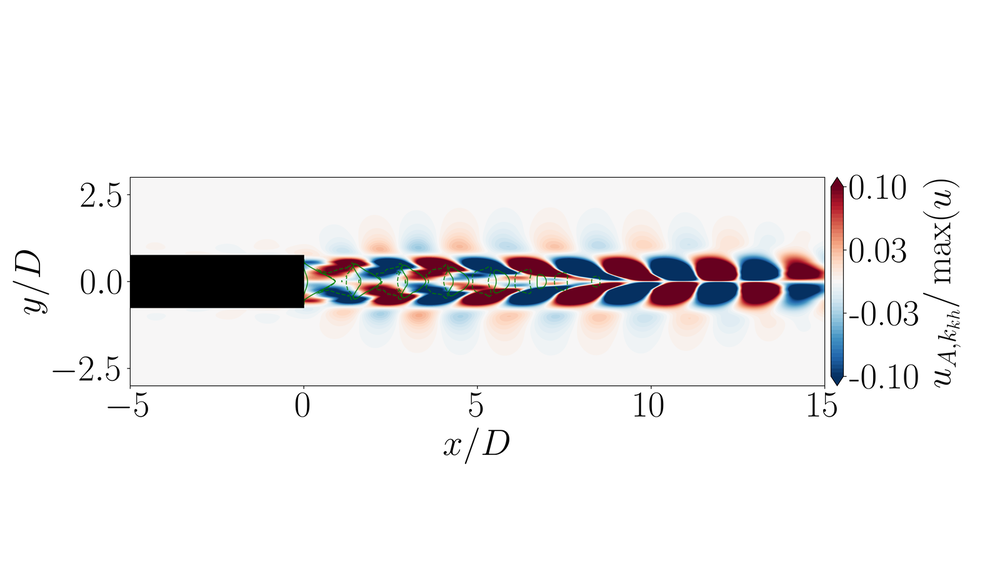}
	\end{subfigure}
	\begin{subfigure}[c]{0.49\linewidth}
		\caption{}\label{fig_phikm_B1_Mj145}
		\includegraphics[width=\linewidth,trim=0 100 0 150,clip]{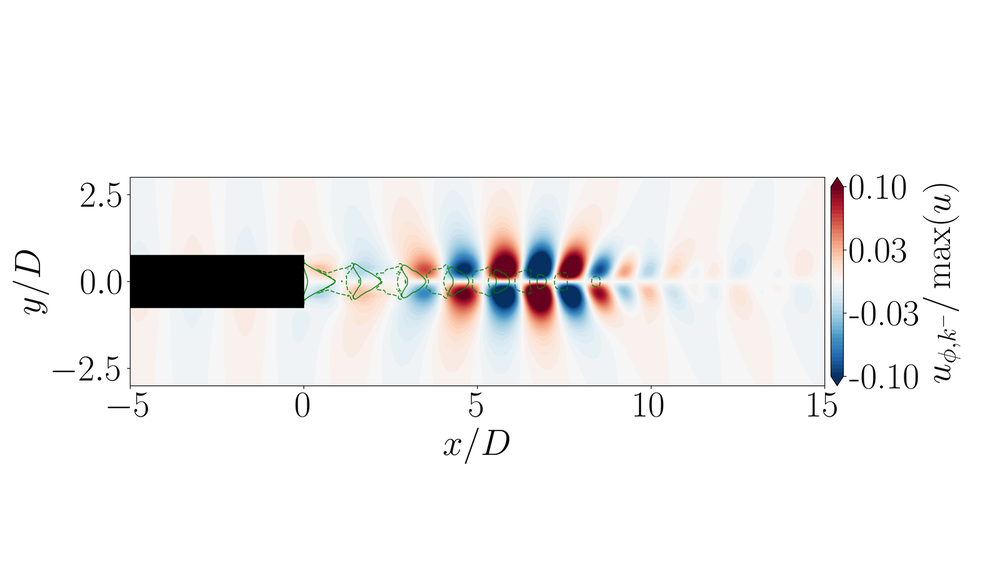}
	\end{subfigure}
	\begin{subfigure}[c]{0.49\linewidth}
		\caption{}\label{fig_Akh_C1_Mj145}
		\includegraphics[width=\linewidth,trim=0 100 0 150,clip]{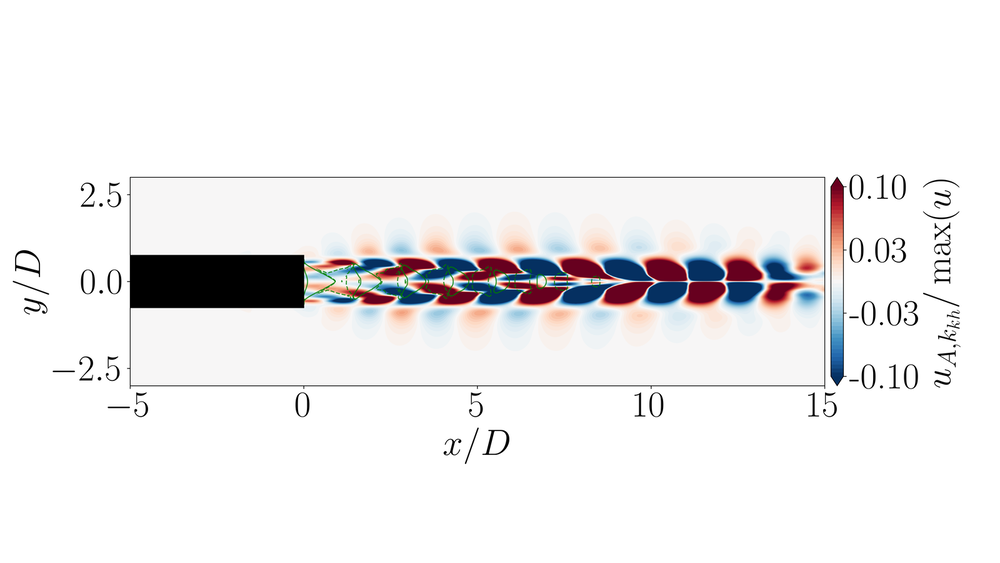}
	\end{subfigure}
	\begin{subfigure}[c]{0.49\linewidth}
		\caption{}\label{fig_phikm_C1_Mj145}
		\includegraphics[width=\linewidth,trim=0 100 0 150,clip]{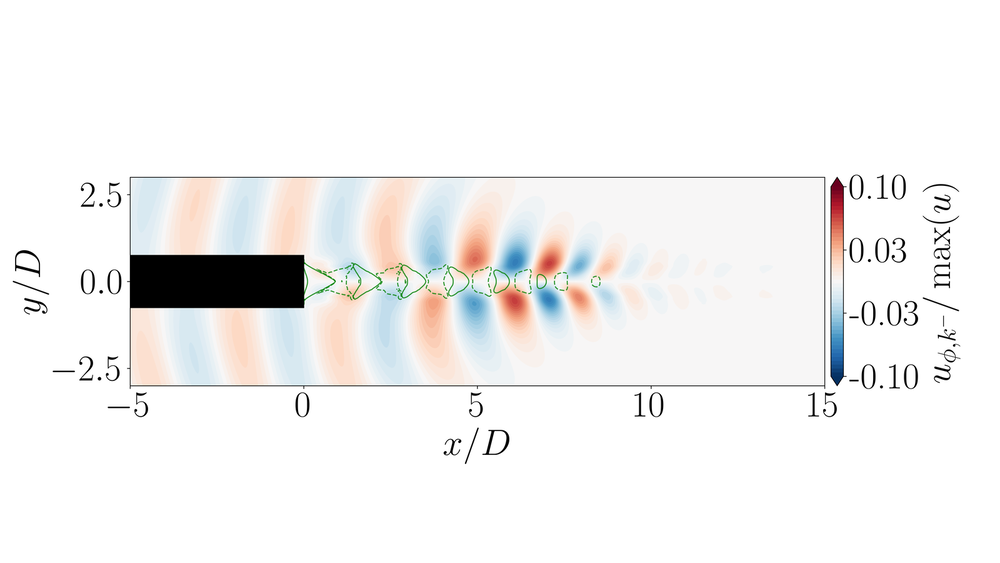}
	\end{subfigure}
	\caption{Filtered wave components extracted from the Helmholtz-decomposed fields. Rows correspond to mode A1 at $M_j=1.1$, mode B at $M_j=1.45$ and mode C at $M_j=1.45$. Left column: downstream-propagating vortical contribution $u_A^{kh}$. Right column: upstream-propagating dilatational contribution $u_{\phi}^{k^-}$.}
	\label{fig_filtered_wave_components}
\end{figure}

\fref{fig_filtered_wave_components} shows the spatial distribution of the retrieved KH and $k^-$ wavepackets for mode A1 at $M_j=1.1$, mode B at $M_j=1.45$ and mode C at $M_j=1.45$. The KH wavepacket is very close to the solenoidal component of the velocity. This shows that most of the vortical part of the perturbation is contained in the downstream-propagating KH wavepacket.

The $k^-$ component has a more specific behaviour inside the dilatational field. After the streamwise-wavenumber filtering, the shock-cell signature is no longer visible, so that the extracted field mainly corresponds to the acoustic perturbation. The $k^-$ wave is mostly concentrated inside the jet core, with a maximum located around the fourth and fifth shock cells. This is consistent with the spectral analysis, where the negative-wavenumber branch was associated with the guided jet mode.

Some differences can also be observed between the modes. For the A1 and B modes, the $k^-$ perturbation remains mainly confined inside the jet core. For the C mode, the perturbation is more spread in the radial direction and a stronger acoustic signature is visible outside the core. This suggests that, for the C mode, a larger part of the upstream-propagating energy is located in the annular acoustic region. On the contrary, for the A1 and B modes, a larger part of the upstream component remains guided inside the jet core. This point is quantified in the next subsection through the energy repartition of the feedback loop.

\subsection{Energy partition of the feedback loop}
\label{sec:energy_feedback_loop}

Most literature on Mach number effects on screech focuses on the staging origin. This section uses the well-identified characteristic waves of the feedback loop and the data's parametric nature. The aim is to provide further insight into how jet Mach number affects the relative kinetic energy distribution between these waves. 

To quantify the relative kinetic energy component of the feedback loop at different jet operating conditions, energy ratios are computed from the isolated wave components. The modal energy is separated into the core contribution (evaluated in a cylindric region where $r/D<1$), and the far-field acoustic contribution  (evaluated for $1<r/D<10$). The core region is defined to quantify effect related to waves propagating in the jet core and the shear layer, while the farfield region is defined to quantify the acoustic kinetic energy. More precisely, the analysis will focus on the energy ratios of the screech related wave components, namely the downstream KH wave, the core $k^-$ contribution and the upstream acoustic field.

\begin{figure}
	\centering
	\begin{subfigure}[c]{0.49\linewidth}
		\caption{}\label{fig_E_km_phi_core}
		\includegraphics[width=\linewidth,trim=0 0 0 0,clip]{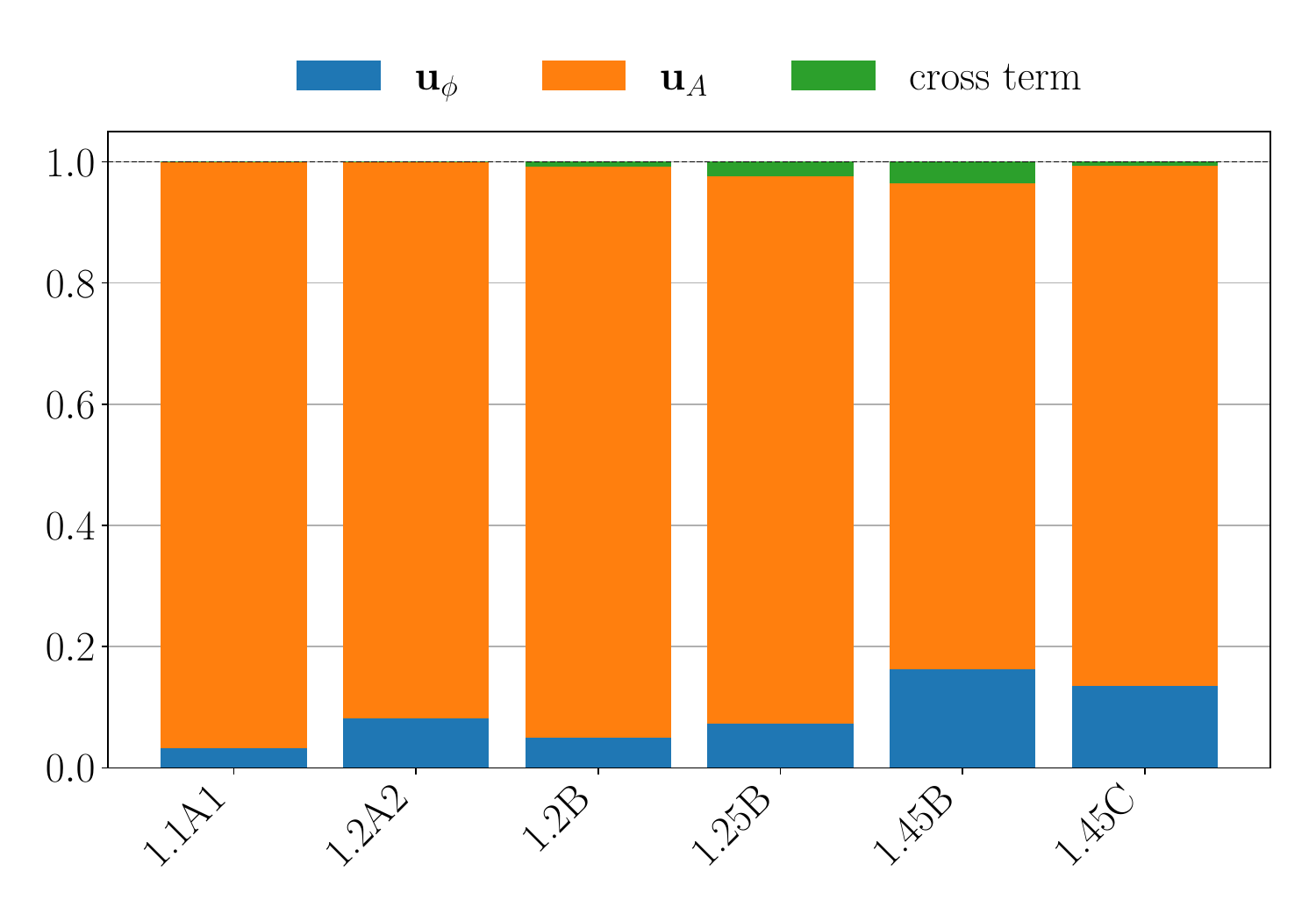}
	\end{subfigure}
	\begin{subfigure}[c]{0.49\linewidth}
		\caption{}\label{fig_E_budget_norm}
		\includegraphics[width=\linewidth,trim=0 0 0 0,clip]{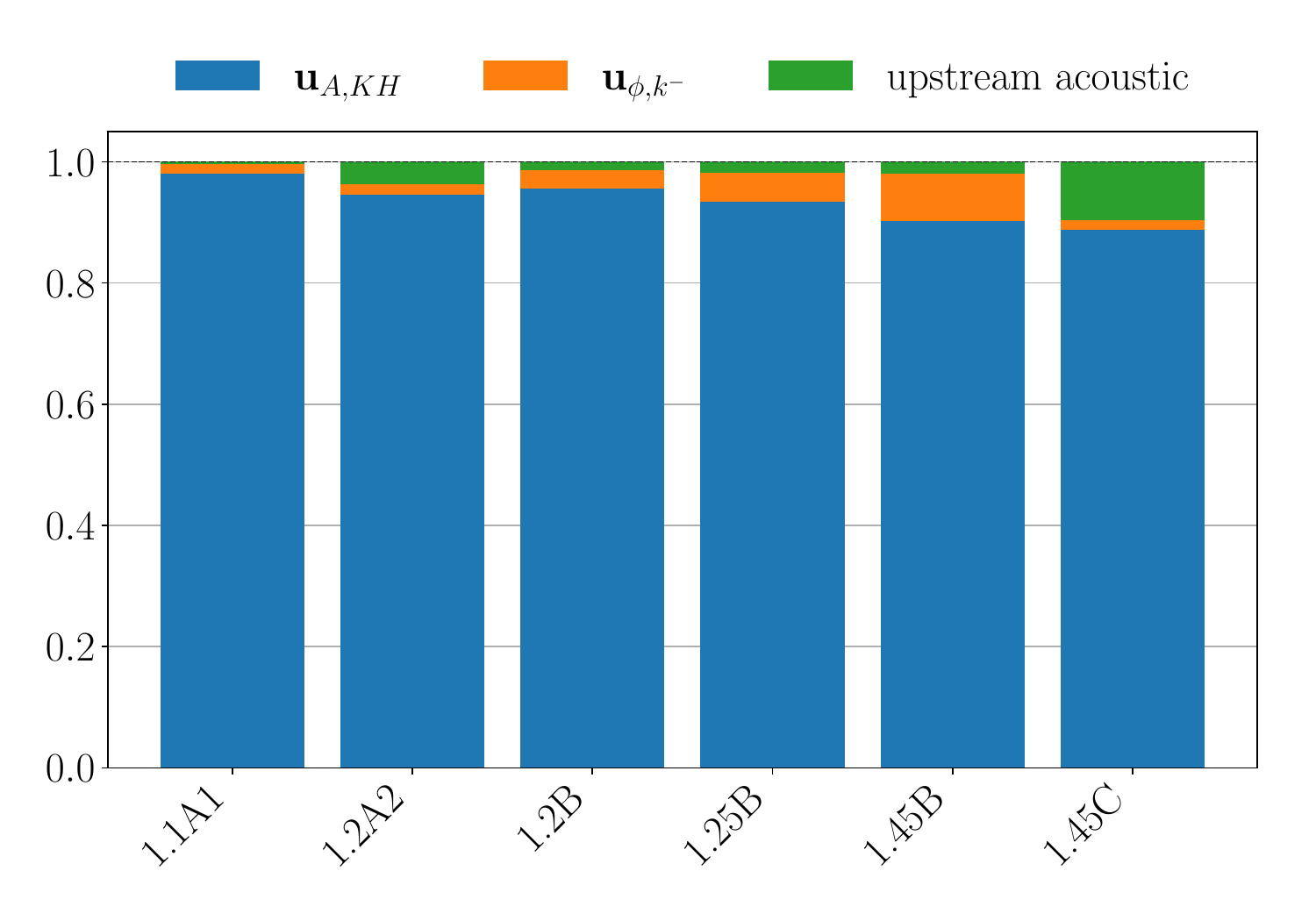}
	\end{subfigure}
	\caption{Energy repartition of the feedback-loop components. (a) Helmholtz decomposition of the core velocity energy into irrotational, solenoidal and cross-term contributions, normalised by the total core velocity energy. (b) Feedback-loop energy budget normalised by the sum of the contributions, including the core KH wave, the core $k^-$ wave and the farfield upstream acoustic field.}\label{fig_energy_budget}
\end{figure}

\fref{fig_energy_budget} first shows the energy repartition between the two Helmholtz components $\mathbf{u}_A$ and $\mathbf{u}_{\phi}$. The solenoidal contribution is dominant for all the modes. This confirms the visual observation made on \fref{fig_helmholtz_global_modes}: the full velocity perturbation is mainly vortical and is therefore mainly associated with the KH wavepacket. The irrotational contribution is smaller, and contains both the acoustic component related to the feedback loop and the farfield acoustic waves. The cross term remains small compared with the total velocity energy, which confirms that the decomposition separates the two components with a good accuracy.

\fref{fig_E_budget_norm} shows the energy distribution that is directly associated with the feedback related waves. The KH component $\mathbf{u}_A^{kh}$ contains most of the core energy, while the extracted $\mathbf{u}_{\phi}^{k^-}$ contribution represents only a small fraction of the total core energy. It is interesting to see that the relative upstream acoustic energy varies within the modes, with the higher ratio of upstream acoustic energy for modes A2 at $M_j=1.2$ and mode C at $M_j=1.45$.
This agrees with the spatial contours of \fref{fig_filtered_wave_components}, where the C mode shows a larger acoustic redistribution outside the jet core.

\begin{figure}
	\centering
	\begin{subfigure}[c]{0.49\linewidth}
		\caption{}\label{fig_E_KH_full}
		\includegraphics[width=\linewidth,trim=0 0 0 0,clip]{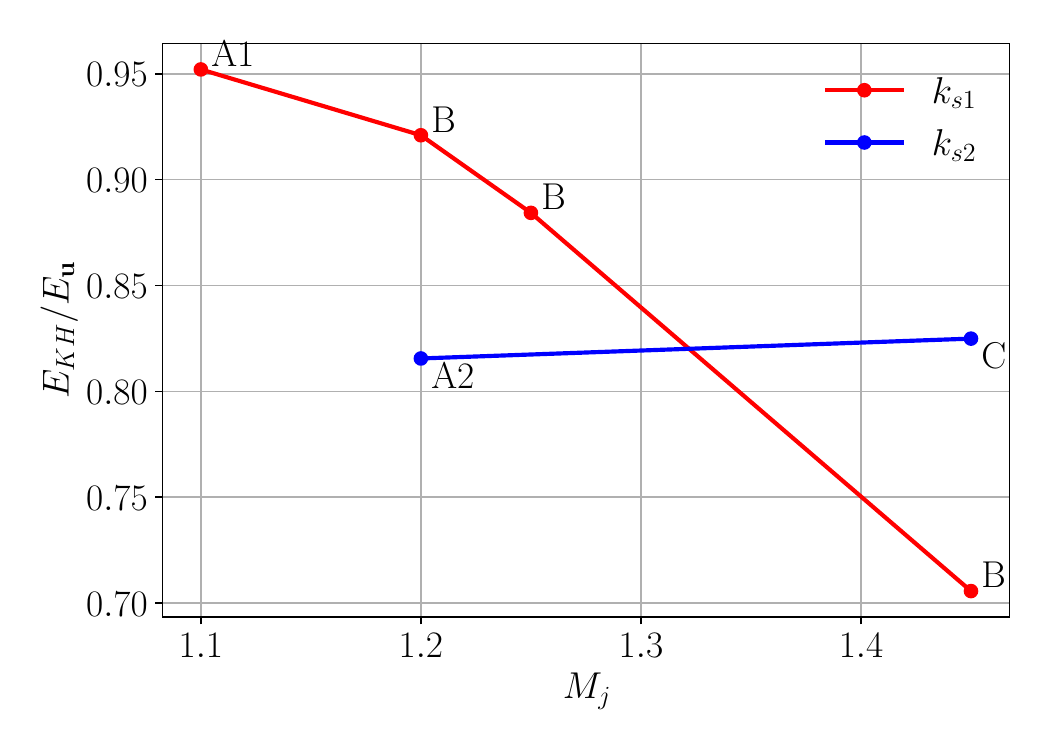}
	\end{subfigure}
	\begin{subfigure}[c]{0.49\linewidth}
		\caption{}\label{fig_E_km_full}
		\includegraphics[width=\linewidth,trim=0 0 0 0,clip]{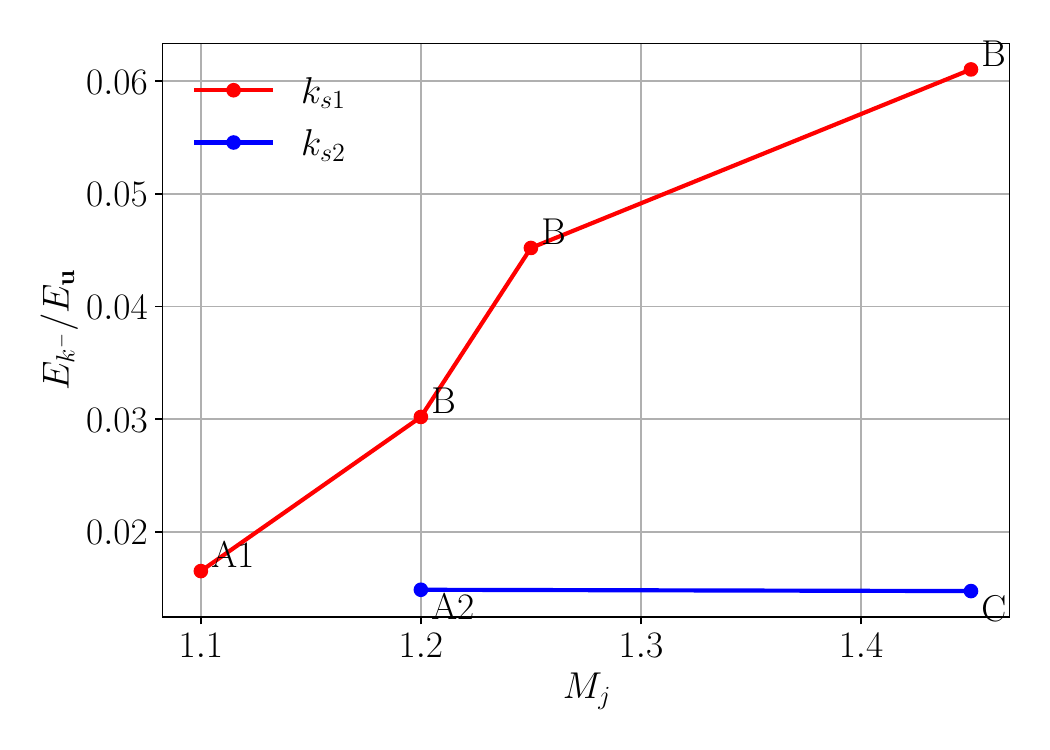}
	\end{subfigure}
	\begin{subfigure}[c]{0.49\linewidth}
		\caption{}\label{fig_E_ac_full}
		\includegraphics[width=\linewidth,trim=0 0 0 0,clip]{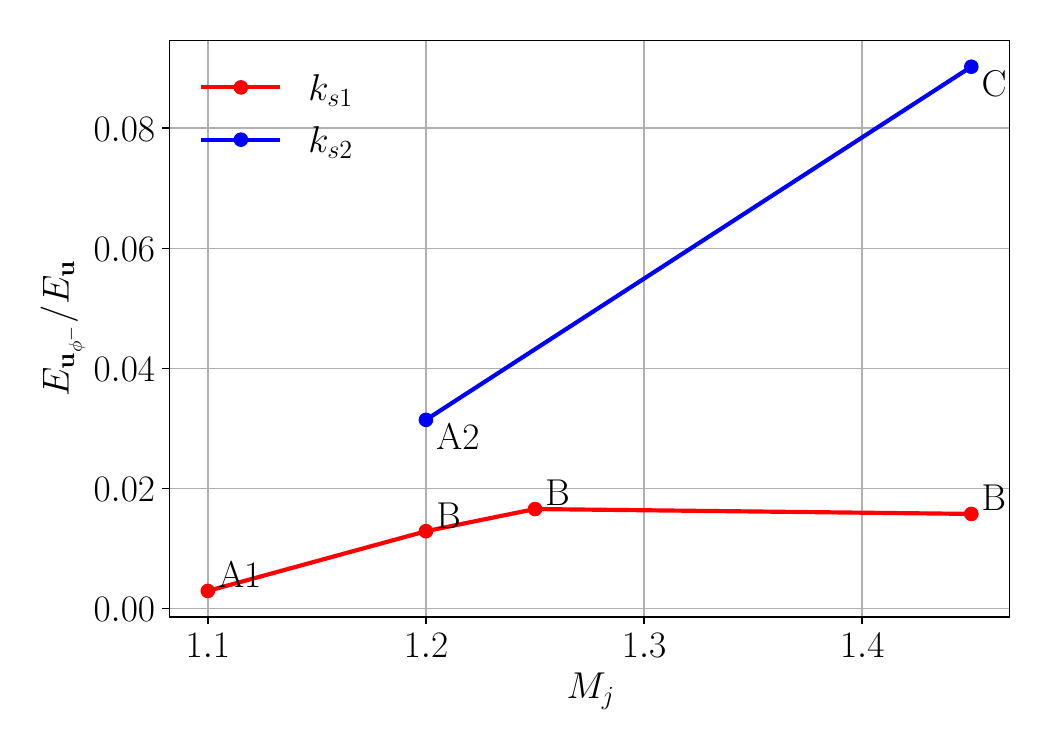}
	\end{subfigure}
	\begin{subfigure}[c]{0.49\linewidth}
		\caption{}\label{fig_E_ac_km}
		\includegraphics[width=\linewidth,trim=0 0 0 0,clip]{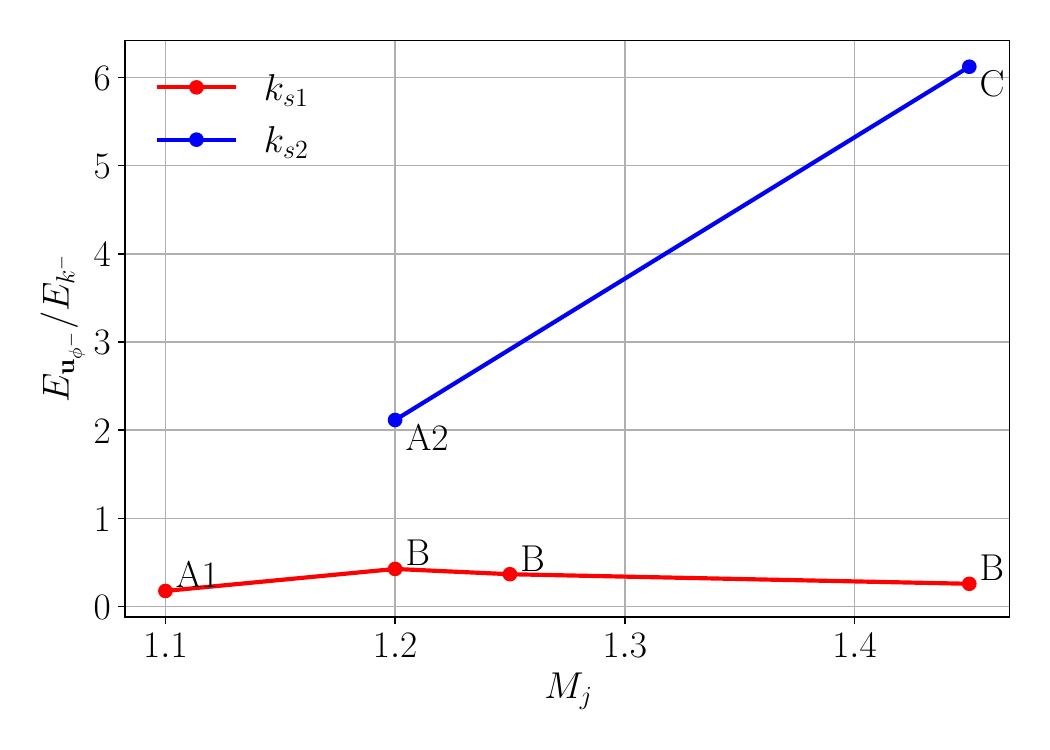}
	\end{subfigure}
	\caption{Energy ratios of the isolated wave components. (a) KH energy in the core normalised by the total modal energy, (b) core $k^-$ irrotational energy normalised by the total modal energy, (c) farfield upstream acoustic energy normalised by the total modal energy, (d) farfield upstream acoustic energy  and core $k^-$ irrotational energy ratio. The red and blue curves correspond to the two interaction families identified from the shock-cell wavenumber analysis.}
	\label{fig_energy_ratios}
\end{figure}

\fref{fig_energy_ratios} shows the energy repartition as a function of $M_j$, and highlights the modes that interact with the first and second wavenumber peaks of the shock cell structures, respectively $k_{s_1}$ and $k_{s_2}$. The KH energy in the core, shown in \fref{fig_E_KH_full}, represents the largest part of the total modal energy for all the modes. For the $k_{s_1}$ family, this ratio decreases when $M_j$ increases. The A1 mode at $M_j=1.1$ is almost entirely dominated by the KH wavepacket, while the B mode at $M_j=1.45$ has a smaller KH contribution. Indicating that at higher under-expansion, a larger fraction of the modal energy is distributed to the irrotational parts of the perturbation. For the $k_{s_2}$ family, it remains almost constant between A2 and C.

The core $k^-$ energy, shown in \fref{fig_E_km_full}, has the opposite trend for the $k_{s_1}$ family. It increases with $M_j$, showing that the guided jet mode becomes relatively more energetic inside the jet core when moving from A1 to B modes. This is consistent with the idea that the upstream-propagating component becomes more visible at higher under-expansion. For the $k_{s_2}$ family, the core $k^-$ energy remains lower, even for the C mode. 

This point is visible in \fref{fig_E_ac_full}, where the farfield upstream acoustic energy is reported. This contribution is weak for the $k_{s_1}$ family, even if it slightly increases with $M_j$. On the contrary, it is much stronger for the $k_{s_2}$ family, and especially for the C mode at $M_j=1.45$. This confirms the observation made from the filtered contours: the C mode has an upstream-propagating irrotational field that is more spread outside the jet core.

Finally, \fref{fig_E_ac_km} compares the farfield upstream acoustic energy with the core $k^-$ energy. For the $k_{s_1}$ family, this ratio remains below one, meaning that the upstream-propagating irrotational energy is mainly located in the core. For the $k_{s_2}$ family, the ratio is larger than one and becomes very large for the C mode. This shows that, for the modes of the $k_{s_2}$ family, the energy associated with the upstream acoustic field is mainly found outside the core rather than in the guided part of the jet.

The jet Mach-number dependence of the energy partition between the characteristic waves of the feedback loop shows two distinct scenarios, depending on the wavenumber of the shock-cell structure considered. For the mode of the first family ($k_{s_1}$ modes), a redistribution of the fluctuating kinetic energy from the solenoidal K-H wave to the dilatational wave content is observed as the Mach number is increased. This is consistent with compressibility effects on shear-layer instability growth rate. For this family of modes, the relative kinetic energy of the $k^-$ wave increases accordingly. Note that the majority of the dilatational kinetic energy in the jet core is not contained in the $k^-$ wave but rather in the shock region (reflecting oscillations of the shocks) and downstream acoustic. By comparison, the relative far-field upstream acoustic kinetic energy is not affected by the Mach number. The scenario is completely different for the second family modes ($k_{s_2}$ modes). In these cases, both the K-H and the $k^-$ relative kinetic energy content are unaffected by the increase of Mach number, while the far-field upstream acoustic kinetic energy significantly increases with it.

While the energy partition evolution for the first-family modes may be expected, the second scenario is less classical. These energy-partition scenarios complement the discussion in the literature, which is largely limited to the modes' origins (staging). It provides insights that may impact future understanding of receptivity mechanisms at the nozzle lips and shocks (not addressed in the present study), which are necessary to anticipate observed acoustic emission in jets.

\section{Conclusion}
Three-dimensional stability analyses are performed on turbulent compressible supersonic jet flows for different configurations of $M_j$, within an URANS framework using the $k-\log(\omega)$ SST turbulence model. Linearised URANS equations are obtained with the automatic differentiation tool \texttt{Tapenade} coupled with the generative computational framework \texttt{dNami}. This methodology is used to study and characterise the resonance behaviour of the screech tone in under-expanded jet flows with a computationally efficient linear framework. A time-stepping Newton--Krylov method is used to compute the fixed-point solutions of the URANS system of equations, which compare well with available experimental data in terms of shock-cell spacing, shock strength and dominant shock-cell wavenumbers. Global stability analyses are then performed over these fixed points using a time-stepping Krylov--Schur method.

The screech behaviour is captured by the linear analysis. The eigenvalues correctly reproduce the experimental acoustic frequency of screech within the variability of the available measurements, while the associated eigenvectors provide detailed information about the spatial distribution and azimuthal symmetry of the perturbation over the fixed-point solution. The analysis recovers the main families observed experimentally, namely the axisymmetric $m=0$ A modes and the $m=\pm 1$ B and C modes. In this sense, the global modes do not only provide a frequency prediction, but also a coherent spatial description of the acoustic and hydrodynamic structures involved in the resonance.

By applying a Fourier transform in the streamwise direction, the downstream-propagating Kelvin--Helmholtz wavepacket, the positive-wavenumber interaction wave and the negative-wavenumber guided jet mode are recovered from the stability modes. Their wavenumbers and spatial distributions are in good agreement with the structures extracted from experimental POD data, confirming that the linear modes reproduce the main wave content of the screech feedback mechanism. This comparison is particularly relevant because it validates not only the predicted screech frequency, but also the physical nature of the associated eigenmodes.

The same spectral validation also shows that the computed modes recover the staging behaviour observed experimentally. The upstream-propagating wave associated with the modes belonging to the first screech family is linked to the interaction $k^+_{kh}-k_{s_1}$ between the Kelvin--Helmholtz wave and the first dominant shock-cell wavenumber. In contrast, the modes belonging to the second family are associated with the interaction $k^+_{kh}-k_{s_2}$. The agreement between these interactions, the spectral content of the URANS fixed point and the experimental interpretation of screech staging confirms that the global stability analysis captures the correct wave-selection mechanism.

A Helmholtz decomposition of the velocity perturbation is then used to separate the vortical and dilatational parts of the global modes. The solenoidal component reproduces most of the spatial structure of the full velocity perturbation and is mainly associated with the downstream Kelvin--Helmholtz wavepacket. The dilatational component contains the compressible part of the perturbation, including the shock-cell modulation, the guided jet mode and the upstream acoustic field. Filtering the decomposed fields in streamwise wavenumber space makes it possible to reconstruct separately the vortical KH wavepacket and the dilatational $k^-$ wave, without mixing hydrodynamic and acoustic contributions.

The corresponding energy budget gives a quantitative description of the feedback loop. The solenoidal contribution is dominant for all the modes, confirming that most of the kinetic energy of the global perturbation is contained in the downstream vortical wavepacket. The dilatational contribution is smaller, but it contains the part of the mode that is directly related to the upstream acoustic closure of the loop. The core $k^-$ energy and the farfield upstream acoustic energy show different trends depending on the screech stage. For the modes associated with the first shock-cell wavenumber $k_{s_1}$, the upstream-propagating dilatational energy remains mainly located inside the jet core. For the modes associated with the second shock-cell wavenumber, and especially for the C mode at $M_j=1.45$, a larger part of the acoustic energy is found in the farfield acoustic region, rather then in the $k^-$ wave. The proposed energy-partition scenarios, based on Mach-number variation, broaden the existing literature, which primarily focuses on the origins of the modes. These findings offer new perspectives that could inform future studies of receptivity mechanisms at nozzle lips and shocks—key factors for anticipating acoustic emissions in jets.

Although the flow features necessary for screech to occur are nonlinear in nature, such as shock waves and turbulent fluctuations, the present results show that the underlying resonance mechanism is captured by a linear model built around a URANS fixed point. The use of URANS is advantageous because it provides fixed-point solutions in which strong shocks can be accurately resolved at high Reynolds number and high Mach number. This is particularly useful for configurations in which linear stability analysis around experimental or mean-flow data is difficult to apply. At the same time, the real part of the eigenvalues remains sensitive to the turbulence modelling, because the eddy-viscosity introduces a dissipative mechanism that can affect whether the feedback loop is self-sustained. The frequency selection and the wave structure, however, remain robust across the cases considered.

The proposed framework therefore provides a fully three-dimensional, matrix-free methodology for studying screech in turbulent jets. Even though the present configuration is axisymmetric, the method does not rely on this simplification and can be applied to more complex geometries. The results show that global stability analysis of URANS fixed points can recover not only the screech frequencies and mode shapes, but also the wave interactions, the staging mechanism and the energy repartition of the feedback loop.

	\section*{Acknowledgements}
	This work was granted access to the HPC resources of TGCC and IDRIS under the allocations 2024-[A0152A06362] and 2025-[A0172A06362] made by GENCI.
	This research work has been supported by the DGAC through the MAMBO project. 
	We thank Drs. Grégoire Pont and Jérôme Huber for the fruitful discussions. A.F. thanks Dr. N.Ciola for the interesting discussions. 
	Declaration of Interests. The authors report no conflict of interest.
	
	\appendix

	\section{\texorpdfstring{$\boldsymbol{k-\log(\omega)}$ SST model}{k-omega SST model}}\label{sec:komega}
The $k\text{-}\omega$ model~\citep{menter1994} is a two-equation turbulence model widely used in modern RANS calculations. The model has undergone several modifications over the years; the version considered here is the $k\text{-}\omega$ SST model~\citep{menter2003}. In the present work, an additional modification is applied by solving for the variable
\begin{equation}
	\widehat{\omega} = \log(\omega),
\end{equation}
in order to avoid large gradients of $\omega$ near the wall, which may lead to numerical errors. This formulation is commonly referred to as the $k\text{-}\log(\omega)$ SST model.

The transport equations for $k$ and $\widehat{\omega}$ then read:
\begin{equation}\label{eq:k_eq}
	\frac{\partial (\overline{\rho} k)}{\partial t}
	+ \frac{\partial (\overline{\rho} \widetilde{u}_j k)}{\partial x_j}
	= P - \overline{\rho} \beta^* e^{\widehat{\omega}} k
	+ \frac{\partial}{\partial x_j}
	\left[
	\left( \mu + \sigma_k \mu_t \right)
	\frac{\partial k}{\partial x_j}
	\right]
\end{equation}

\begin{multline}
	\frac{\partial (\overline{\rho} \widehat{\omega})}{\partial t}
	+ \frac{\partial (\overline{\rho} \widetilde{u}_j \widehat{\omega})}{\partial x_j}
	=
	\frac{\gamma}{\nu_t} e^{-\widehat{\omega}} P
	- \beta \overline{\rho} e^{\widehat{\omega}} \\
	+ \frac{\partial}{\partial x_j}
	\left[
	\left( \mu + \sigma_{\omega} \mu_t \right)
	\frac{\partial \widehat{\omega}}{\partial x_j}
	\right]
	+ 2(1-F_1)\frac{\overline{\rho}\sigma_{\omega_2}}{e^{\widehat{\omega}}}
	\frac{\partial k}{\partial x_j}
	\frac{\partial \widehat{\omega}}{\partial x_j} \\
	+ \left( \mu + \sigma_{\omega} \mu_t \right)
	\frac{\partial \widehat{\omega}}{\partial x_j}
	\frac{\partial \widehat{\omega}}{\partial x_j}
	\label{eq:omega_eq}
\end{multline}
The term
\[
\left( \mu + \sigma_{\omega} \mu_t \right)
\frac{\partial \widehat{\omega}}{\partial x_j}
\frac{\partial \widehat{\omega}}{\partial x_j},
\]
arises from the variable transformation and is present only in the $k\text{-}\log(\omega)$ formulation.
The remaining model terms are defined as follows:
\begin{equation}\label{eq:production}
	\tau_{ij} =
	\mu_t
	\left(
	2 S_{ij}
	- \frac{2}{3}
	\frac{\partial \widetilde{u}_k}{\partial x_k}
	\delta_{ij}
	\right)
	- \frac{2}{3} \overline{\rho} k \delta_{ij},
\end{equation}

\begin{equation}\label{eq:production_limiter}
	\widetilde{P} = \tau_{ij} \frac{\partial \widetilde{u}_i}{\partial x_j},
	\qquad
	P = \min \left( \widetilde{P},
	10 \beta^* \overline{\rho} e^{\widehat{\omega}} k \right),
\end{equation}

\begin{equation}\label{eq:S_ij}
	S_{ij} = \frac{1}{2}
	\left(
	\frac{\partial \widetilde{u}_i}{\partial x_j}
	+ \frac{\partial \widetilde{u}_j}{\partial x_i}
	\right),
	\qquad
	S = \sqrt{2 S_{ij} S_{ij}} .
\end{equation}
The turbulent eddy viscosity is computed as
\begin{equation}
	\mu_t =
	\frac{\overline{\rho} a_1 k}
	{\max \left( a_1 e^{\widehat{\omega}}, S F_2 \right)} .
	\label{eq:mu_t}
\end{equation}
Each constant without a numerical subscript is obtained by blending an inner (subscript 1) and outer (subscript 2) value according to
\begin{equation}
	\phi = F_1 \phi_1 + (1 - F_1) \phi_2 .
	\label{eq:blend}
\end{equation}
The blending and auxiliary functions are defined as
\begin{equation}
	F_1 = \tanh \left( \text{arg}_1^4 \right),
	\label{eq:F1}
\end{equation}

\begin{equation}
	\text{arg}_1 =
	\min \left[
	\max \left(
	\frac{\sqrt{k}}{\beta^* e^{\widehat{\omega}} d},
	\frac{500 \nu}{d^2 e^{\widehat{\omega}}}
	\right),
	\frac{4 \overline{\rho} \sigma_{\omega_2} k}
	{\text{CD}_{k\omega} d^2}
	\right],
	\label{eq:arg1}
\end{equation}

\begin{equation}
	\text{CD}_{k\omega} =
	\max \left(
	2 \overline{\rho} \sigma_{\omega_2}
	\frac{\partial k}{\partial x_j}
	\frac{\partial \widehat{\omega}}{\partial x_j},
	10^{-10}
	\right),
	\label{eq:CD_komega}
\end{equation}

\begin{equation}
	F_2 = \tanh \left( \text{arg}_2^2 \right),
	\label{eq:F2}
\end{equation}

\begin{equation}
	\text{arg}_2 =
	\max \left(
	2 \frac{\sqrt{k}}{\beta^* e^{\widehat{\omega}} d},
	\frac{500 \nu}{d^2 e^{\widehat{\omega}}}
	\right).
	\label{eq:arg2}
\end{equation}

The boundary conditions recommended in the original references and adapted to the $k\text{-}\log(\omega)$ formulation are
\begin{equation}
	\log\!\left(\frac{U_{\infty}}{L}\right)
	< \widehat{\omega}_{\text{farfield}}
	< \log\!\left(10 \frac{U_{\infty}}{L}\right),
	\label{eq:omega_farfield}
\end{equation}

\begin{equation}
	\frac{10^{-5} U_{\infty}^2}{Re_L}
	< k_{\text{farfield}}
	< \frac{0.1 U_{\infty}^2}{Re_L},
	\label{eq:k_farfield}
\end{equation}

\begin{equation}
	\widehat{\omega}_{\text{wall}} =
	\log \left(
	10 \frac{6 \nu}{\beta_1 (\Delta d_1)^2}
	\right),
	\label{eq:omega_wall}
\end{equation}

\begin{equation}
	k_{\text{wall}} = 0.
	\label{eq:k_wall}
\end{equation}

Here, $L$ denotes the approximate length of the computational domain. The combination of the farfield values should yield a free-stream turbulent viscosity between $10^{-5}$ and $10^{-2}$ times the free-stream laminar viscosity; consequently, the farfield turbulence boundary conditions retain some degree of flexibility.
The model constants are
\begin{equation}
	\gamma_1 = \frac{5}{9}, \quad
	\sigma_{k1} = 0.85, \quad
	\sigma_{\omega 1} = 0.5, \quad
	\beta_1 = 0.075,
	\label{eq:constants1}
\end{equation}

\begin{equation}
	\gamma_2 = 0.44, \quad
	\sigma_{k2} = 1.0, \quad
	\sigma_{\omega 2} = 0.856, \quad
	\beta_2 = 0.0828,
	\label{eq:constants2}
\end{equation}

\begin{equation}
	\beta^* = 0.09, \quad
	\kappa = 0.41, \quad
	a_1 = 0.31.
	\label{eq:constants3}
\end{equation}

\subsection{Compressibility correction}

The $k\text{-}\omega$ turbulence model may fail to predict the experimentally observed reduction in spreading rate of compressible mixing layers in supersonic flows~\citep{wilcox1992}. In~\cite{sarkar1991}, the total dissipation $\epsilon$ is expressed as a function of the turbulent Mach number $M_t$,
\begin{equation}
	M_t^2 = \frac{2k}{c^2},
\end{equation}
where $c$ is the speed of sound. Following~\cite{wilcox1992}, this correction is incorporated into the $k\text{-}\omega$ model by modifying the coefficient $\beta^*$ as
\begin{equation}
	\beta^* = \beta^*_0 \left( 1 + \xi^* F(M_t) \right),
\end{equation}
with
\begin{equation}
	\xi^* = 1, \qquad F(M_t) = M_t^2 .
\end{equation}

Here, $\beta^*_0 = 0.09$ denotes the value of the coefficient in the unmodified $k\text{-}\omega$ model. This compressibility correction is adopted in the present study, as it has been shown to outperform the baseline formulation for supersonic shear-layer flows~\citep{sarkar1991}.
	\section{Sensitivity study}\label{sec:sensitivity}

To explore the sensitivity of the spectra of the stability analysis problem to the URANS model, the two-dimensional planar screech problem has been studied using two different turbulence models: the negative Spalart--Allmaras model \citep{Spalart1992,Allmaras2012} and the $k-\log(\omega)$ SST turbulence model described in \secref{sec:komega}. 

\begin{figure}
	\centering
	\begin{subfigure}{0.49\linewidth}
		\caption{}\label{spectrum_sa}
		\includegraphics[width=\linewidth,trim=0 0 0 0,clip]{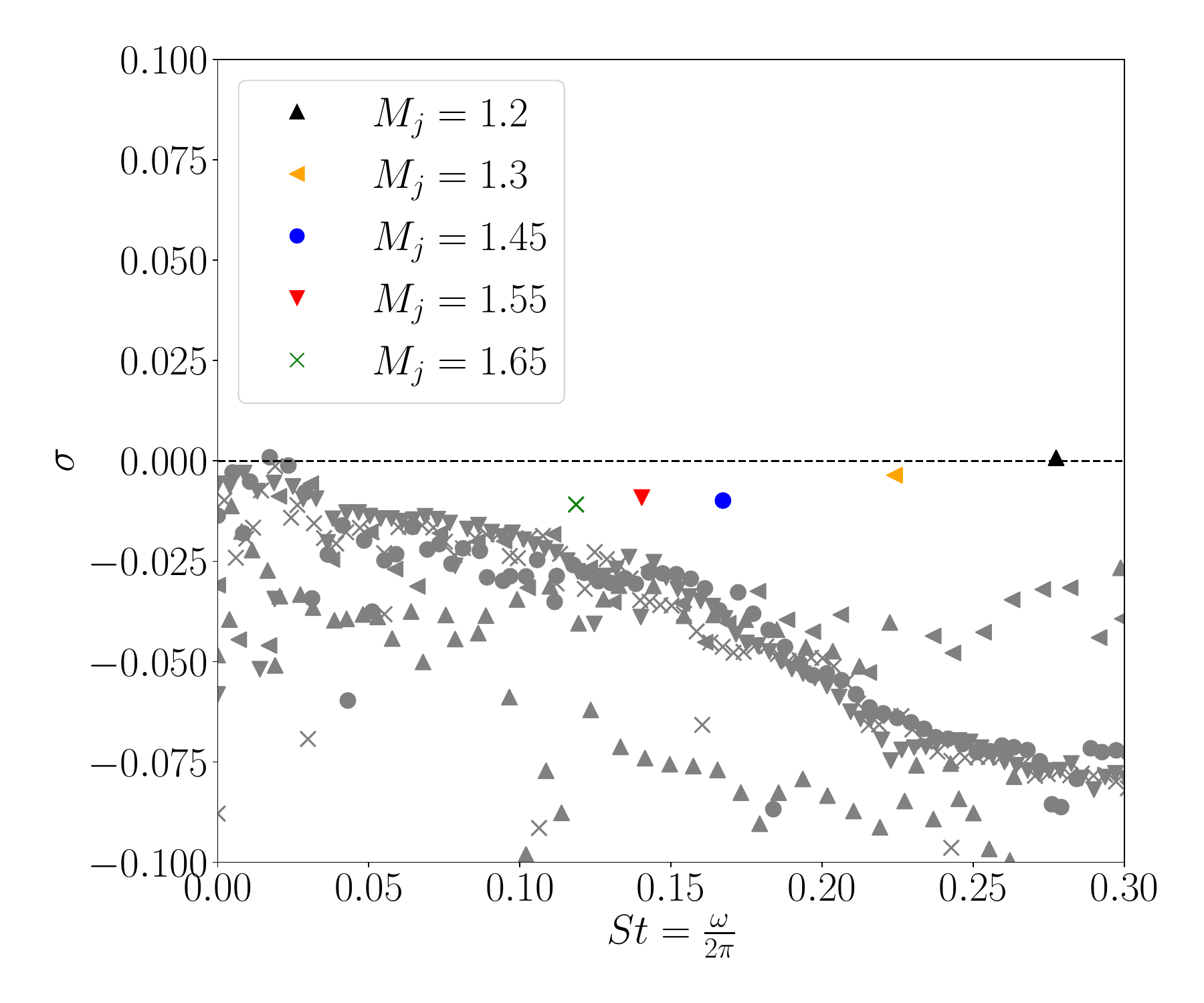}
	\end{subfigure}
	\begin{subfigure}{0.49\linewidth}
		\caption{}\label{spectrum_kom}
		\includegraphics[width=\linewidth,trim=0 0 0 0,clip]{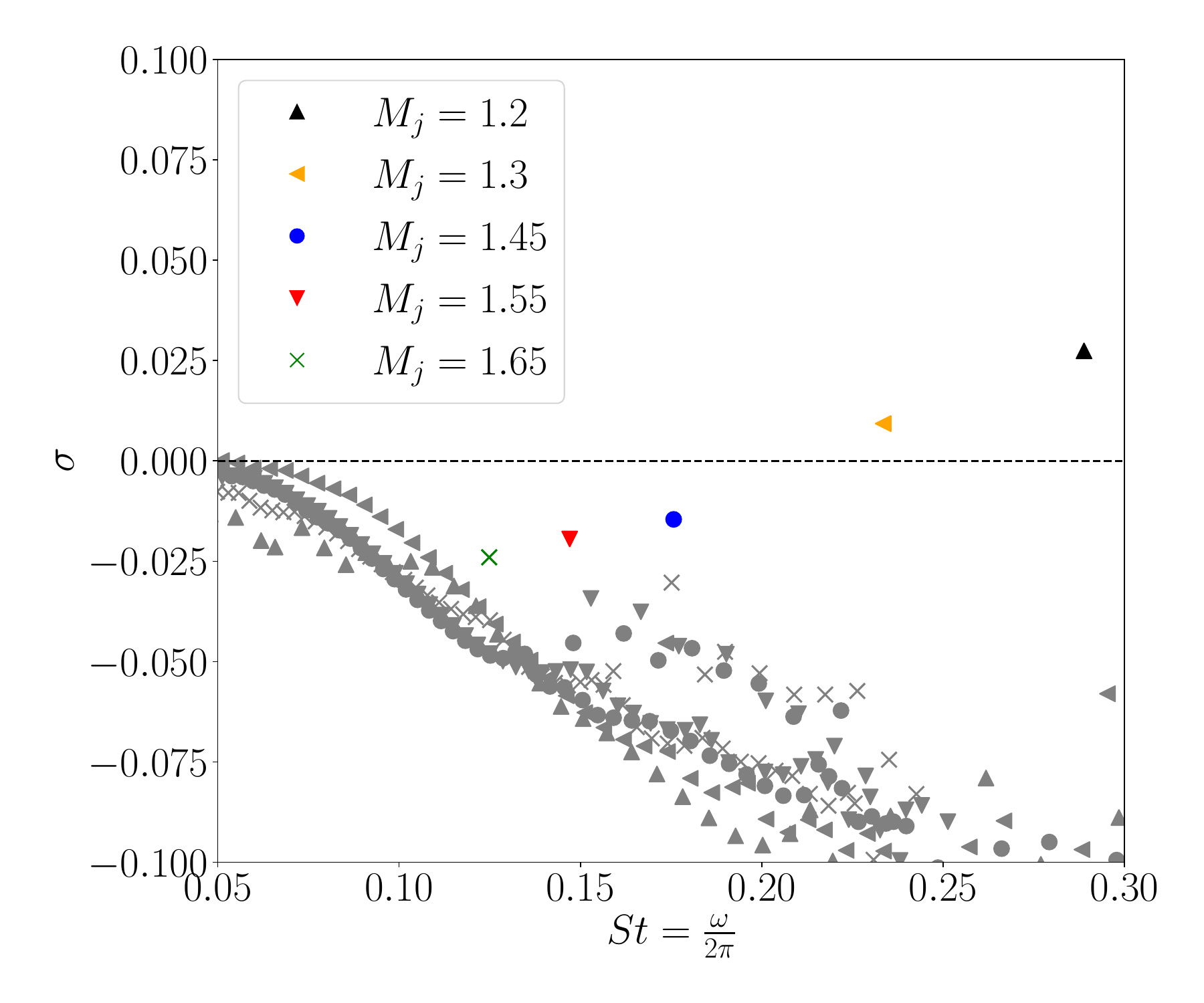}
	\end{subfigure}
	\caption{Global stability analysis spectra for different $M_j$ configurations. (a) Spalart--Allmaras model. (b) $k-\log(\omega)$ SST model. Grey eigenvalues are poorly converged: $\|\epsilon\| > 10^{-4}$.}\label{spectra_sa_kom_MJ}
\end{figure}

\fref{spectra_sa_kom_MJ} shows a comparison between the spectra computed with the two models. The computations are performed using identical numerical schemes, Krylov parameters, and grids for both turbulence models.

The two spectra exhibit a similar overall structure. Convective branches of poorly converged eigenvalues are present, together with a series of well-converged isolated modes for each $M_j$ configuration. In the spectrum computed using the Spalart--Allmaras turbulence model, only one unstable mode is observed, corresponding to the case $M_j = 1.2$.

\begin{figure}
	\centering
	\begin{subfigure}{0.48\linewidth}
		\caption{}\label{compare_sigma_kom_sa}
		\includegraphics[width=\linewidth,trim=0 0 0 0,clip]{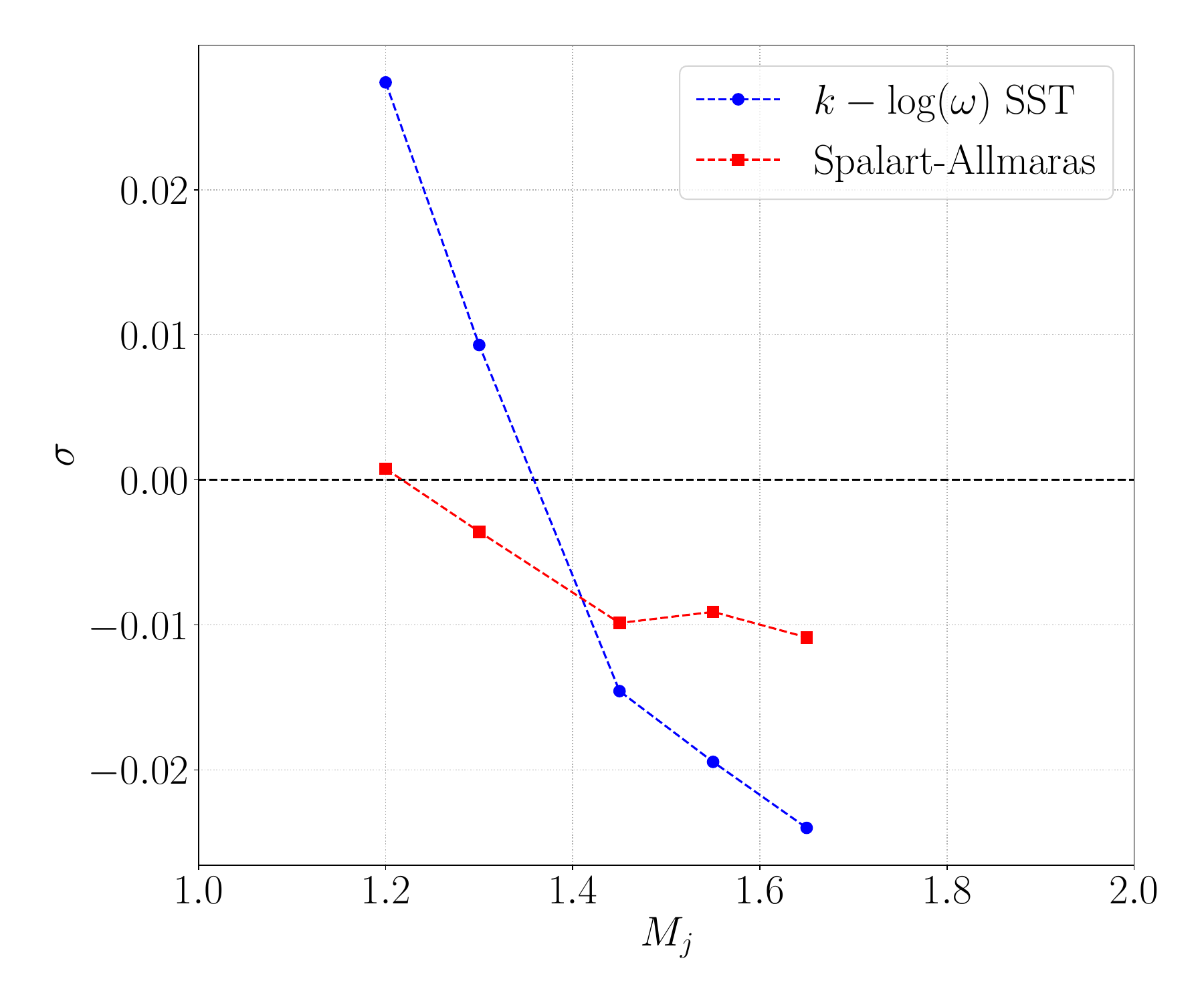}
	\end{subfigure}
	\begin{subfigure}{0.48\linewidth}
		\caption{}\label{compare_st_kom_sa}
		\includegraphics[width=\linewidth,trim=0 0 0 0,clip]{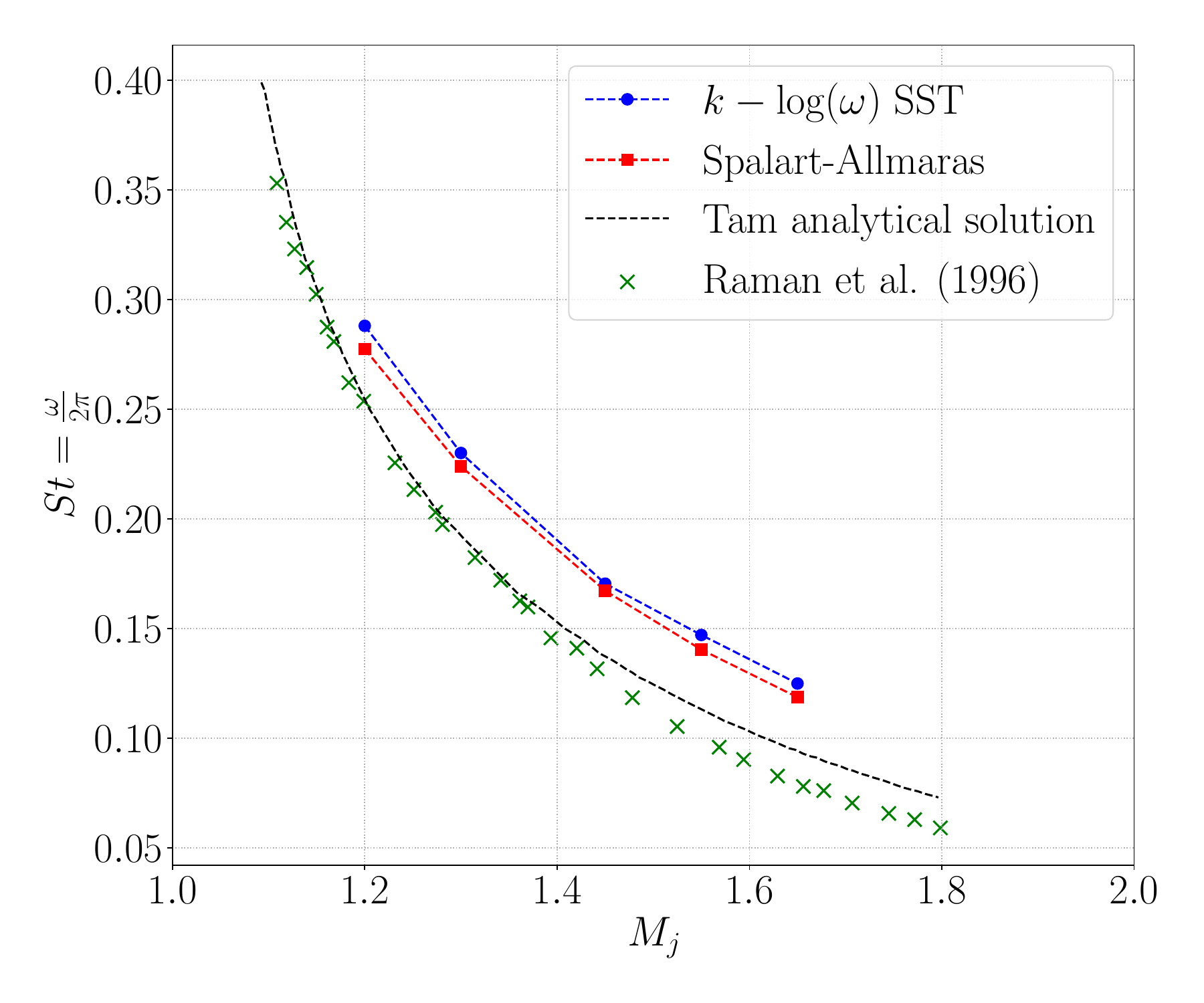}
	\end{subfigure}
	\caption{(a) Comparison of the real part of the isolated eigenvalue $\sigma$ for both turbulence models. (b) Comparison of the Strouhal number of the isolated eigenvalue for both turbulence models, experimental data \citep{raman1997}, and Tam's analytical estimate.}
	\label{compare_isolated}
\end{figure}

In \fref{compare_sigma_kom_sa}, a comparison of the growth rates of the isolated eigenvalues for each $M_j$ configuration is shown for the two turbulence models. The dependence of the growth rate on $M_j$ is much steeper for the $k-\log(\omega)$ SST model, which predicts a significantly higher growth rate for the unstable configuration, but smaller growth rates for the stable cases. \fref{compare_st_kom_sa} shows that the turbulence model has only a minor impact on the predicted frequency of the mode, with both models accurately capturing the instability frequency.

To further investigate the differences observed in the stability analysis, the case $M_j = 1.3$ is examined, for which the $k-\log(\omega)$ SST model predicts instability, while the Spalart--Allmaras model yields a stable solution.

\begin{figure}
	\centering
	\begin{subfigure}{0.49\linewidth}
		\caption{}\label{compare_mut_fp_Mj_1.3}
		\includegraphics[width=\linewidth,trim=0 0 0 0,clip]{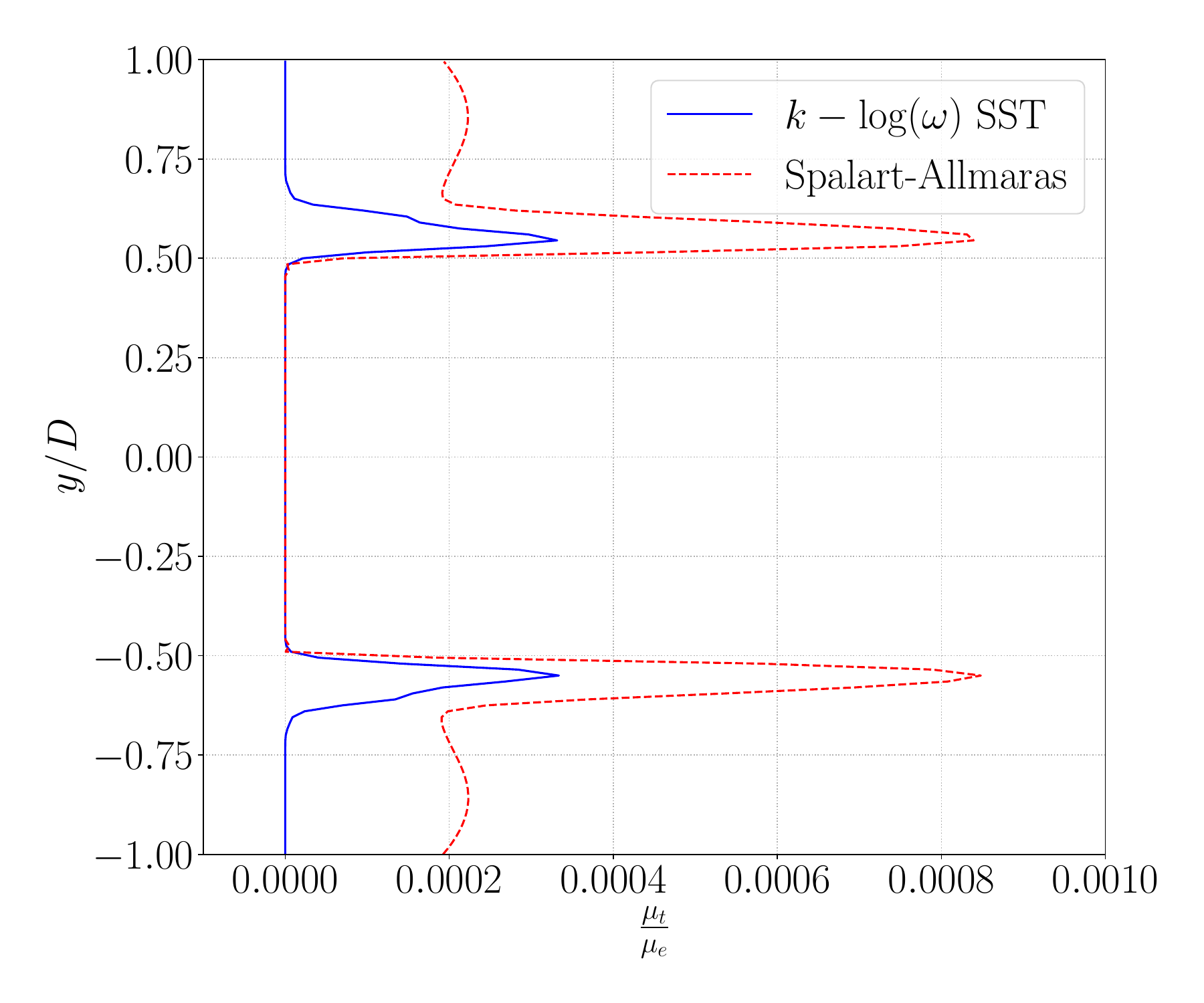}
	\end{subfigure}
	\begin{subfigure}{0.49\linewidth}
		\caption{}\label{compare_pressure_fp_Mj_1.3}
		\includegraphics[width=\linewidth,trim=0 0 0 0,clip]{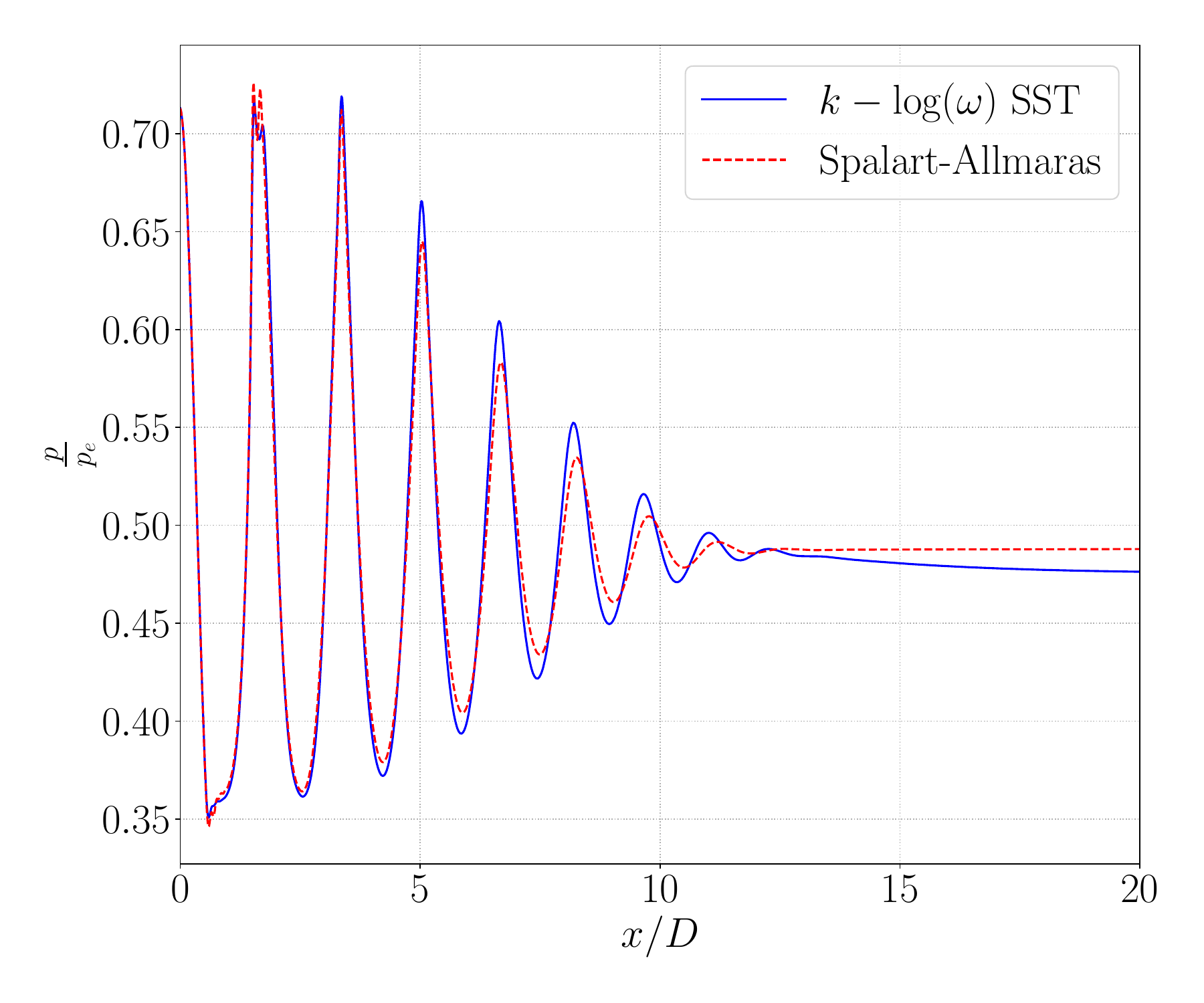}
	\end{subfigure}
	\caption{(a) Comparison of the turbulent eddy-viscosity $\mu_t$ at $x/D = 0.25$ for $M_j = 1.3$ using the two turbulence models. (b) Comparison of the normalised pressure distribution along the jet axis ($y/D = 0$) for $M_j = 1.3$.}
	\label{compare_fp_Mj_1.3}
\end{figure}

\fref{compare_pressure_fp_Mj_1.3} shows the pressure distribution along the jet centreline for $M_j = 1.3$ for both turbulence models. The two curves overlap closely for the first two shock cells, while further downstream the Spalart--Allmaras model exhibits stronger dissipation of the pressure peaks. Nevertheless, the shock locations remain in good agreement between the two models.

This observation helps explain the good agreement in the predicted frequency, since the screech frequency is known to be strongly dependent on shock spacing \citep{Tam1986}. Furthermore, experimental evidence suggests that sound generation occurs approximately at the fourth shock, and that shock strength can influence the self-sustained nature of screech \citep{EdgingtonMitchell2019}.

Another crucial element of the screech feedback loop is the receptivity mechanism at the shear layer near the nozzle lip. In this region, the upstream-propagating wave excites the shear layer, converting its energy into a downstream-propagating wave and thereby closing the feedback loop. \fref{compare_mut_fp_Mj_1.3} shows a comparison of the turbulent eddy viscosity near the nozzle at $x/D = 0.25$. The Spalart--Allmaras model tends to overestimate the eddy-viscosity levels in this region, which may lead to excessive dissipation and reduce both the energy carried by the upstream-propagating wave and the spatial amplification of the downstream-propagating wave. As a result, the instability may fail to sustain itself.

All these effects are likely interconnected and, when considered individually, cannot fully explain the variability observed in the real part of the isolated eigenvalues. Nevertheless, these results demonstrate that the linear information provided by global stability analysis can be a valuable tool for assessing the sensitivity of URANS solutions to turbulence modelling choices and model parameters.

\section{Helmholtz decomposition}
\label{app:helmholtz}

A Helmholtz decomposition is used in this work to separate the vortical and dilatational contributions of the perturbation velocity field. According to the Helmholtz theorem, a sufficiently smooth vector field can be decomposed into the sum of a divergence free component $ \mathbf{u}_{A}  $ and a curl free component $ \mathbf{u}_{\phi} $  \citep{schoder2020postprocessing,sharma2023effect}. The velocity perturbation is therefore written as
\begin{equation}
	\mathbf{u} = \mathbf{u}_{A} + \boldsymbol{u}_{\phi},
\end{equation}
Since these two fields satisfy:
\begin{equation}
	\nabla \cdot \mathbf{u}_{A} = 0,
	\qquad
	\nabla \times \boldsymbol{u}_{\phi} = \mathbf{0}.
\end{equation}
There exist one scalar potential $\phi$ and one vector potential $\mathbf{A}$ such that:
\begin{equation}
	\mathbf{u}_{A} = \nabla \times \mathbf{A} ,
	\qquad
	\mathbf{u}_{\phi} = \nabla \phi
\end{equation}
The solenoidal component is associated with the vortical part of the perturbation, while the dilatational component contains the compressible part of the velocity field.
Taking the divergence of the decomposition and using $\nabla \cdot \mathbf{u}_A=0$ gives the Poisson equation
\begin{equation}\label{app:poisson}
	\nabla^2 \phi = \nabla \cdot \mathbf{u}.
\end{equation}
The source term of the Poisson problem is therefore the divergence of the velocity field. Once the scalar potential has been obtained, the dilatational velocity is reconstructed from $\boldsymbol{u}_{\phi}=\nabla \phi$, and the solenoidal contribution is then obtained by subtraction,
\begin{equation}
	\mathbf{u}_{A} = \mathbf{u} - \mathbf{u}_{\phi} .
\end{equation}
The Possion problem is solved by iterate the discretised one-dimensional linear problem associated with the Poisson equation \ref{app:poisson}.
Following the formulation proposed by \citet{schoder2020postprocessing}, homogeneous Neumann boundary conditions are imposed on solid walls,
\begin{equation}
	\frac{\partial \phi}{\partial n} = 0,
\end{equation}
while homogeneous Dirichlet boundary conditions are imposed at the inlet and outlet boundaries,
\begin{equation}
	\phi = 0.
\end{equation}
The Poisson equation is solved on the same grid used for the stability analysis. 
The boundary condition are well posed in this case, given that the global mode has compact support. The sponge layers imposes the perturbation to be zero in the outlet boundaries, and the supersonic inlet condition imposes a null velocity perturbation. This boundary value problem is solved in dNami, with a GMRES method for solving the discrete linear solver, with a residual of $\varepsilon_{\text{GMRES}} < 10^{-6}$.  providing a decomposition of each global mode into a vortical part $\mathbf{u}_A$, and an dilatational part $\mathbf{u}_{\phi}$.

The numerical implementation of the Helmholtz decomposition was validated by measuring the degree of orthogonality between the dilatational and solenoidal components of the reconstructed velocity field. This was quantified through the cross term
\begin{equation}
	C_{\phi A} =
	\left\langle \mathbf{u}_{\phi},\mathbf{u}_A \right\rangle ,
\end{equation}
where $\langle \cdot,\cdot\rangle$ denotes the $L^2$ inner product used in the energy calculation. The magnitude of this term is compared with the total velocity energy,
\begin{equation}
	E_u = \left\lVert \mathbf{u} \right\rVert_2^2 .
\end{equation}
For an exact continuous Helmholtz decomposition, the solenoidal and dilatational components are orthogonal under the chosen boundary conditions \citep{schoder2020postprocessing}. Therefore, the quantity $\left| C_{\phi A} \right|/E_u$ provides a direct measure of the residual numerical coupling between the two components.

\begin{table}
	\centering
	\caption{Absolute value of the cross term between the dilatational and solenoidal components, together with the same quantity normalised by the total velocity energy.}
	\begin{tabular}{c|c|c|c}
		Mode & $M_j$ & $\left| C_{\phi A} \right|$ & $\left| C_{\phi A} \right| / E_u$ \\
		\hline
		A1 & 1.1 & $3.876 \times 10^{-10}$ & $1.571 \times 10^{-4}$ \\
		\hline
		A2 & 1.2 & $6.484 \times 10^{-9}$ & $1.906 \times 10^{-3}$ \\
		B & 1.2 & $5.895 \times 10^{-8}$ & $9.226 \times 10^{-3}$ \\
		\hline
		B & 1.25 & $1.885 \times 10^{-7}$ & $2.403 \times 10^{-2}$ \\
		\hline
		B & 1.45 & $4.682 \times 10^{-7}$ & $3.512 \times 10^{-2}$ \\
		C & 1.45 & $8.170 \times 10^{-8}$ & $6.261 \times 10^{-3}$ \\
		\hline
	\end{tabular}
	\label{tab:helmholtz_validation}
\end{table}

The values reported in \tabref{tab:helmholtz_validation} show that the cross term between $\mathbf{u}_{\phi}$ and $\mathbf{u}_A$ remains small compared with the total velocity energy for all the modes considered. The residual contribution is below a few percent in all cases, indicating that the numerical decomposition preserves the expected orthogonality between the dilatational and solenoidal fields with sufficient accuracy for the present analysis.

\newpage

\bibliographystyle{jfm}
\bibliography{bibliography}

\end{document}